\documentclass{aa}  
\usepackage{graphicx}

\usepackage{txfonts}

\usepackage[T1]{fontenc}
\usepackage{graphicx}
\usepackage{amsmath}
\usepackage{amssymb}
\usepackage{xcolor}
\usepackage[normalem]{ulem}

\newcommand{\dynb}{\texttt{DynBaS}}
\newcommand{\tgas}{\texttt{TGASPEX}}
\newcommand{\logMt}{$\log (M/M_\odot)$}
\newcommand{\logMl}{$\log (M_*/M_\odot)$}
\newcommand{\mwa}{$<\log age (yr)>_{M}$}
\newcommand{\lwa}{$<\log age (yr)>_{L}$}
\newcommand{\mwz}{$<[Z/Z_\odot]>_{M}$}
\newcommand{\lwz}{$<[Z/Z_\odot]>_{L}$}
 
\begin{document} 

\title{J-PLUS: Detecting and studying extragalactic globular clusters - the case of NGC\,1023}
\titlerunning{GC candidates through J-PLUS filters}

\authorrunning{D. de Brito Silva et al.}

   \author{Danielle de Brito Silva \inst{1}
          \and Paula Coelho \inst{2}
          \and Arianna Cortesi \inst{3}
          \and Gustavo Bruzual \inst{4}
          \and Gladis Magris C. \inst{5}
          \and Ana L. Chies-Santos\inst{6,7}
          \and Jose A. Hernandez-Jimenez \inst{8}
          \and Alessandro Ederoclite \inst{2,9}
          \and Izaskun San Roman \inst{9}
          \and Jes\'us Varela \inst{9}
          \and Duncan A. Forbes \inst{10}
          \and Yolanda Jim\'enez-Teja \inst{11}
          \and Javier Cenarro \inst{9}
          \and David Crist\'obal-Hornillos \inst{9}
          \and Carlos Hern\'andez-Monteagudo \inst{12,13}
          \and Carlos L\'opez-Sanjuan \inst{9}
          \and Antonio Mar\'in-Franch \inst{9}
          \and Mariano Moles \inst{9}
          \and H\'ector V\'azquez Rami\'o \inst{9}
          \and Renato Dupke  \inst{14}
          \and Laerte Sodr\'e Jr.  \inst{2}
          \and Raul E. Angulo \inst{15}
          }

   \institute{N\'ucleo de Astronom\'ia, Universidad Diego Portales, Ej\'ercito 441, Santiago, Chile\\
              \email{danielle.debrito@mail.udp.cl}
         \and
             Universidade de S\~ao Paulo, Instituto de Astronomia, Geof\'isica e Ci\^encias Atmosf\'ericas, R. do Mat\~ao 1226, S\~ao Paulo, SP, 05508-090, Brazil
          \and
             Observat\'orio do Valongo, Universidade Federal do Rio de Janeiro, Ladeira do Pedro Ant\^onio 43, Rio de Janeiro, RJ, 20080-090 Brazil
          \and
             Instituto de Radioastronom\'ia y Astrof\'isica, Universidad Nacional Aut\'onoma de M\'exico, Morelia, Michoac\'an, 58089, M\'exico
           \and Centro de Investigaciones de Astronom\'ia (CIDA), M\'erida, 5101, Venezuela
          \and Instituto de F\'isica, Universidade Federal do Rio Grande do Sul, Av. Bento Gonçalves 9500, Porto Alegre, R.S. 90040-060, Brazil
          \and Shanghai Astronomical Observatory, Chinese Academy of Sciences, 80 Nandan Road, Shanghai 200030, China
           \and Universidade do Vale do Para\'iba, Av. Shishima Hifumi, 2911, S\~ao Jos\'e dos Campos, SP, 12244-000, Brazil
         \and
             Centro de Estudios de F\'isica del Cosmos de Arag\'on, Unidad Asociada al CSIC, Plaza San Juan 1, 44001 Teruel, Spain.
         \and Centre for Astrophysics \& Supercomputing, Swinburne University, Hawthorn VIC 3122, Australia
         \and Instituto de Astrof\'isica de Andaluc\'ia–CSIC, Glorieta de la Astronom\'ia 
s/n, E–18008 Granada, Spain
         \and Instituto de Astrof\'isica de Canarias, Calle V\'ia L\'actea SN, ES38205 La Laguna, Spain
         \and Departamento de Astrof\'isica, Universidad de La Laguna, ES38205, La Laguna, Spain
         \and Observat\'orio Nacional - MCTI (ON), Rua Gal. Jos\'e Cristino 77, S\~ao Crist\'ov\~ao, 20921-400, Rio de Janeiro, Brazil
         \and Ikerbasque, Basque Foundation for Science, E-48013 Bilbao, Spain
             }

   \date{Received ; accepted}
 
  \abstract
   {Extragalactic globular clusters (GCs) are key objects for studying the history of galaxies. The arrival of wide-field surveys such as the Javalambre Photometric Local Universe Survey (J-PLUS) offers new possibilities for the study of these systems.}
   {We perform the first study of GCs in J-PLUS to recover information about the history of NGC\,1023 taking advantage of wide-field images and 12 filters.}
   {We develop the semiautomatic pipeline \texttt{GCFinder} that detects GC candidates in J-PLUS images and can also be adapted to similar surveys. We study the stellar population properties of a sub-sample of GC candidates using spectral energy distribution (SED) fitting.}
   {We find 523 GC candidates in NGC\,1023, of which  $\sim$300 are new.
   We identify subpopulations of GC candidates, where age and metallicity distributions have multiple peaks.
   By comparing our results with simulations, we report a possible broad age-metallicity relation, evidence that NGC\,1023 experienced accretion events in the past. The dominating age peak is at $10^{10}$ yr. We report a correlation between masses and ages that suggests that massive GC candidates are more likely to survive the turbulent history of the host galaxy. Modeling the light of NGC\,1023, we find two spiral-like arms and detect a displacement of the galaxy's photometric center  with respect to the outer isophotes and center of GC distribution ($\sim$700\,pc and $\sim$1600\,pc, respectively), which could be the result of ongoing interaction between NGC\,1023 and NGC\,1023A.}
   {By studying the GC system of NGC\,1023 with J-PLUS we showcase the power of multi-band surveys for this kind of study and find evidence of a complex accretion history of the host galaxy.}

\keywords{galaxies: star clusters: general -- galaxies: individual (NGC\,1023) -- surveys}

\maketitle

\section{Introduction}

Globular Clusters (GCs) are ubiquitous compact stellar systems found in most galaxies with stellar masses M$_{star}$ > 10$^{6.8}$ M$_\odot$ (e.g. \citealt{eadie2022clearing}).
Some of these objects are among the oldest objects in the universe \citep{Larsen2001} with typical ages larger than 10\,Gyr \citep[]{strader2005extragalactic, chies2011optical}. The investigation of GCs can shed light on how galaxies form and evolve through time, since these objects can be used to study galaxy assembly, star formation history, and galaxy chemical evolution, among other topics \citep{Brodie2006, beasley20}. As discussed in  \cite{Brodie2006}, the typical mass of GCs is between 10$^{4}$ and 10$^{6}$ M$_\odot$ and the size of the GC population in a galaxy is a function of galaxy luminosity, ranging from none to a few in dwarf galaxies up to more than 10000 in cD galaxies \citep{alamo2017specific}.

A well-described 
property of the globular cluster population of a massive galaxy is its optical color bimodality, showing that there are subpopulations of this class of objects in most massive galaxies \citep{peng2006acs}.
The bimodality in GC colors is believed to occur due to differences in metallicities. However, age effects and a combination of age and metallicities effects might also play an important role
\citep{Brodie2006,lee2018nonlinear}. It might also be important to take into account non-linear effects in the color-metallicity relations brought by the horizontal-branch morphology in the optical-bands \citep{richtler2005some,yoon2006explaining,cantiello2007metallicity,yoon2011nonlinearb,chung2016nonlinear,lee2018nonlinear, villaume2019new,lee2020nonlinear,kim2021nonlinear}.
Spectroscopic studies have
shown that the blue subpopulations of GCs are more metal-poor than the red populations \citep{beasley20082df,usher2012sluggs}. From chemical evolution models of galaxies as well as from observations, it is known that in dwarf irregular galaxies and low mass galaxies GCs tend to be metal-poor and blue \citep{lotz2004colors}.
The fact that   
most galaxies tend to have sub-populations of GCs can be explained by a hierarchical formation: to form a massive system, many small systems are merged throughout time. 
We note that \textit{optical/NIR} colors of GC candidates do not have such bimodal distribution in most galaxies, except for NGC 3115, which seems bimodal (\citealt{brodie2012},\citealt{Cantiello2014}) in any color and metallicity studied (see \citealt{cantiello2007metallicity,Blakeslee2012,Cantiello2014,Cho2016}).

Considering how much colors, ages, and metallicities of GCs are important to understanding the assembly of galaxies and their evolution, investigating new photometric bands and colors as well as the interaction of new colors with stellar models and libraries can be interesting to build more detailed spectral energy distributions (SEDs) for these systems.
The Javalambre Photometric Local Universe 
Survey\footnote{\url{www.j-plus.es}} (J-PLUS; \citealt{cenarro2019j}) 
operates with a set of 5 broad-band filters based on SDSS (\citealt{york2000sloan,strauss2002spectroscopic}) and 7 narrow--band filters that cover the main stellar indices from 370 to 900 nm ([OII], Ca H+K,D4000, H$\delta$, Mgb, H$\alpha$ and CaT). 
This filter set makes it possible to study GCs with novel colors and more detailed SEDs.  
Other surveys that can also add new colors to the study of extragalactic GCs are J-PAS (Javalambre Physics of the Accelerating Universe Astrophysical Survey; \citealt{benitez2014j}) and S-PLUS (Southern Photometric Local Universe Survey; \citealt{mendes2019southern}). J-PAS is composed of 56 narrow--band filters in the optical, while S-PLUS has similar properties to J-PLUS, employing a twin filter system.

To explore the J-PLUS filter set to study extragalactic GCs, we use as a test case the galaxy NGC\,1023. NGC\,1023 is a SB0 galaxy, located at 11.1\,Mpc away \citep{brodie2014sages}, with an effective radius of 48$\arcsec$ \citep{dolfi2021sluggs}. This galaxy is located at the center of a small group of galaxies \citep{Tully1980}, and it is currently undergoing a minor merger with NGC\,1023 A \citep{barbon75,hart80,capaccioli86}. NGC\,1023 is characterized by a complex and extended HI cloud, whose densest clump is associated with the companion galaxy \citep{Sancisi1984, Morganti2006}.  NGC\,1023 is consistent with being composed of a nearly classical bulge and a fast rotating disk, as extracted from its planetary nebulae system \citep{Noordermeer2008, Cortesi2011}. Its star cluster system has been explored before in the literature, including spectroscopy (e.g. \citealt{Larsen2001, Chies2013,Forbes2014}), and it presents complex kinematics, characterized by rotation in the inner disk-dominated region, and gradually turning into a pressure supported system in the outer halo-dominated part \citep{Cortesi2016}. The photometric studies trace the GC system  up to 8 effective radii \citep{Kartha2014} and the spectroscopic sample is selected from the photometric sample to maximize the construction of the MOS (Multi-Object Spectroscopy) masks, leading to a non-uniform catalog of GCs ages, metallicities and velocities.
 
NGC\,1023 was observed with the Javalambre Auxiliary Survey Telescope
(JAST80). The data was obtained with a Director’s Discretionary Time (DDT) proposal observed during science verification.
The Globular Cluster Luminosity Function (GCLF) peak is at M$_V$ = -7.5  \citep{harris2001globular}, which corresponds to V $\approx$ 22.7 at 11.1\,Mpc. J-PLUS reaches $g = 21.5$ with a signal-to-noise ratio $S/N = 5$  \citep{cenarro2019j}, allowing us to study the brightest part of the GCLF at these distances.
On the other hand, the observations of NGC\,1023 used in this article were not done with the standard exposure times of J-PLUS and we are able to also detect faint objects. In particular, we detected about 50\% of expected objects with magnitudes between the peak of GCLF and 1 sigma below it and no objects between 2 and 3 sigmas below the peak of GCLF.
As a result, our sample represents the majority of GC candidates expected for NGC\,1023.

In this work, we propose a methodology to detect and select GC candidates from images obtained with JAST80, aiming at exploiting data from J-PLUS, which can be easily adapted for other photometric surveys such as J-PAS and S-PLUS (see also \citealt{buzzo2022new} and \citealt{chies2022j}). With a catalog of GCs, we are set to investigate the stellar population content of GC candidates in NGC\,1023 to create an unbiased magnitude limited catalog.
This fact makes this galaxy an excellent case of study as we can explore new methodologies and colors, and compare such 
results to those in literature.

In Section \ref{sec:data} we describe the data used in this article. 
In Section \ref{sec:methods_gcfinder} we present our methodology to detect and select GC candidates, as well as our pipeline \texttt{GCFinder}. In Section \ref{sec:results} we show our results. In Section \ref{sec:discussion} we discuss our findings. Finally, in Section \ref{sec:conclusion} we present our conclusions.

\section{Data}\label{sec:data}

NGC\,1023 was observed in July 2017 through 
the DDT 
proposal 1600101 (P.I. Ana Chies Santos) using JAST80 and T80Cam, a 
the panoramic camera of 9.2k $\times$ 9.2k pixels that provides a $2\deg^2$ 
field of view (FoV) with a pixel scale of 0.55 arcsec pix$^{-1}$ \citep{t80cam}.

\begin{figure*}
\centering
\includegraphics[width=12cm]{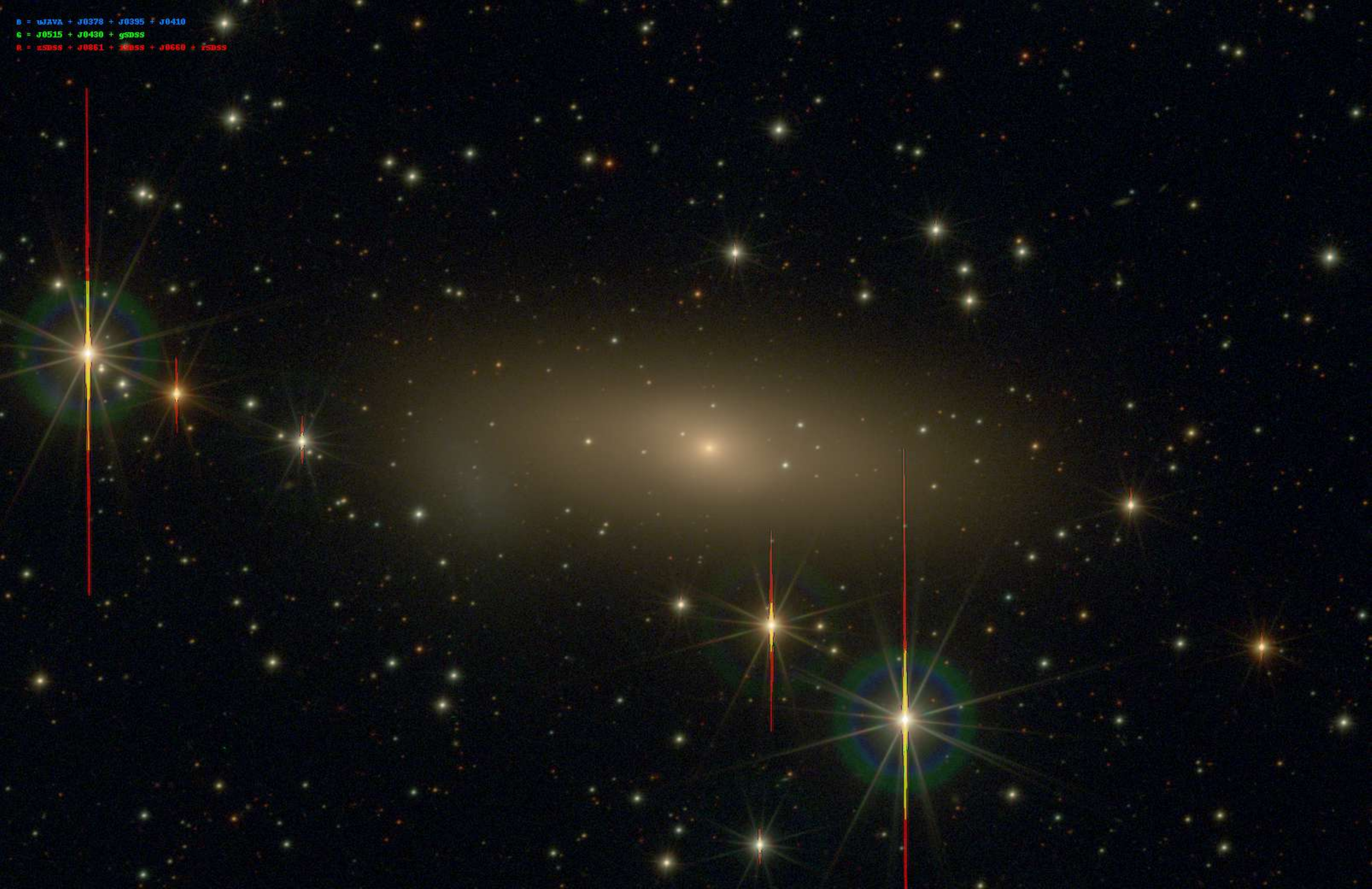}
\caption{J-PLUS color image of NGC\,1023 obtained with the software Trilogy \citep{Coe2012}. In order to  build the color image, the combination of filters ($u$, $J0378$, $J0410$),  ($J0430$, $J0515$, $g$) and ($J0660$, $J0861$, $r$, $i$, $z$) were used to compose the blue, green and red components, respectively. The FoV is $\approx$ 0.1 $\deg^2$.}  
\label{fig:NGC1023}
\end{figure*}

These data are not part of 
J-PLUS, but they were observed as part of the commissioning period to test the survey capabilities for extragalactic GC science.  
This galaxy was observed using all filters available at the JAST80 telescope, namely broad-bands $u$, $g$, $r$, $i$, $z$ and narrow--bands $J0378$, $J0395$, $J0410$, $J0430$, $J0515$, $J0660$, $J0861$. However, due to problems related to the calibration of the image referring to the $J0395$ band, we do not use this band in our work. The data are publicly available \footnote{\url{https://tacdata.cefca.es/application?id=101}}. The full width at half maximum (FWHM) is presented in Table \ref{tab:seeing}, as well as the exposure times. A color image of NGC\,1023 is presented in Figure \ref{fig:NGC1023}, built using the software Trilogy \citep{Coe2012}.

\begin{table}
\caption{\texttt{FWHM} values and exposure times for NGC\,1023 data in each filter. 
} 
\centering 
\begin{tabular}{l c c}
\hline\hline
Filter & \texttt{FWHM} (arcsec) & Exposure time (s)\\ \hline
$u$ & 1.181 $\pm$ 0.034  & 1257 \\
$J0378$ & 1.146  $\pm$ 0.044 & 1245\\
$J0410$ & 1.151  $\pm$ 0.025 & 231\\
$J0430$ & 1.174  $\pm$ 0.024 & 219\\
$g$ & 1.400  $\pm$ 0.043 & 651\\
$J0515$ & 1.094  $\pm$ 0.035 & 231\\
$r$ & 1.264  $\pm$ 0.034 & 764\\
$J0660$ & 1.050  $\pm$ 0.041 & 1311\\
$i$ & 1.085  $\pm$ 0.032 & 216\\
$J0861$ & 1.084  $\pm$ 0.031 & 651\\
$z$ & 0.997  $\pm$ 0.025 & 582\\
\hline
\label{tab:seeing}
\end{tabular}
\end{table}

To perform this study 
we crop the original wide-field images to a smaller region of approximately 2 000 X 1 500 pixels ($\approx$ 0.1 $\deg^2$) 
around NGC\,1023 using IRAF \citep{iraf1993} tasks.

The images have been reduced with the standard pipeline developed by the Data Processing and Archiving Unit (\textit{Unidad de Procesado y Archivo de Datos}; hereafter UPAD) of CEFCA. In short, the process includes a correction for bias, flat-field, fringing, handling contaminants (cosmic rays, satellite traces, calibrating astrometry, and combining individual images of the same region into deeper images \citep[for further details see section 2.5 in][]{cenarro2019j}.

The photometric calibration procedure adopted in J-PLUS (see \citealt{lopez2019j} for details) 
was not available at the time that our data was processed, and the photometric zero-points (ZPs) have been obtained separately. ZPs for the bands $g$, $r$, $i$, and $z$ were obtained from stars in the field observed with the Panoramic Survey Telescope and Rapid Response System (Pan-STARRS; \citealt{chambers2016pan}).
ZPs for the remaining bands were computed using standard and secondary stars in the field.
The bands $g$ and $r$ had the ZPs derived from both procedures, yielding small offsets of 0.087 and 0.069, respectively.  
Due to the impossibility of inferring the offsets to all bands we do not apply corrections to the magnitudes, using them as provided. These offsets act as adding a small difference between the broad and narrow--bands when SEDs are built, which we expect to cause an increase in $\chi^2$ when performing the SED fitting (see section \ref{sec:color_sed_results}).

In the current work, we assume that the field is locate at high galactic latitude (as in the case of J-PLUS) and thus is not significantly affected by extinction. Throughout this work, no correction for line-of-sight extinction was applied.  The line-of-sight extinction in the direction of NGC\,1023 is estimated to be E(B-V) = 0.052 (following \citealt{schlafly_finkbeiner11}).

\section{Methodology}\label{sec:methods_gcfinder}

We analyze the data presented in Section \ref{sec:data} in two processes, first by compiling a list of GC candidates around NGC\,1023, and then by analyzing their stellar population properties. 

\subsection{Detection of candidates with \texttt{GCFinder}}
To detect GC candidates in J-PLUS images, we develop the semiautomatic pipeline \texttt{GCFinder}. \texttt{GCFinder} consists of a python code that detects compact sources in a white image (a sum of frames of the 4 broad-bands $g$, $r$, $i$, and $z$) and performs a selection of GC candidates based on data quality, shape, color and magnitude criteria. The code uses Source Extractor \citep{Bertin1996} and Montage \citep{Berriman2004}, and run inside a support folder that also contains necessary files to use \texttt{GCFinder}.

A detailed description of \texttt{GCFinder} and technical requirements are  given in Appendices \ref{sec:methods_gcfinder_appendix} and \ref{sec:append}. In Appendix \ref{different_methods} we present the main different methodologies tested for the detection of GC candidates in J-PLUS data before choosing the strategy deployed in \texttt{GCFinder}.

An advantage of \texttt{GCFinder} is that it is a very light code that does not require the modeling of the host galaxy or smoothing filters before the detection of sources. The code is flexible and can be easily adapted to other photometric surveys besides J-PLUS, including the detection of other stellar systems such as Ultra Compact Dwarf Galaxies (UCDs, \citealt{phillipps2001ultracompact}). 

We run \texttt{GCFinder} on the data described in Section \ref{sec:data} and obtained a set of GC candidates, that are presented in Table \ref{tab:detections}. We note that a larger number of GC candidates are recovered in redder bands, in agreement with what is expected from old stellar objects. We also note that the S/N of blue bands, if compared to red bands tends to be lower in J-PLUS.

\begin{table}
\caption{Number of GC candidates selected per band after the selection criteria are applied.}
\centering 
\begin{tabular}{l c}
\hline\hline
Filter & Number of candidates \\ \hline
$u$ & 395 \\ 
$J0378$ & 383  \\
$J0410$ & 373  \\
$J0430$ & 397   \\
$g$ & 523   \\
$J0515$ & 450   \\
$r$ & 522   \\
$J0660$ & 516   \\
$i$ & 522   \\
$J0861$ & 505   \\
$z$ & 523   \\
\hline
\label{tab:detections}
\end{tabular}
\end{table}

To measure the detection efficiency, i.e. how many GCs the method can recover in the extended light region of the host galaxy's halo, we compare our detections with the catalog from \cite{Kartha2014}, used as reference. 

\begin{figure}
\centering
\includegraphics[width=9cm]{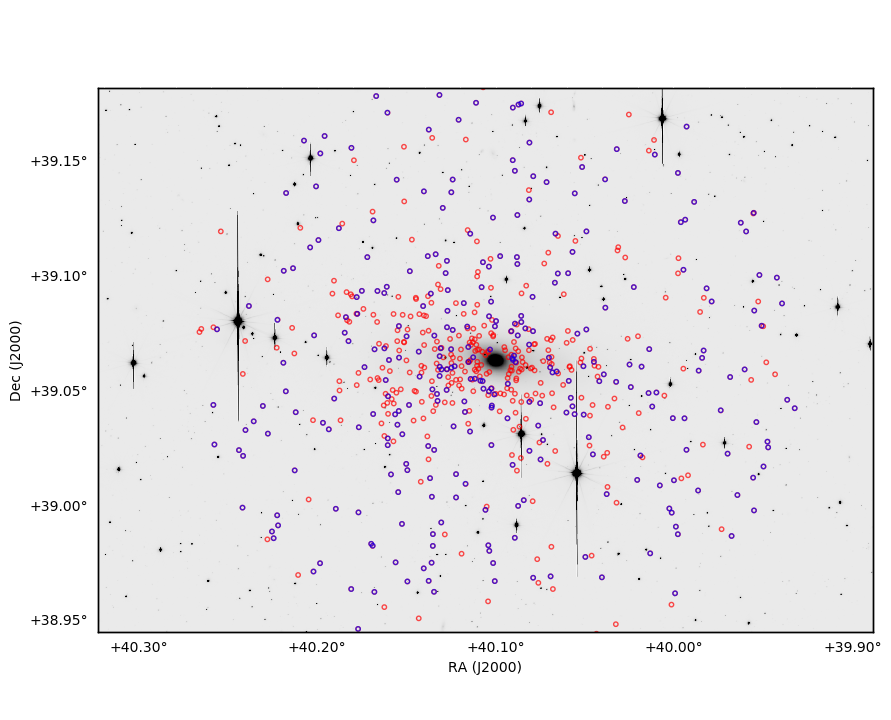}
\caption{In the background we show the white image of  NGC\,1023 used for the detection of sources. The FoV is $\approx$ 0.1 $\deg^2$. Red circles represent GC candidates from the reference catalog \citep{Kartha2014}. Blue circles represent GCs in common between \texttt{GCFinder} selection and the reference catalog.} 
\label{fig:detection}
\end{figure}

The NGC\,1023 reference catalog of \cite{Kartha2014} originally had 627 objects including faint sources. We note that the work done in \cite{Kartha2014} use data from MegaCam \citep{lenzen2003naos} at Canada-France-Hawaii Telescope (\texttt{CHFT}), which has a limiting magnitude fainter than T80Cam. In the $g$-band, we are able to detect objects up to a magnitude of around 24 mag and, for comparison purposes, we exclude from the reference catalog objects fainter than the limiting magnitude of our data.

This results in about 200 GC candidates that could be detected within the limiting magnitude of J-PLUS. 

Comparing the positions of our GC candidates with the reference catalog (see Figure \ref{fig:detection}), we retrieve 188 GCs in common which means that \texttt{GCFinder} detected a large fraction (about 93 \%) of possible GC candidates. 

Based on the numbers presented, we estimate that \texttt{GCFinder} does not detect about 7 \% 
of expected GCs when compared to the literature. 

We attribute this difference to the fact that \cite{Kartha2014} detect many globular clusters near the central region of the galaxy since its light central body hampers our ability to detect compact sources, even if modeled (see Appendix \ref{different_methods}).

On the other hand, we detect 335 new GC candidates with \texttt{GCFinder} due to: 1) new outer halo GCs were identified taking advantage of the wide field of J-PLUS images and 2) reduction in the  
contamination from Milky Way stars, which was possible from the adopted methodology (see Section \ref{sec:methods_gcfinder_appendix}). From these 335 new GC candidates, 48\% of them are located beyond the studied region in previous papers such as \cite{Kartha2014} and the rest of them are in the same halo region explored in the literature but are on average fainter than the GC candidates located at same region reported in \cite{Kartha2014} .

\subsection{Stellar population properties}\label{sec:methods_sed}

We obtain stellar population properties for our GC candidates via SED
fitting. We use the codes \dynb\  and \tgas\ \citep{magris+15,mejia+17}, adapted to work with the J-PAS, J-PLUS and S-PLUS filter systems as illustrated in \cite{gd+21} for mini-JPAS, where \tgas\ was used. Both \dynb\  and \tgas\ are non-parametric fitting codes, in which the star formation history is expressed as an arbitrary superposition of different simple stellar populations. 

This is the first time that the stellar population properties of GC candidates are derived from J-PLUS data. To investigate the effect of adding narrow--bands to the SED fitting, we perform the fits for 3 different combinations of filters: {\it (a)} using only the broad-band filters, {\it (b)} using only the narrow--band filters, and {\it (c)} using all available filters.

We use the version of the \citet{Bruzual2003} stellar population synthesis models described in \cite{plat2019}, C\&B models hereafter. The C\&B models 
follow the PARSEC evolutionary tracks \citep{marigo2013,chen2015} and use the MILES \citep{sanchez-blazquez2006,falcon-barroso2011,prugniel2011} and IndoUS \citep{valdes2004,sharma2016} stellar libraries in the spectral range covered by the J-PLUS data.
The C\&B models are available for 15 different metallicities ranging from $\log \left(Z_\star/Z_\odot \right) = -2.23$ to $0.55$
(using $Z_\odot$ = 0.017) and run in age from $\sim$\,0 to $14$\,Gyr. 
In this paper we discard C\&B models of super solar metallicity and use the following 12 metallicities:
$\log \left(Z_\star/Z_\odot \right) =  -2.23$, $-1.93$, $-1.53$, $-1.23$, $-0.93$, $-0.63$, $-0.45$, $-0.33$, $-0.23$, $-0.08$, $0$ and $0.07$.
All the models were computed for the \citet{chabrier2003} initial mass function. 

In \dynb\ \ and \tgas\ the {\it best}-the fitting solution is obtained by computing the non-negative values of the coefficients $x_{tZ}$ that minimize the merit function

\begin{equation}
\chi^2=\sum_{\lambda,tZ}^N \frac{[F^\mathrm{obs}_\lambda - \sum_{t,Z} x_{tZ}\  f_{\lambda, tZ}(\tau_V)]^2}{\sigma_\lambda^2}
\label{chi2}
\end{equation}

\noindent used to measure the goodness-of-fit. In Equation \ref{chi2}, $F^\mathrm{obs}_\lambda$ and $f_{\lambda,tZ}$ are the observed and model flux in each of the bands, respectively, and $\sigma_l^2$ is the corresponding uncertainty. The sum is done over all the filters (index $\lambda$) and the $N$ models (indices $t$ and $Z$).
$N$ is equal to the number of time steps $\times$ the number of metallicities used in the fit.

As described by \cite{magris+15}, in \dynb\  we use $N = 3$ (hereafter \dynb3), and minimize $\chi^{2}$ in Equation \ref{chi2} requiring that the $3$ derivatives
$\frac{\partial}{\partial x_{tZ}}\chi^{2} = 0$.
This results in a system of $3$ equations with $3$ unknowns that we solve using Cramer's rule for all possible combinations of $3$ model spectra.
The \dynb3 solution is then the one with the minimum $\chi^{2}$, subject to the condition $x_{tZ} \ge 0$.
In \tgas\ we use the non-negative least squares (NNLS) algorithm \citep{lawson1974} to find the vector $x_{tZ}$ that minimizes $\chi^2$.
In both codes, we use an outer loop to minimize by the dust attenuation $\tau_V$.

We remark that both \dynb3 and \tgas\ provide independent, deterministic solutions, as opposed to statistical solutions, to the SED fitting problem. The \tgas\ solution, in general, with $N \ge 3$, contains the \dynb3\ solution. As has been shown in
\cite{magris+15} and \cite{mejia+17}, the \dynb3\ and \tgas\ solutions are consistent within errors. \cite{gd+21} show that the \tgas\ solution is consistent with the solutions obtained by other SED fitting codes, including codes that use a fully Bayesian approach. Prieto et al. (2022, in prep) show that the deterministic \dynb3\ and \tgas\ solutions are consistent with those derived following a Bayesian treatment for both methods. In this paper, we opt for the deterministic approach for simplicity and because we know from the cited papers that the Bayesian approach does not add new insight into our problem.

Following \cite{gd+21}, we use the frequentist approach of Monte Carlo-ing the input (by adding Gaussian noise with observationally defined amplitudes) and repeating the fit many times ($\approx$\,1000) assuming that the errors in the different bands are uncorrelated, to perform a statistical analysis based on the observed photon-noise and its impact on the results.
A probability distribution function (PDF) for each stellar population property is then built by weighting the results from each iteration by the likelihood $\mathcal{L} \propto \exp(-\chi^2/2)$. The inferred value for the property of each GC is then obtained directly from the corresponding marginalized PDF. In the end, each population property is characterized by its {\it best} value derived directly by \dynb3 and \tgas, and the {\it mean}, the {\it median} and the {\it percentiles} defining the confidence interval in the distribution. The {\it best} value is then considered as the best estimate of the unknown {\it true} value, and its precision is obtained by averaging the precision determined for each cluster from the PDF built as indicated above. The following properties are given as result:

\begin{itemize}
\item \emph{Total stellar mass ($M$)}: Total mass of the stellar population. It is calculated directly from the mass converted into stars according to our solutions for the GC, reported as \logMt.

{\item \hspace{0.1cm}\emph{Luminous stellar mass ($M_\star$)}: Stellar-mass of the stellar population at present. It is calculated from the mass converted into stars reduced by the mass lost by stars during their evolution}, reported as \logMl.
 
{\item \emph{Age of the stellar population}: We define the mass-weighted logarithmic age (hereafter mass-weighted age) following \citet[][eq.\,9]{cid-fernandes2013} as
\begin{equation} 
\label{eq:at_mass}
\langle \log\ \mathrm{age} \rangle_\mathrm{M} = \sum_{t,Z}^N \mu_{tZ} \times \log t,
\end{equation}
\noindent where $\mu_{tZ}$ is the fraction of mass of the base element with age $t$ and metallicity $Z$, reported as \mwa.
Similarly, the light-weighted logarithmic age (hereafter light-weighted age) is defined as
\begin{equation} 
\label{eq:at_light}
\langle \log\ \mathrm{age} \rangle_\mathrm{L} = \sum_{t,Z}^N \mathcal{F}_{tZ} \times \log t,
\end{equation}
\noindent where $\mathcal{F}_{tZ}$ is the fraction of light in the $r$ filter corresponding to the base element with age $t$ and metallicity $Z$, reported as \lwa.}

{\item \emph{Metallicity of the stellar population}: We define the mass-weighted logarithmic metallicity (hereafter mass-weighted Z) as
\begin{equation} 
\label{eq:at_met}
\langle \log\ \mathrm{Z} \rangle_\mathrm{M} = \sum_{t,Z}^N \mu_{tZ} \times \log Z,
\end{equation}
\noindent reported as \mwz, and the light-weighted logarithmic metallicity (hereafter light-weighted Z) as
\begin{equation} 
\label{eq:at_light2}
\langle \log\ \mathrm{Z} \rangle_\mathrm{L} = \sum_{t,Z}^N \mathcal{F}_{tZ} \times \log Z,
\end{equation}
\noindent reported as \lwz.}

\end{itemize}

We define the precision (mean standard deviation, $\sigma_j$) for the 
$j^{th}$ stellar property as
\begin{equation}
<\sigma_j>=\frac{1}{2N}\sum_i^N(p84_{j,i} - p16_{j,i}), 
\label{eq:sigma}
\end{equation}
where $p84_{j,i}$ and $p16_{j,i}$ are, respectively, the percentiles 84 and 16 of the PDF of the $j^{th}$ property for the $i^{th}$ GC candidate. 

\begin{figure}
\centering
\includegraphics[width=7.3cm]{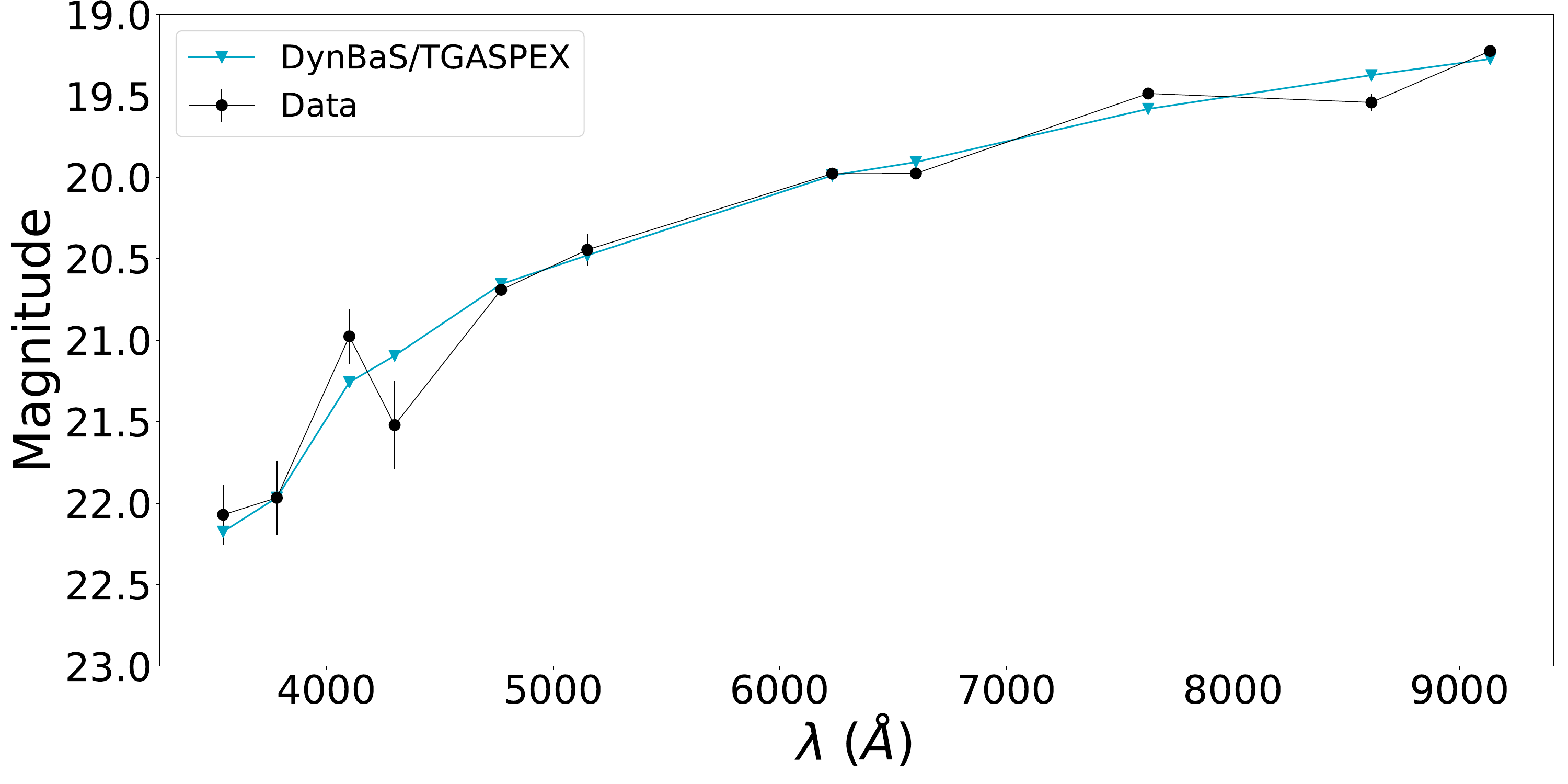}
\includegraphics[width=7.3cm]{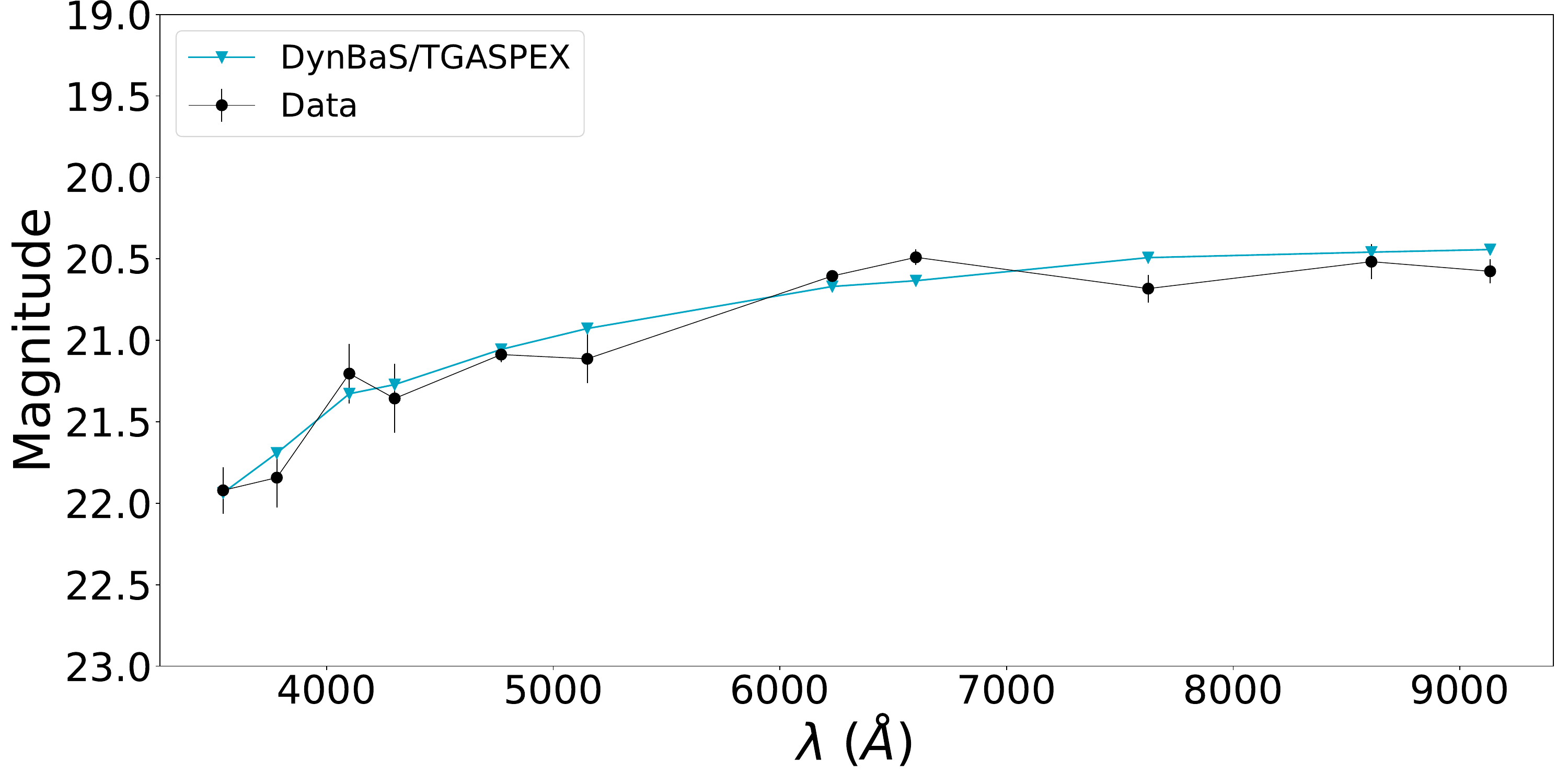}
\includegraphics[width=7.3cm]{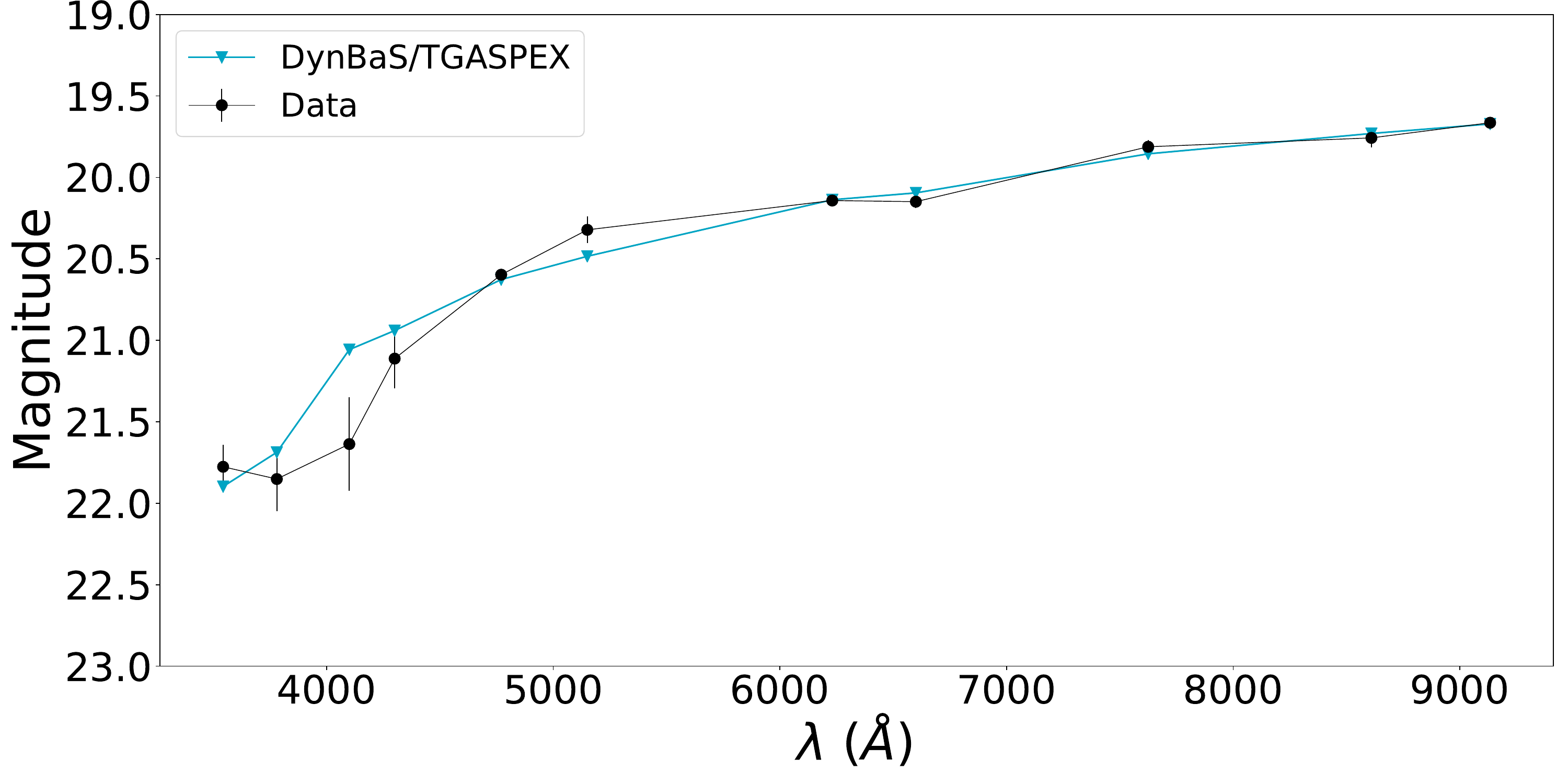}
\caption{Examples of SED fitting for 3 GC candidates. The observed and synthetic SEDs are presented in black and blue, respectively. The model fits of the two codes are very similar, therefore they appear superposed in the figure.}
\label{fig:sed_example}
\end{figure}

Example of SEDs chosen randomly are presented in Figure \ref{fig:sed_example} for illustration purposes.

\section{Results} \label{sec:results}

\subsection{Color distributions}\label{sec:photometry_methodology}

The study of GC systems in massive galaxies has shown that they can have a bimodal \textit{optical} color distribution (e.g. \citealt{larsen2001properties,peng2006acs,kundu2007bimodal}). At the same time, several studies have shown that the color--metallicity relation is highly non-linear (e.g. \citealt{yoon2006explaining},\citealt{kim2021nonlinear}).

\begin{figure*}
\centering
\includegraphics[width=5cm]{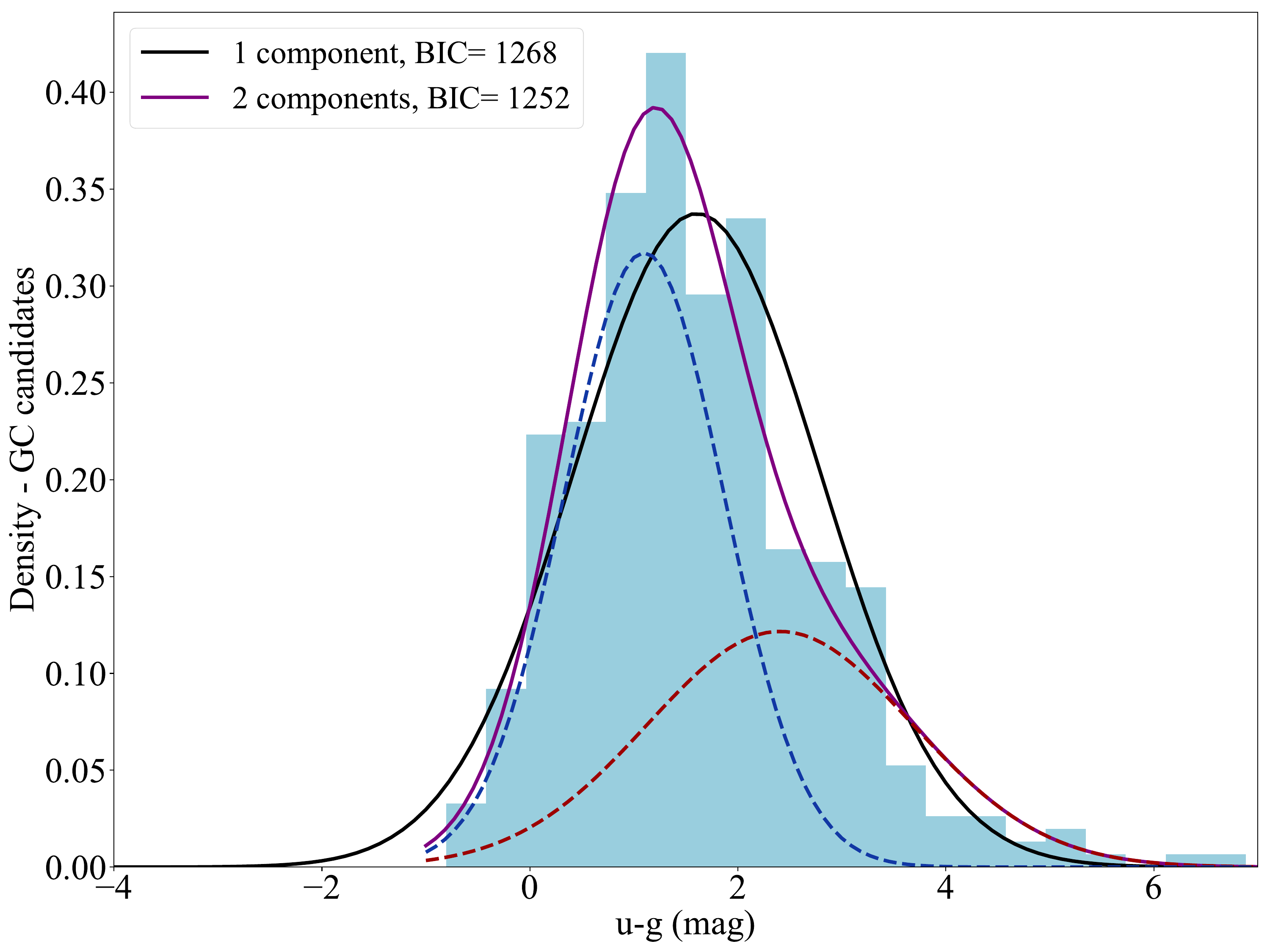}
\includegraphics[width=5cm]{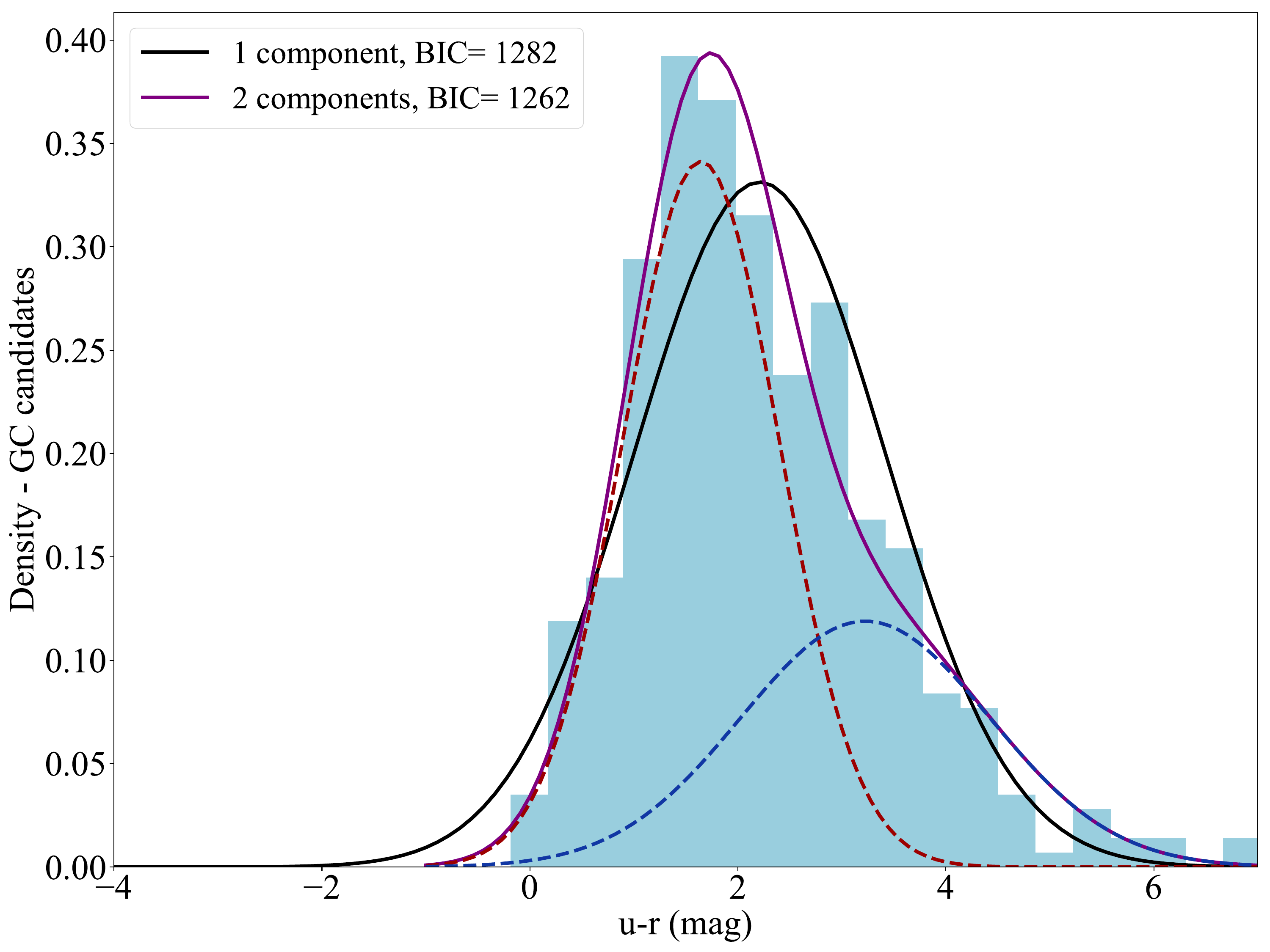}
\includegraphics[width=5cm]{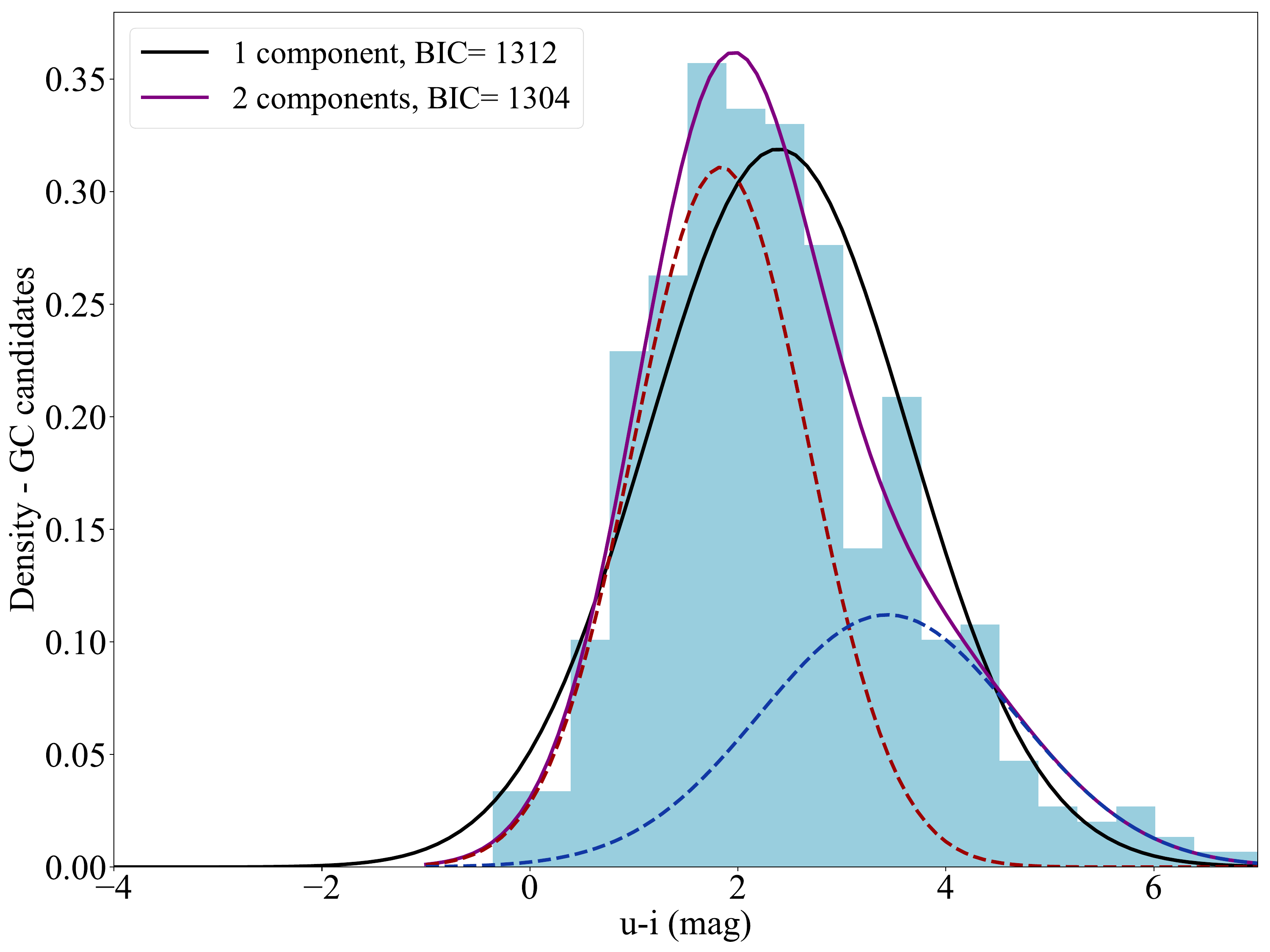}\\
\includegraphics[width=5cm]{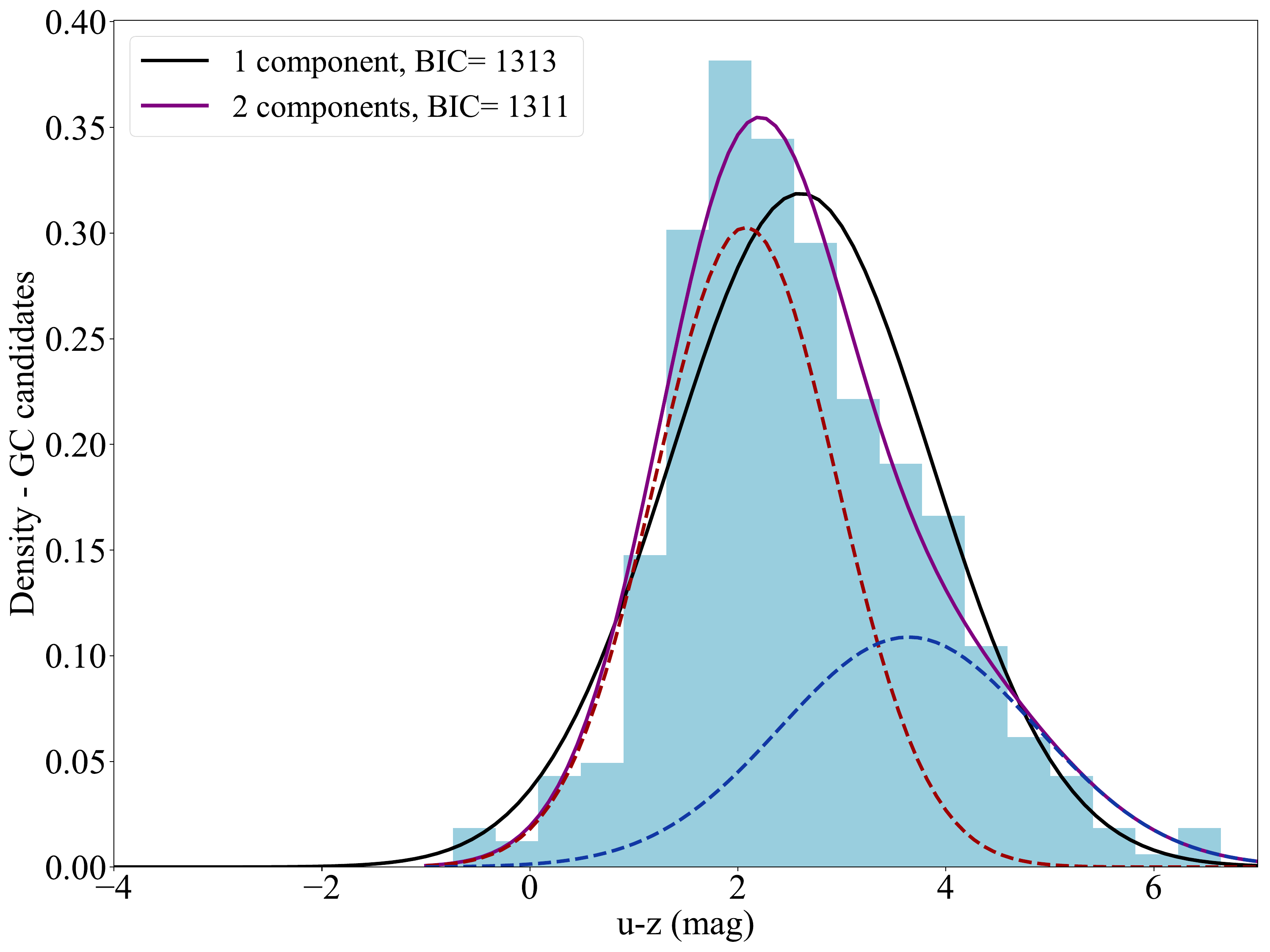}
\includegraphics[width=5cm]{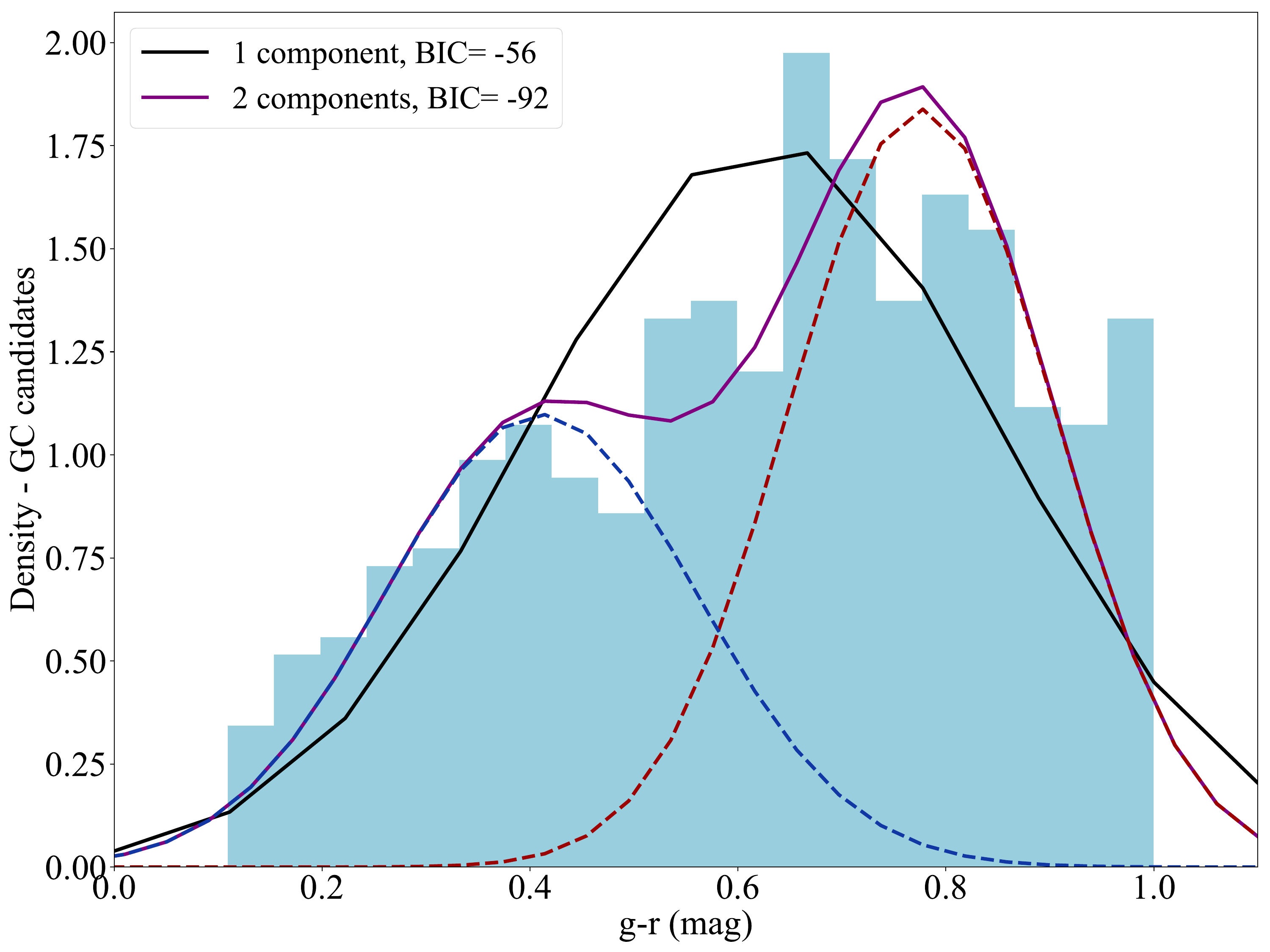}
\includegraphics[width=5cm]{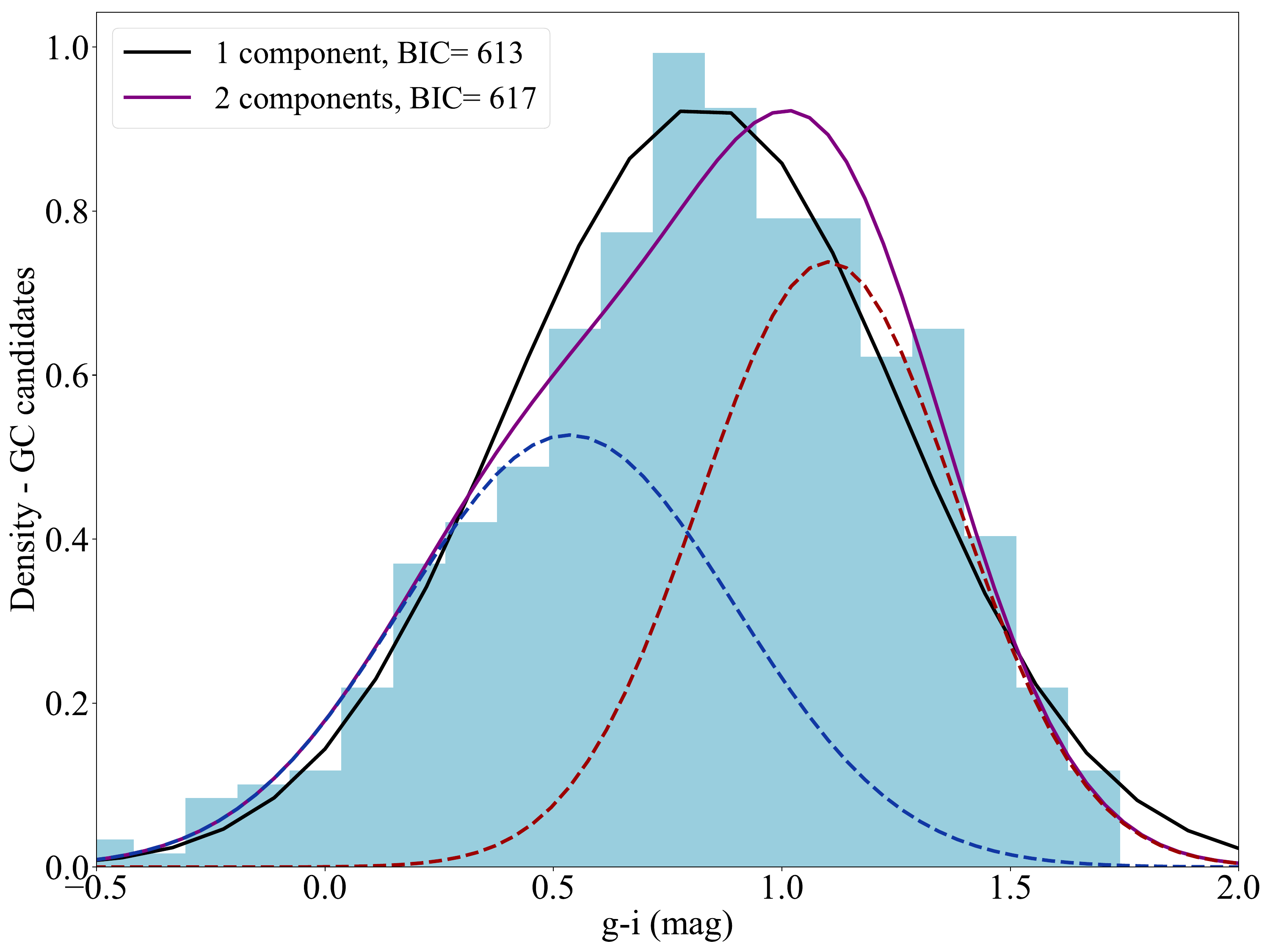}\\
\includegraphics[width=5cm]{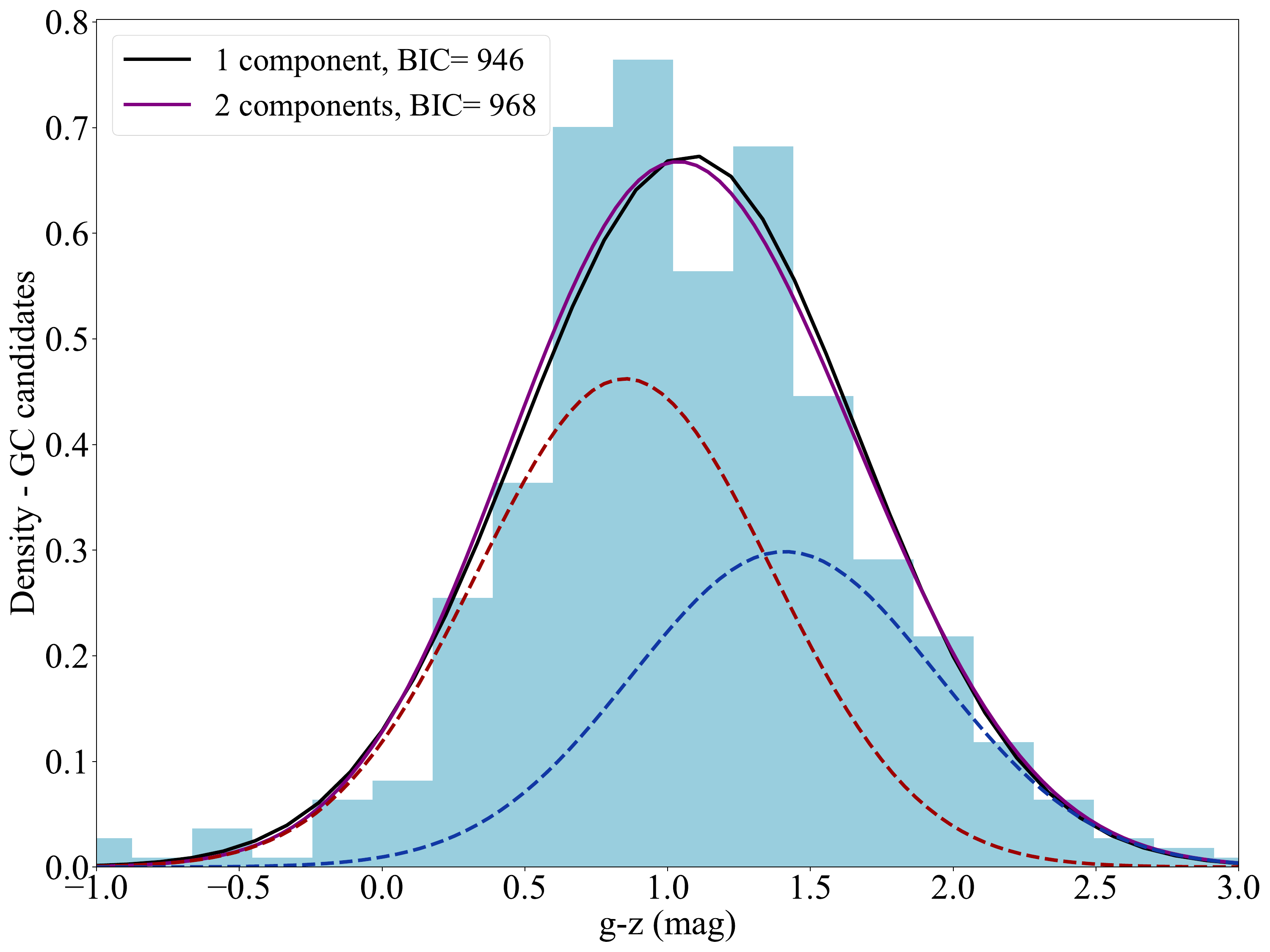}
\includegraphics[width=5cm]{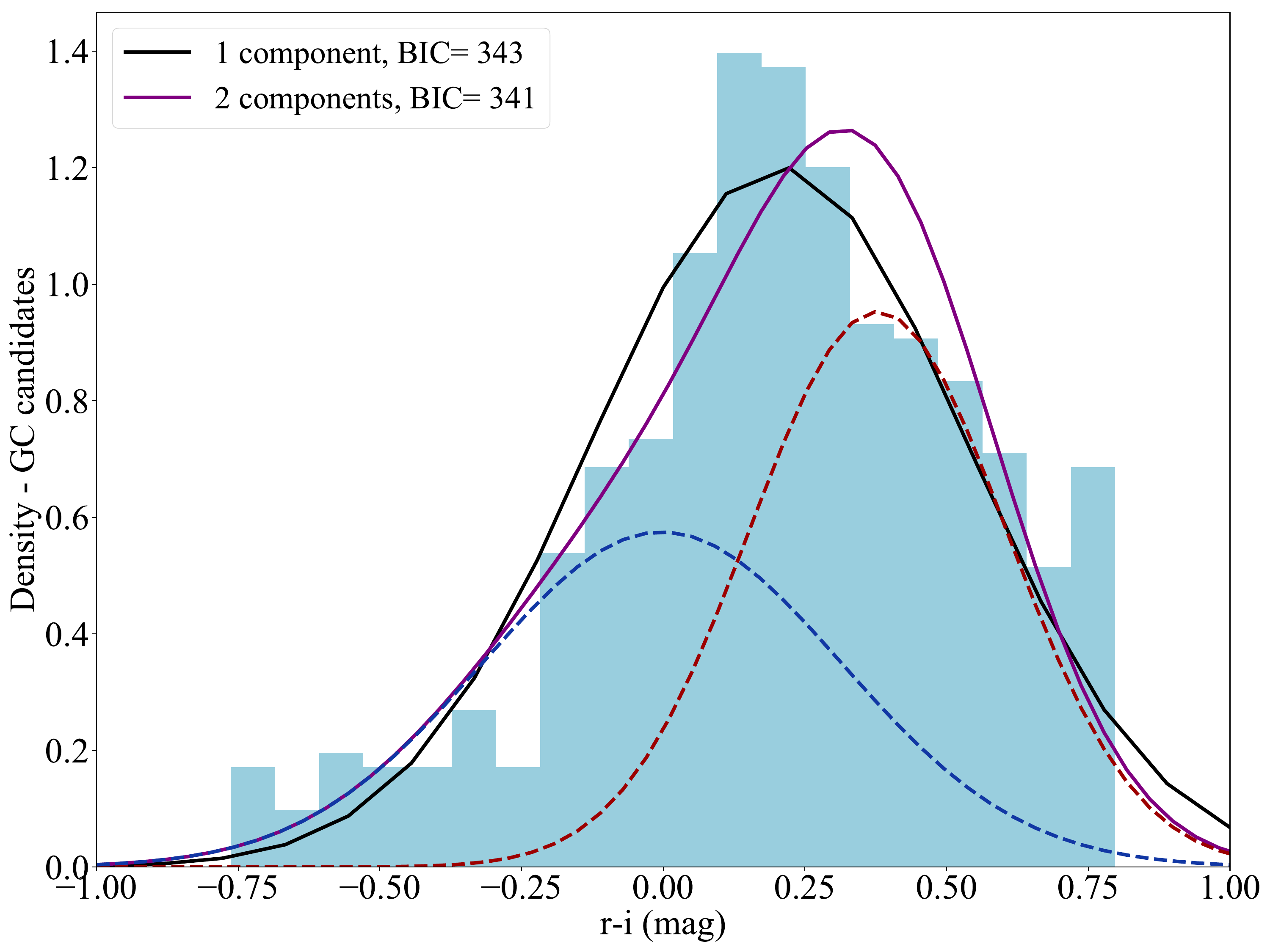}
\includegraphics[width=5cm]{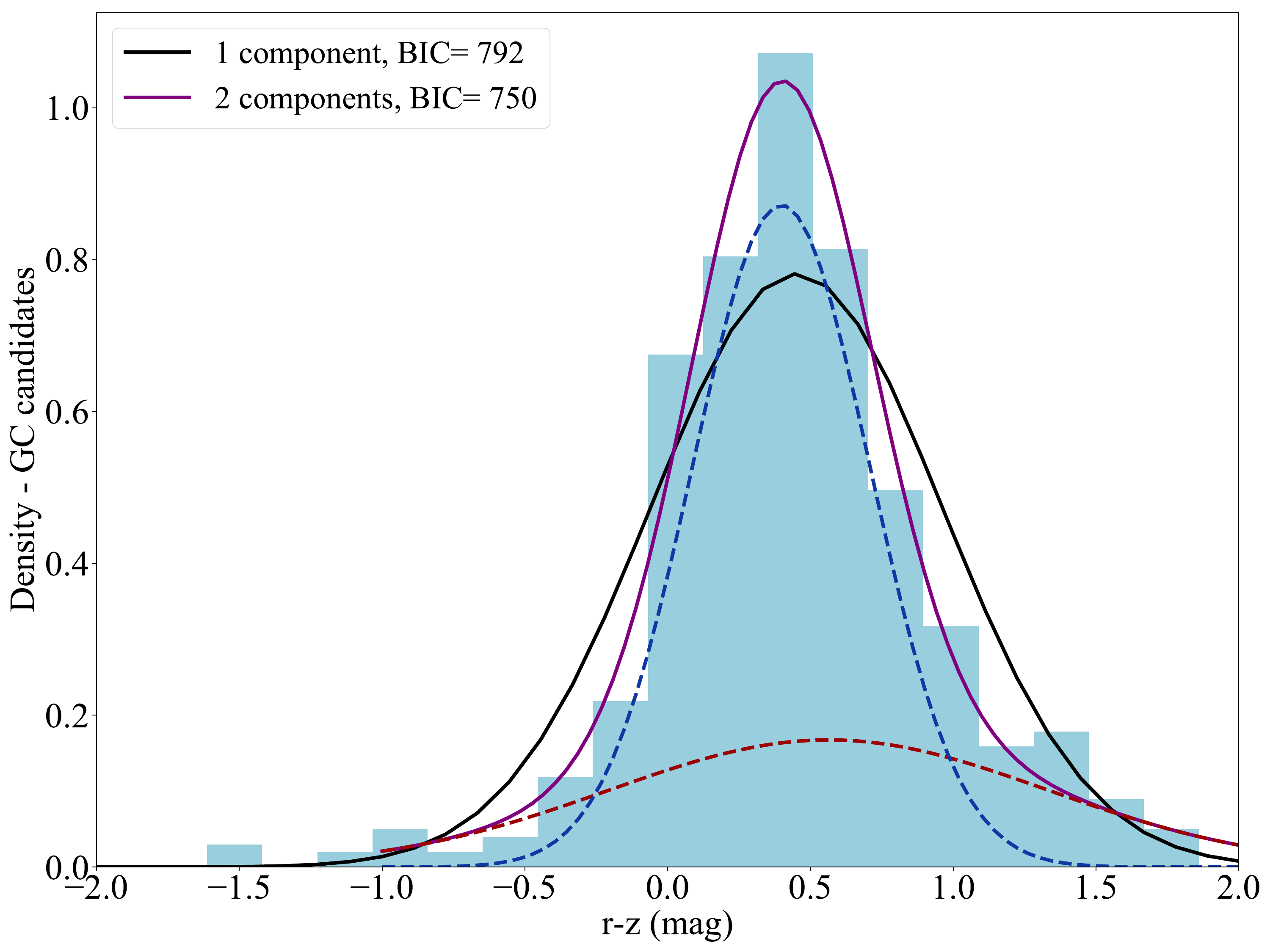}\\
\includegraphics[width=5cm]{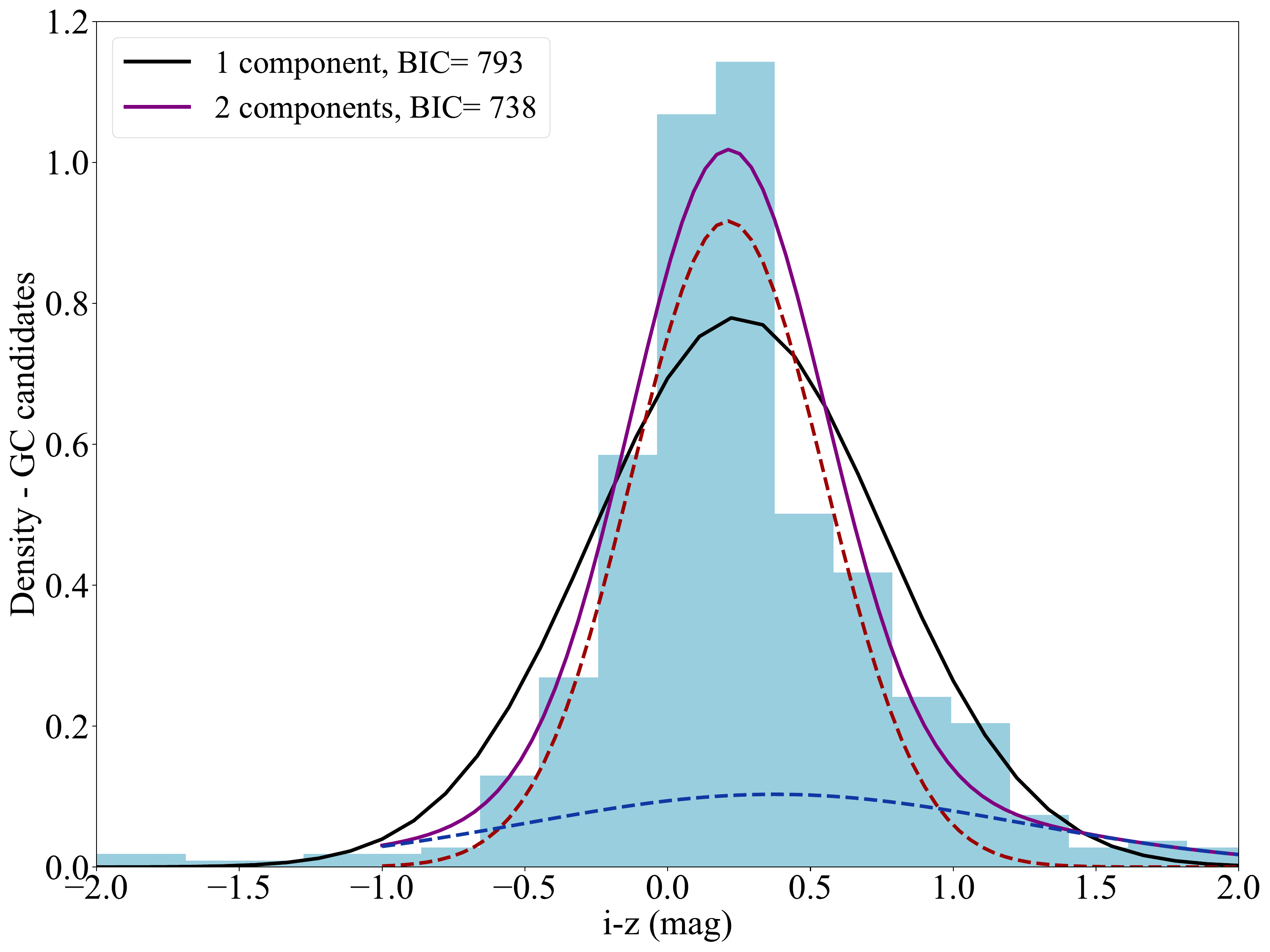}
\caption{Color distribution computed using broad-band filters only. The curves represent the unimodal (black) and bimodal (purple) distributions returned by the GMM analysis. We also show in blue and red dashed lines the two peaks that compose the bimodal distribution. BICs values for GMM with one and two components are shown as an example.}
\label{broad_colors}
\end{figure*}

\begin{figure*}
\centering
\includegraphics[width=5cm]{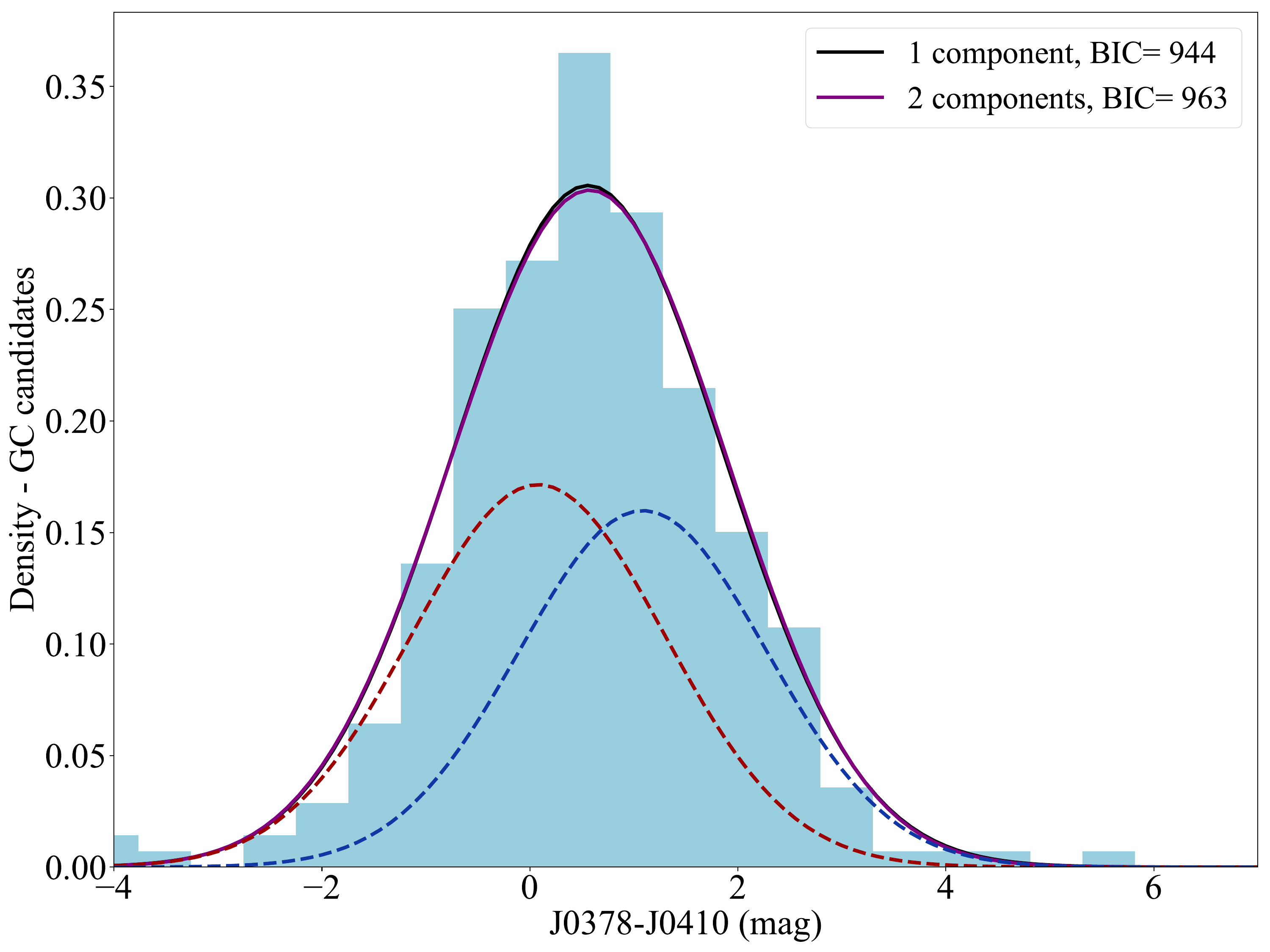}
\includegraphics[width=5cm]{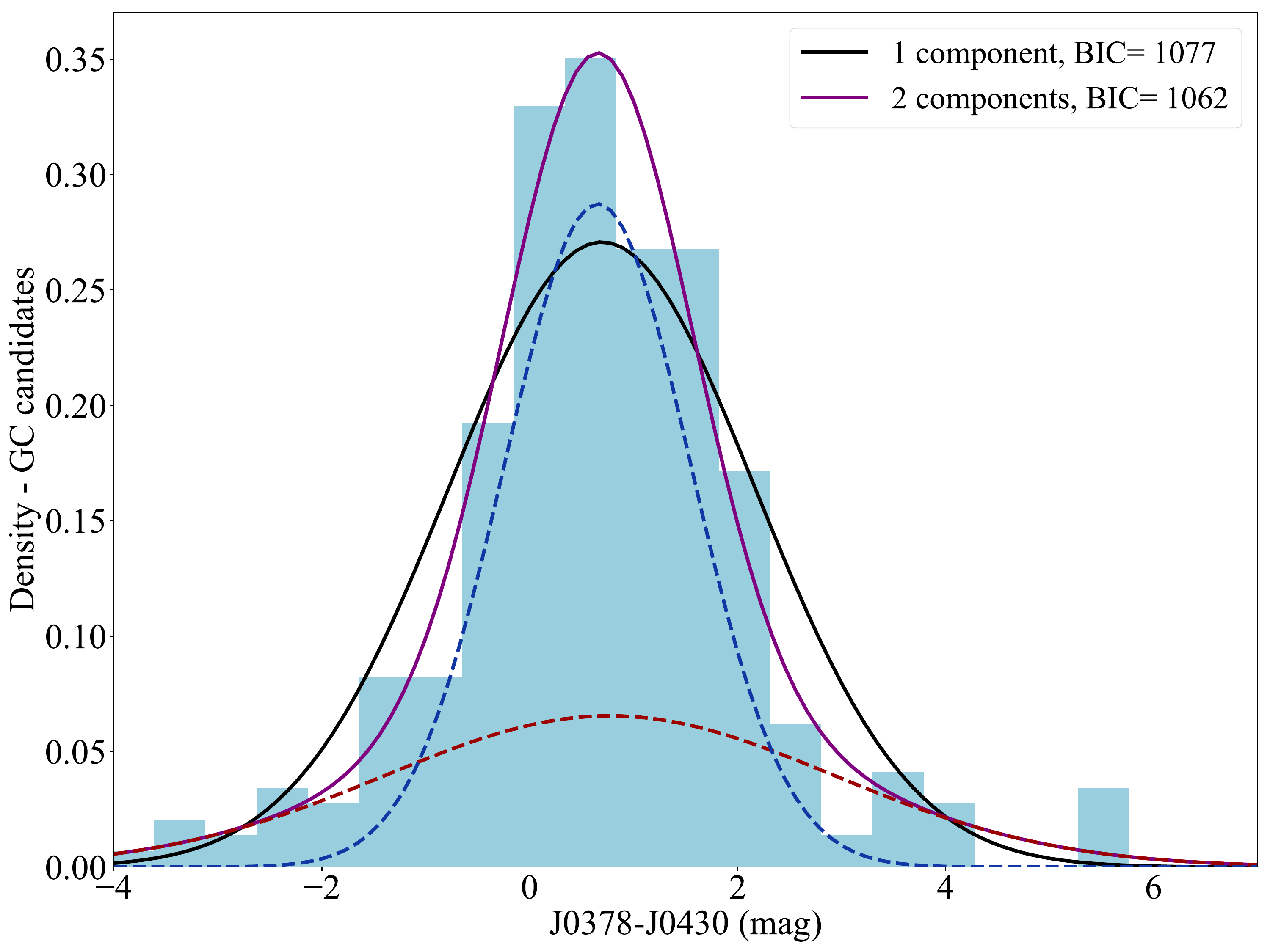}
\includegraphics[width=5cm]{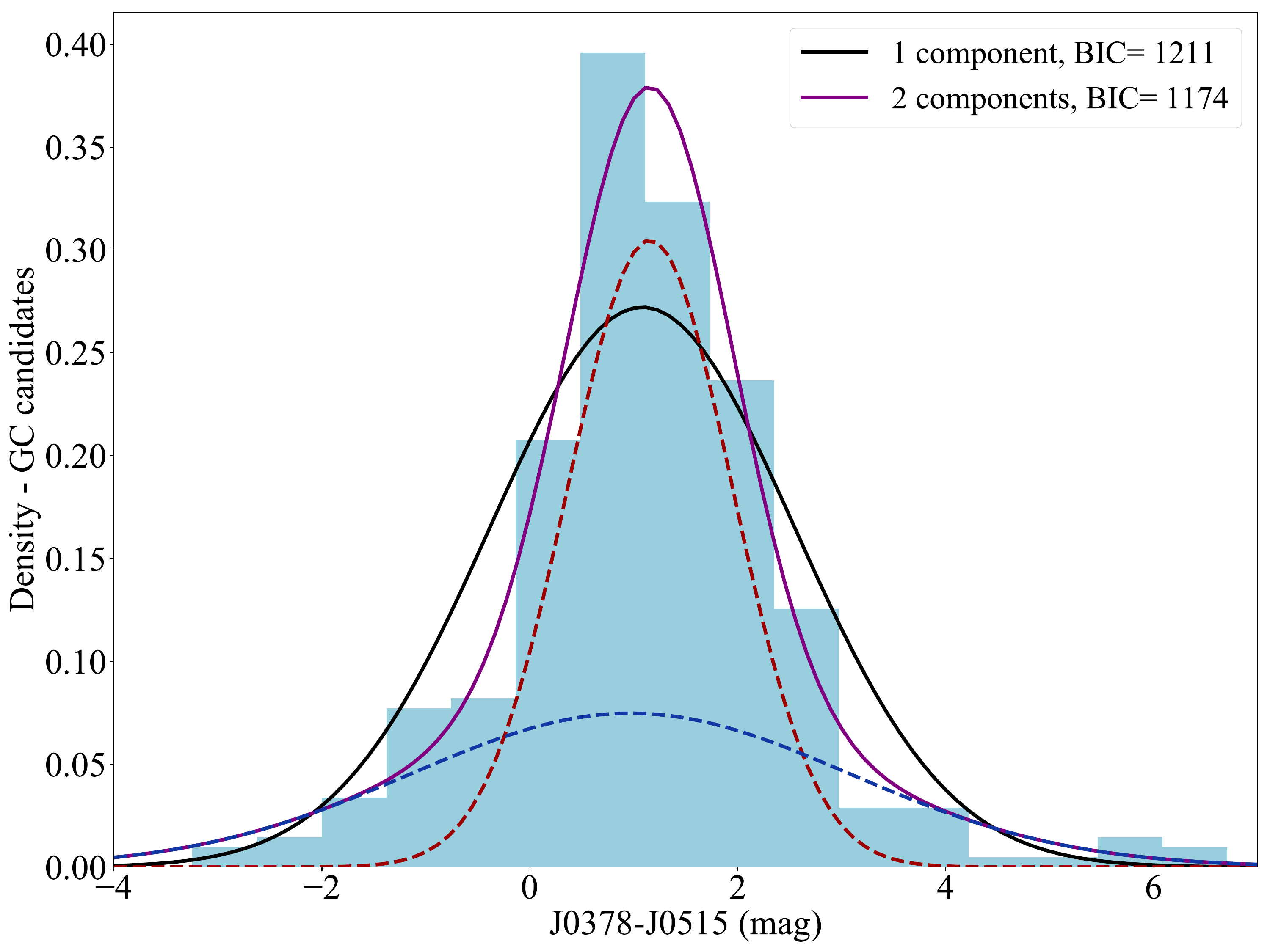}\\
\includegraphics[width=5cm]{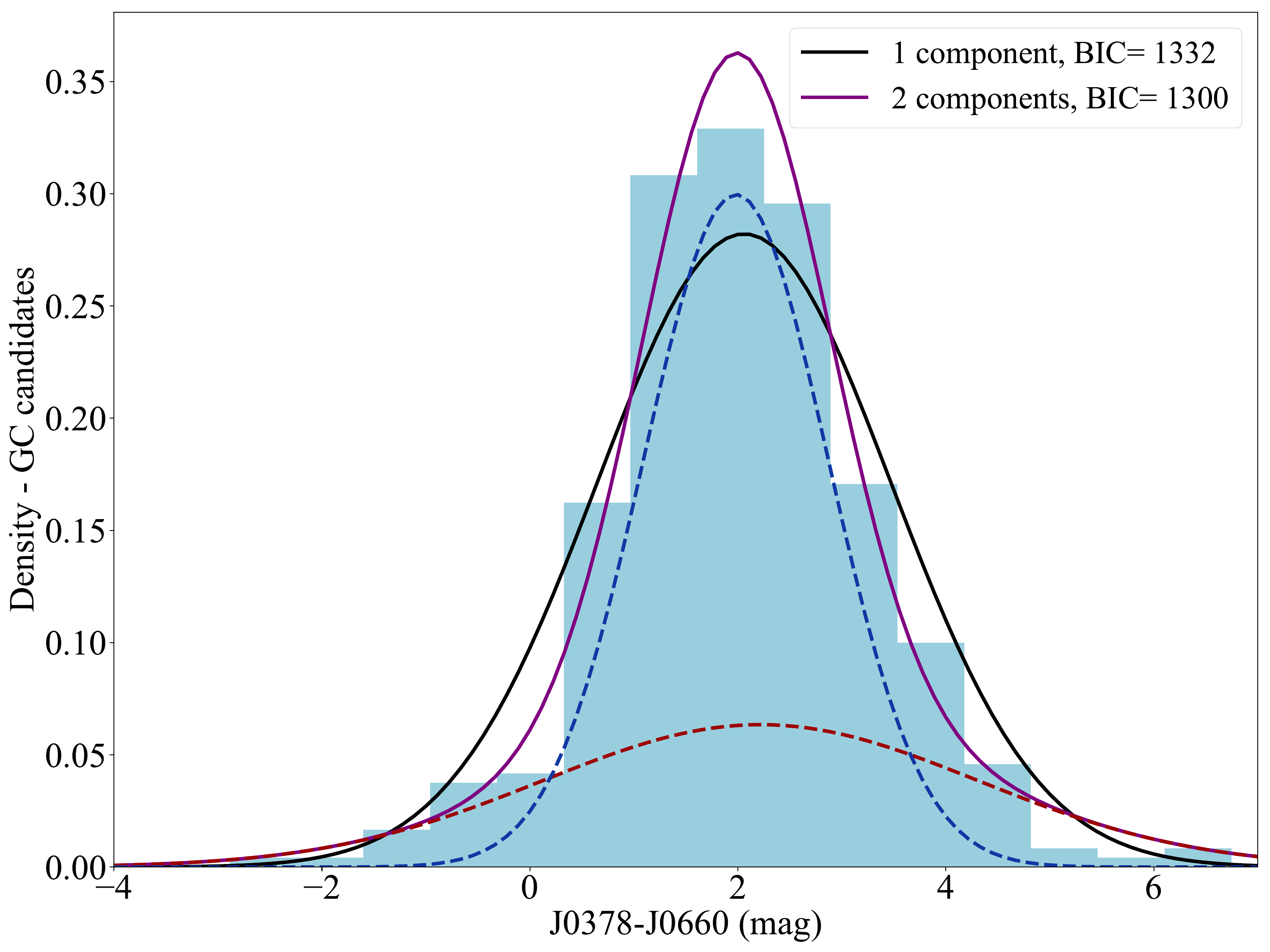}
\includegraphics[width=5cm]{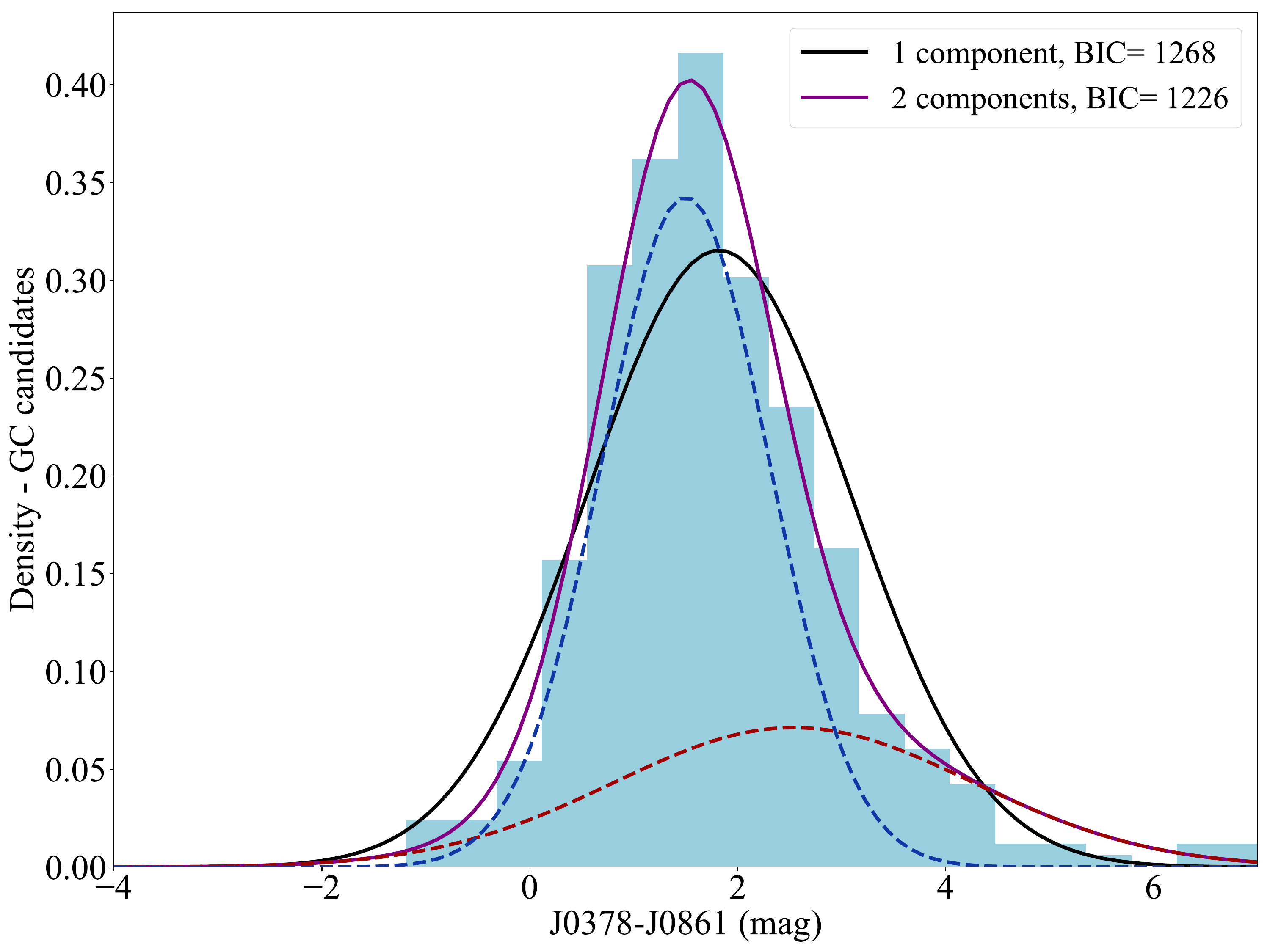}
\includegraphics[width=5cm]{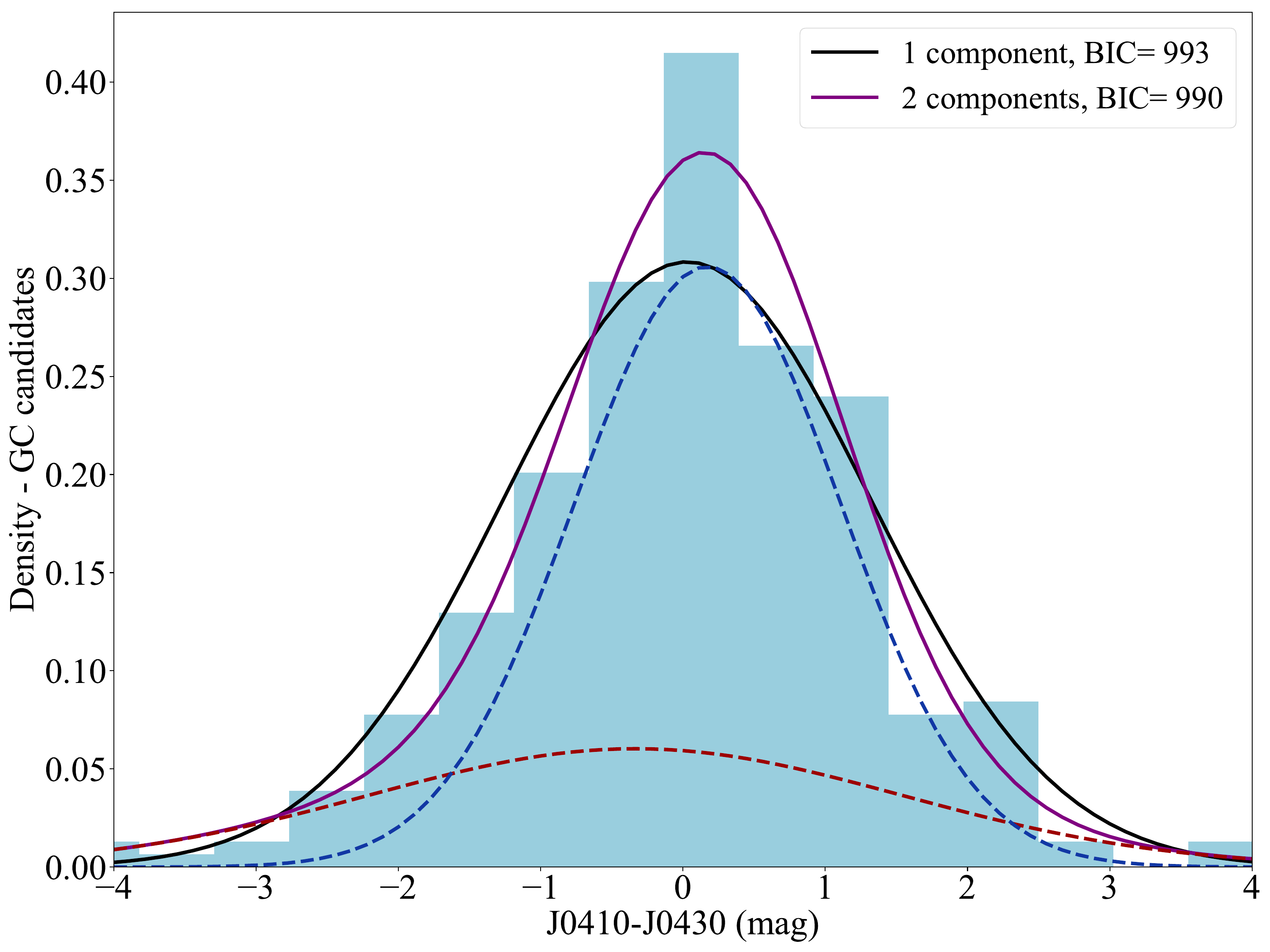}\\
\includegraphics[width=5cm]{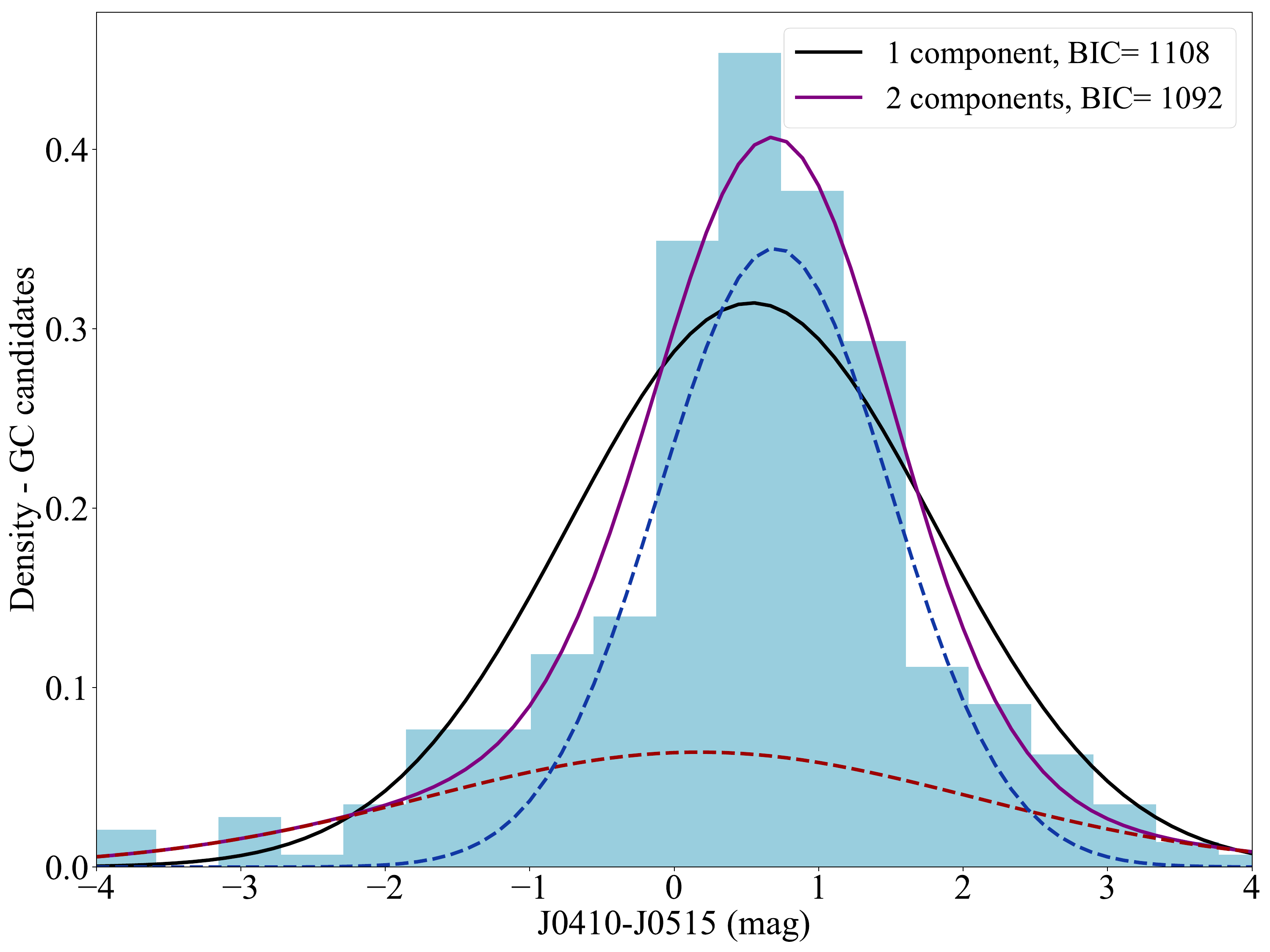}
\includegraphics[width=5cm]{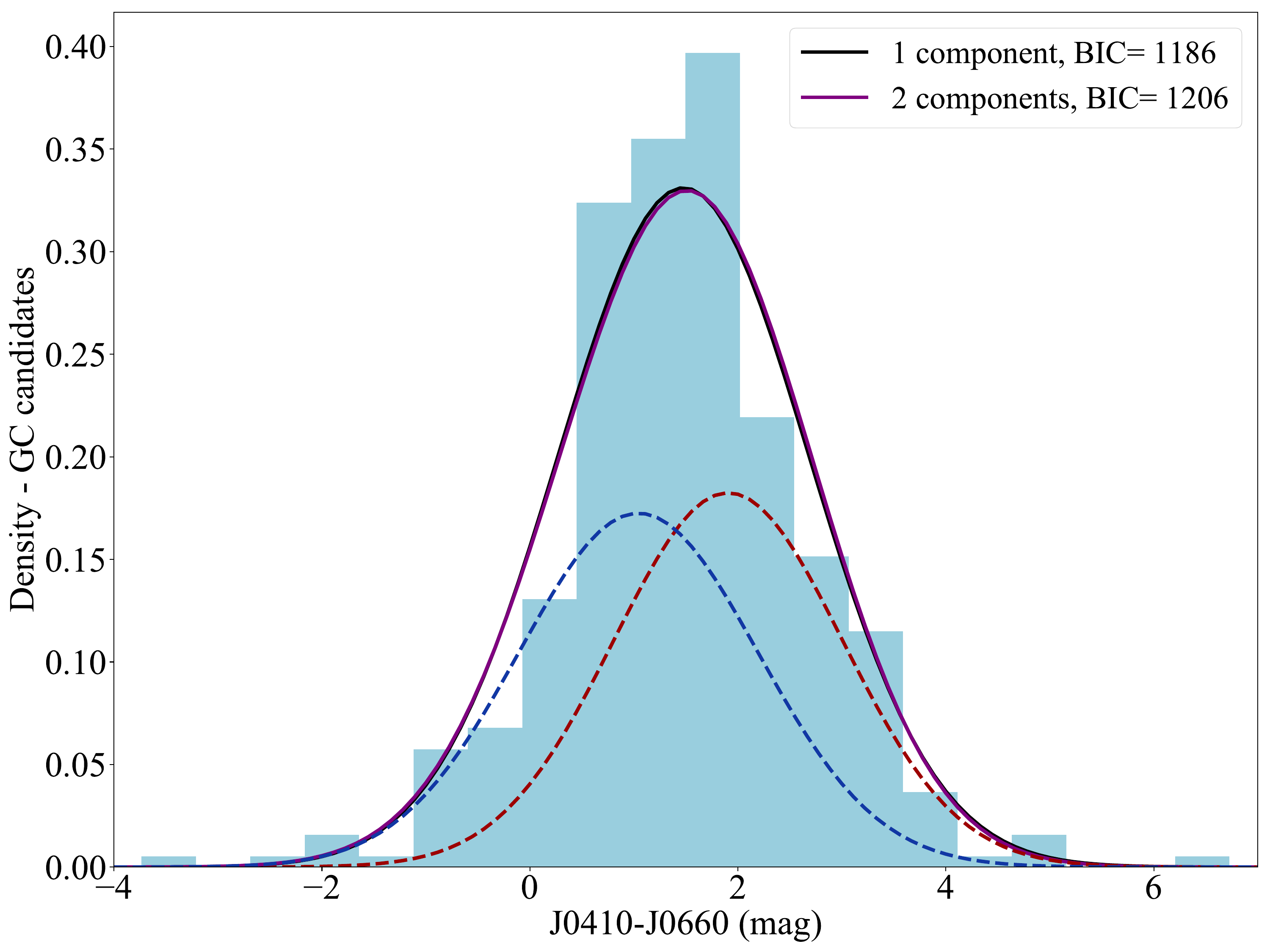}
\includegraphics[width=5cm]{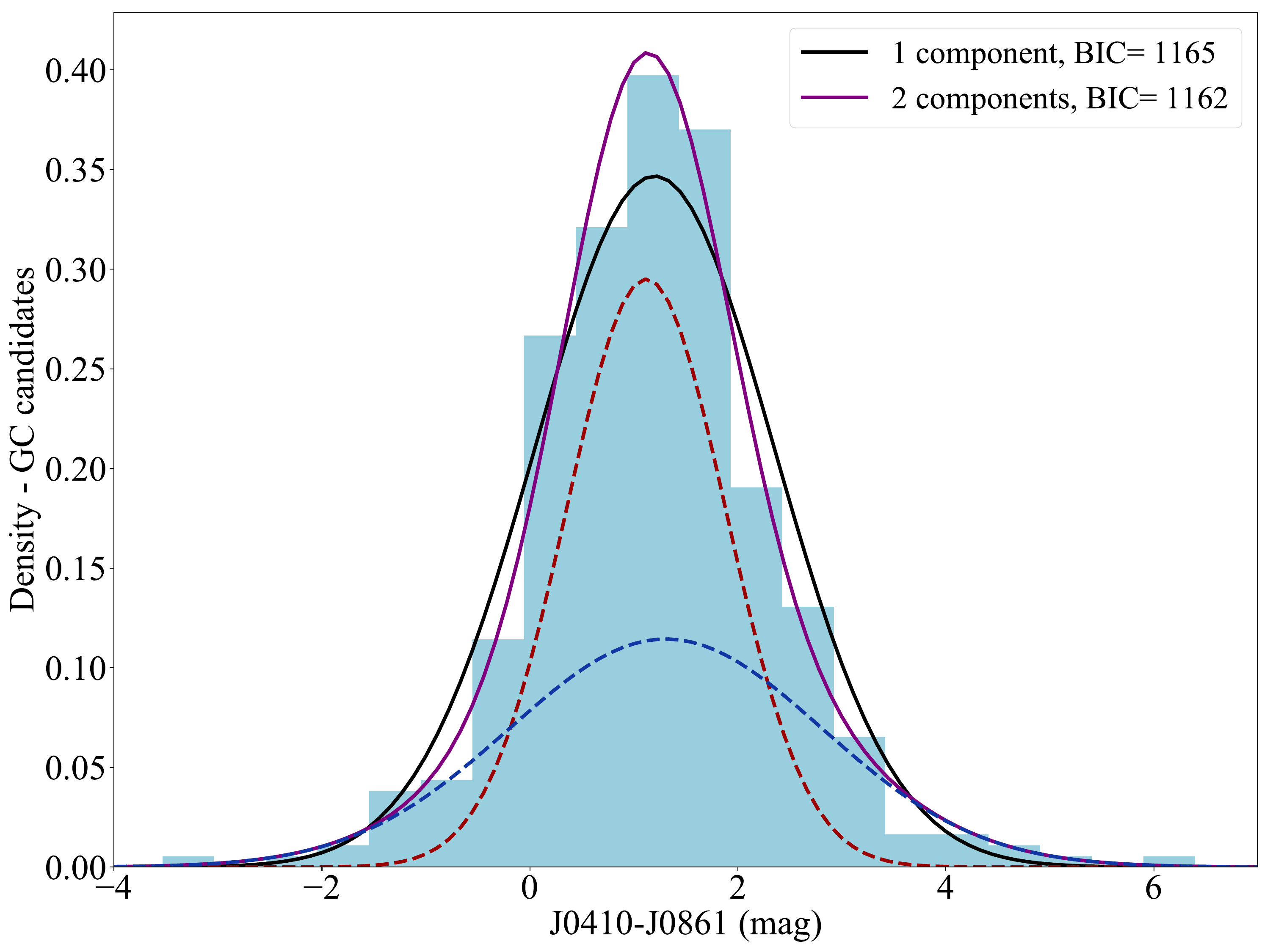}\\
\includegraphics[width=5cm]{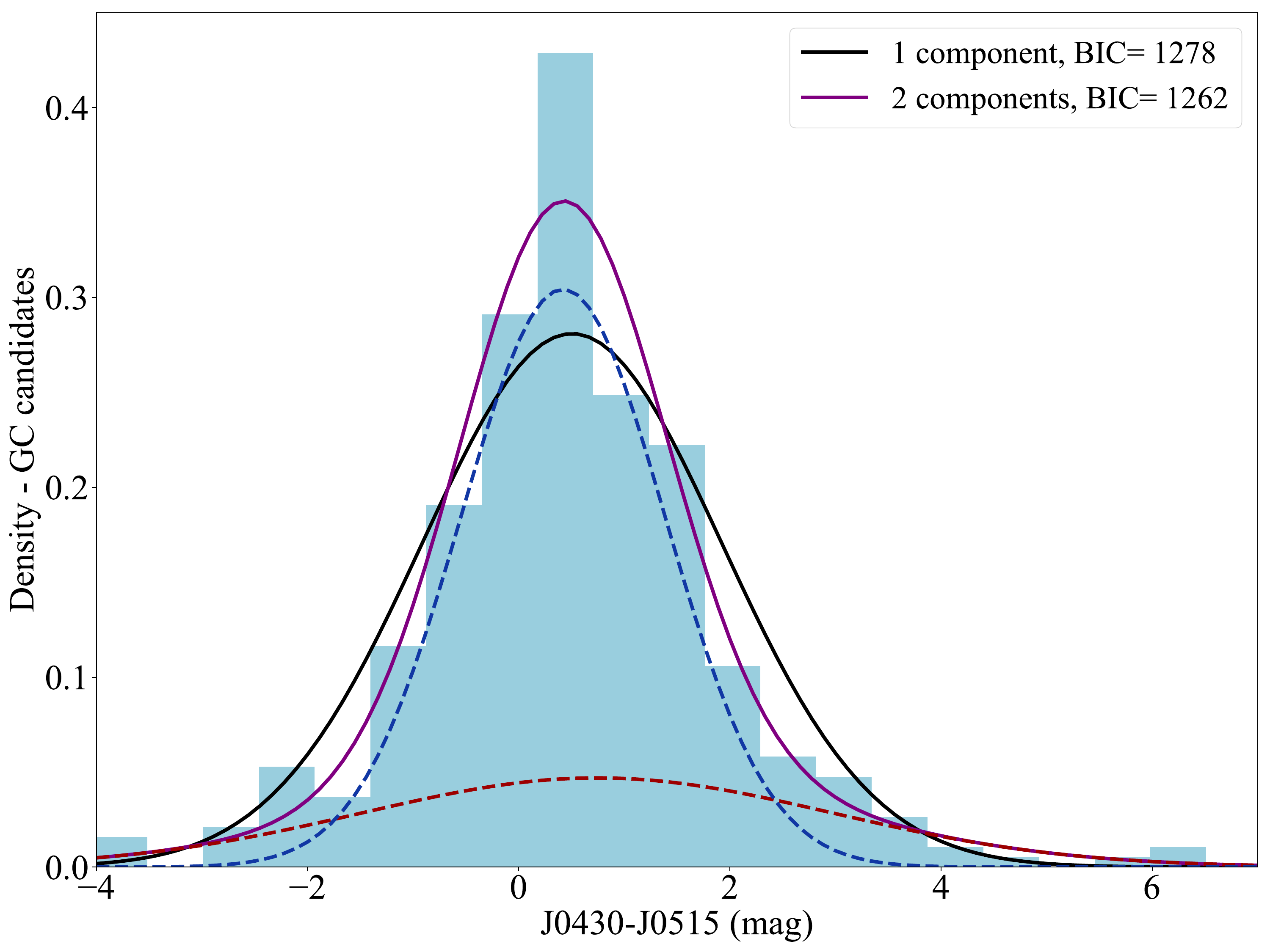}
\includegraphics[width=5cm]{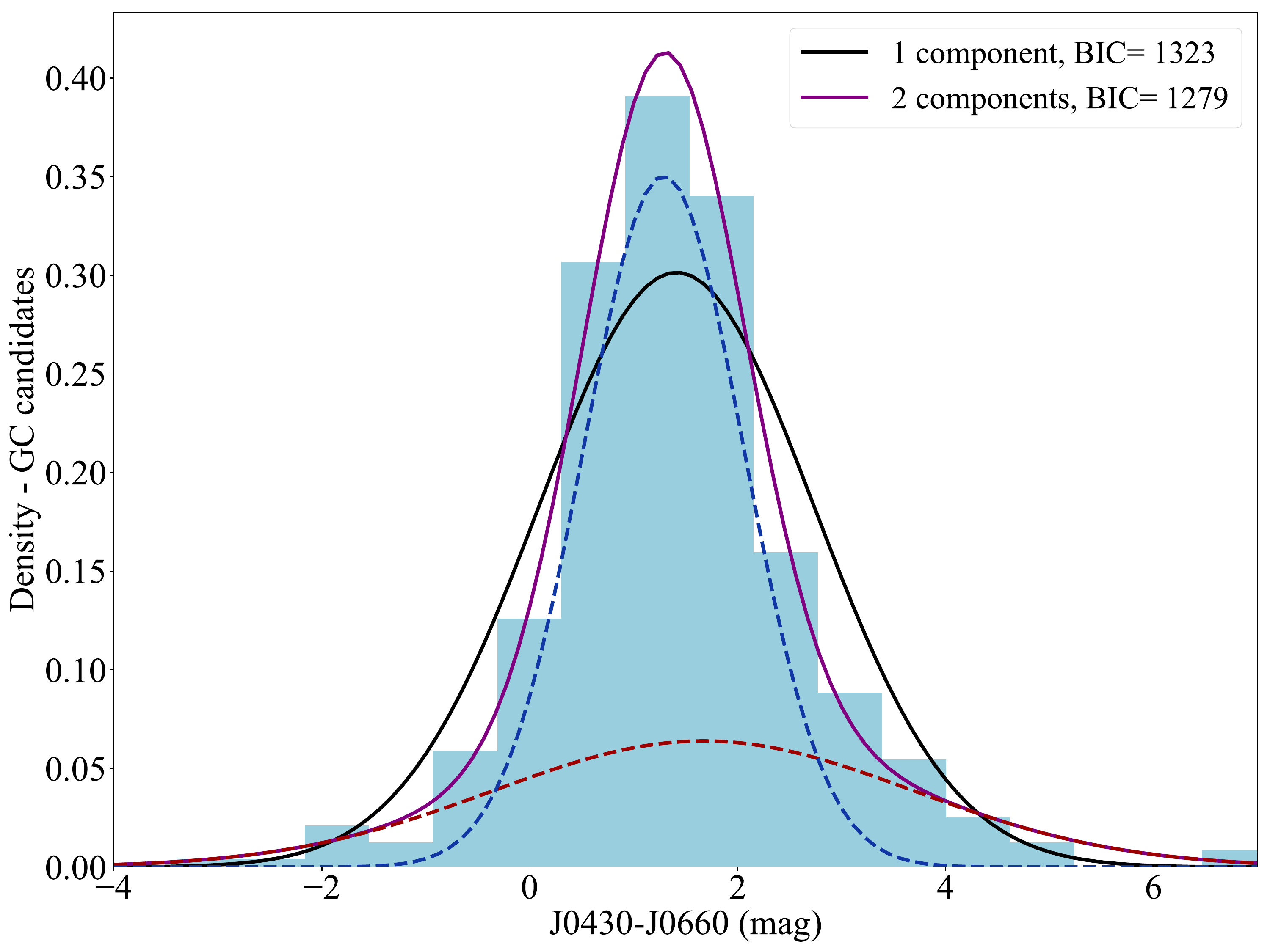}
\includegraphics[width=5cm]{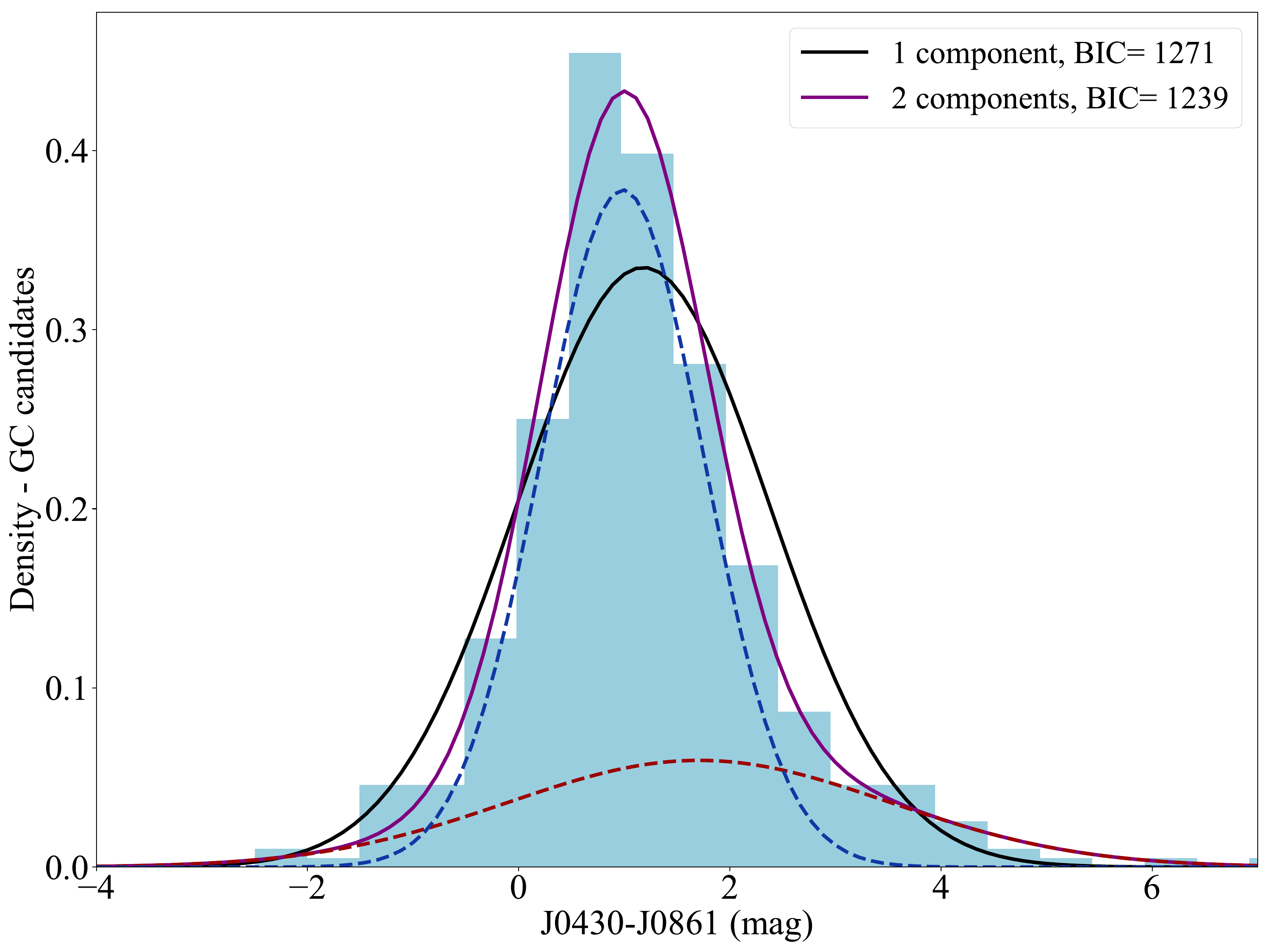}\\
\includegraphics[width=5cm]{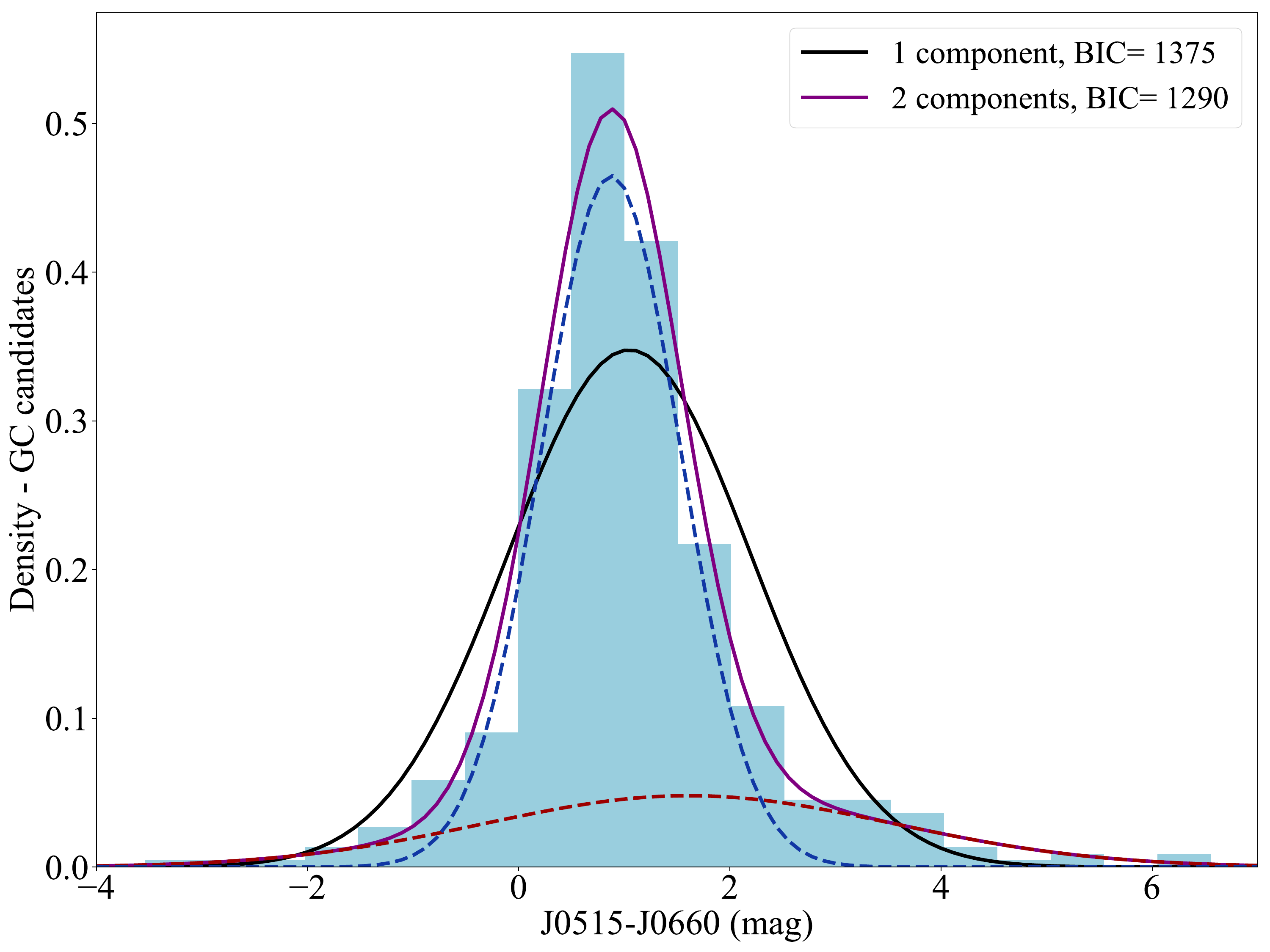}
\includegraphics[width=5cm]{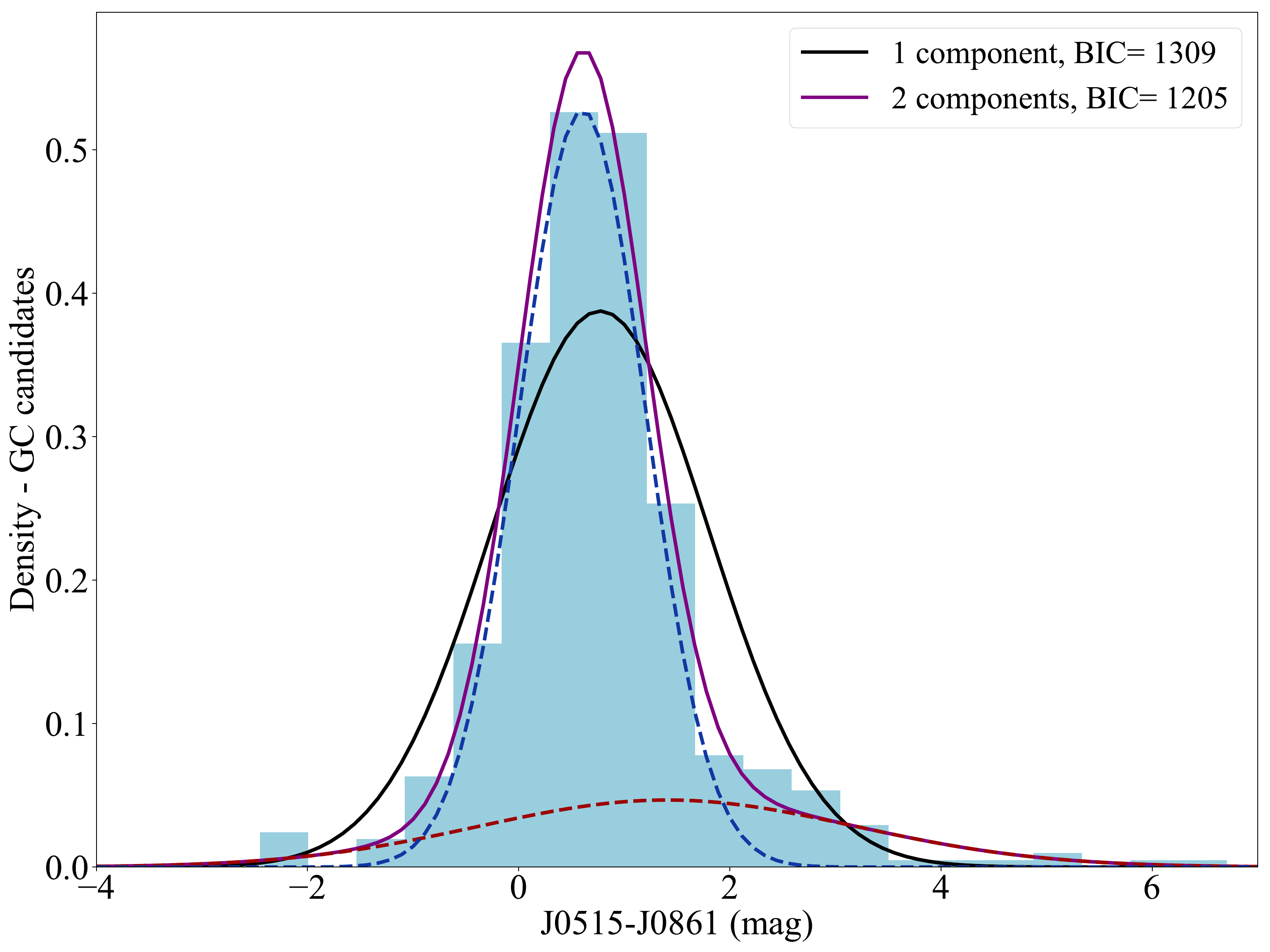}
\includegraphics[width=5cm]{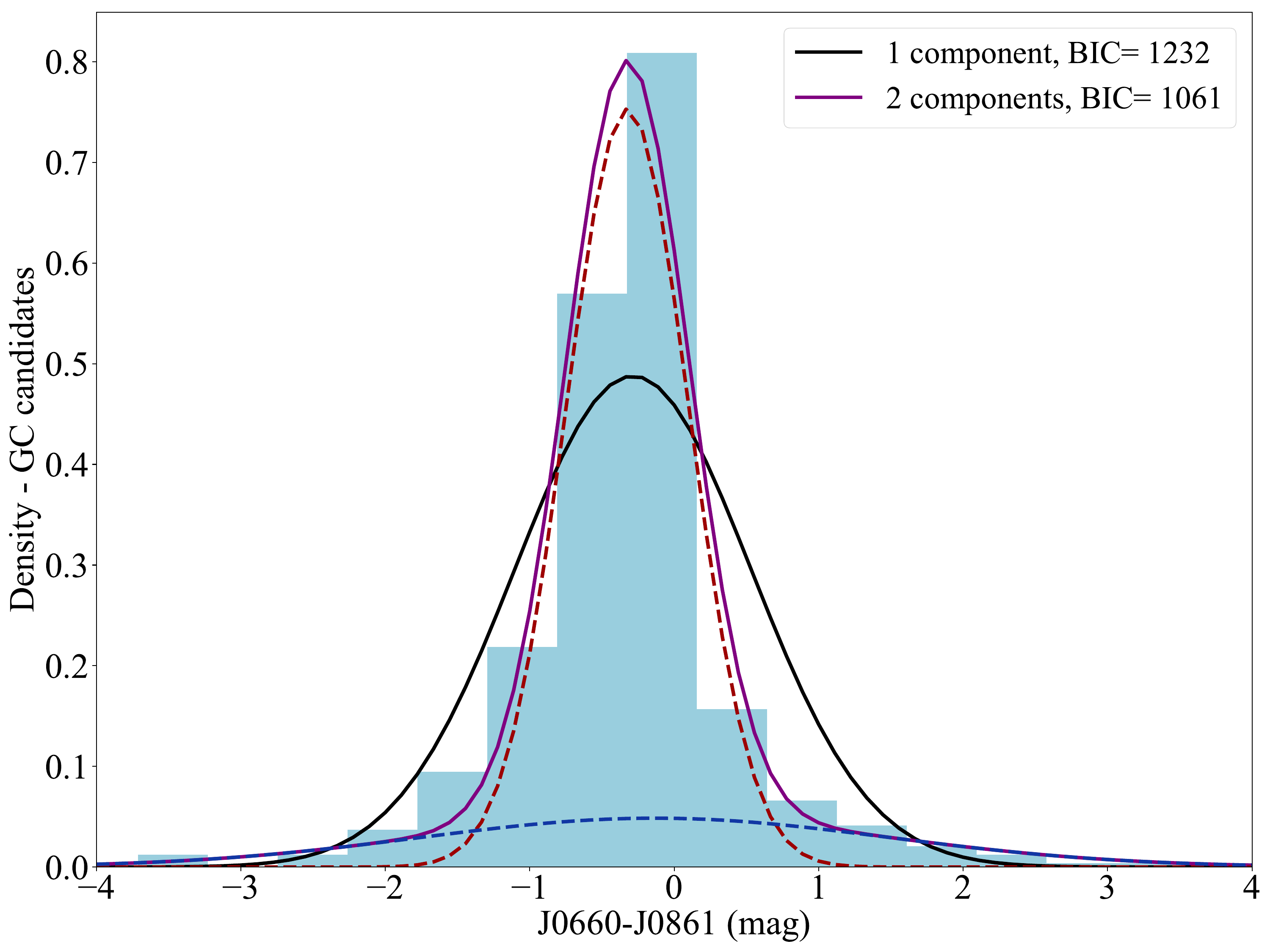}
\caption{Color distribution computed using narrow--band filters only. The curves represent the unimodal (black) and bimodal (purple) distributions returned by the GMM analysis. We also show in blue and red the two peaks that compose the bimodal distribution. BICs values for GMM with one and two components are shown as an example.}
\label{narrow_colors}
\end{figure*}

We study the distributions of all possible color-color diagrams with the J-PLUS filter system without mixing broad and narrow--band filters (as broad-- and narrow--band filters were calibrated using different methodologies, see Section \ref{sec:data}). A total of 25 colors were inspected, 10 using broad-band and 15 colors on narrow--band filters (Figures \ref{broad_colors} and  \ref{narrow_colors}, respectively).
Gaussian mixture modeling (GMM) was performed on the distributions using the Python library Scikit-Learn (\texttt{sklearn}, \citealt{scikit-learn}), following the procedures from \cite{ivezic2014statistics}. We compare the GMM results obtained for one component (black curve) and two components (purple curve), using the Bayesian information criterion (BIC). The BIC makes assumptions about the likelihood that aims to simplify the calculation of the odds ratio and is useful to estimate the statistical significance of clusters found in the data. Therefore we assume that lower BIC values are associated with highly significant clusters, following \citet[][chapter 5.4]{ivezic2014statistics}. 

According to BIC statistics, we find evidence of color bimodality in 17 colors, namely $u-g$, $u-r$, $u-i$,
$g-r$, $r-z$, $i-z$, $J0378-J0430$, $J0378-J0515$, $J0378-J0660$, $J0378-J0861$, $J0410-J0515$, $J0430-J0515$, $J0430-J0660$, $J0430-J0861$, $J0515-J0660$, $J0515-J0861$, $J0660-J0861$. In these cases, the BICs of the GMM with two components have a lower value than the BICs of the GMM with one component. We note that even though the BICs of the GMM with two components are lower than the values for one component in the cases of $u-z$,$r-i$, $J0410-J0430$ and $J0410-J0861$, the differences of the BICs are too small to be conclusive. In Figures \ref{broad_colors} and  \ref{narrow_colors} we show BIC values for GMM with one and two components as an example.

\citet{de2017probabilistic} favors the use of a regularized version of BIC, namely the Integrated Complete Likelihood (ICL). As a sanity check, we repeat the analysis evaluating possible color bimodality using the ICL criterion. According to ICL statistics, we find evidence of color bimodality in 10 colors, namely $g-r$, $r-z$, $i-z$, $J0378-J0515$, $J0378-J0861$, $J0430-J0660$, $J0430-J0861$, $J0515-J0660$, $J0515-J0861$, $J0660-J0861$. In the case of the color $J0378-J0660$, the ICL of the GMM with one component has a lower value than the GMM with two components, but the difference is small, hence we consider this case inconclusive.

A table summarising the results from both statistics is presented in Appendix \ref{appendix:bicicl}, Table \ref{tab:bicaic_info}.

\subsection{Stellar population properties}\label{sec:color_sed_results}

In this section, we present the results of the SED fitting performed on a sub-sample of 171 GC candidates. This sub-sample consists of only GC candidates with measured magnitudes in all bands. 
The codes and models used are those described in Section \ref{sec:methods_sed}.

The uncertainties of the fits were estimated according to Equation \ref{eq:sigma} in Section \ref{sec:methods_sed}, extending over the $N$\,=\,$171$ GC candidates. In Table \ref{tab:sigma} we list the resulting values of $\sigma_j$.
The fits reported were performed for $A_V$\,=\,0, the value of <$\sigma(A_V)$> listed in Table \ref{tab:sigma} gives an indication of the error
of this assumption.

\begin{table}
    \centering
    \caption{Precision (mean standard deviation) of the stellar population properties derived with the \tgas\ code.}
    \begin{tabular}{l  c}
    \hline
    \hline
        Property & <$\sigma$> \\
     \hline
     \logMl & 0.20 \\
     \mwa & 0.32 \\
     \lwa & 0.45 \\
     \mwz & 0.55 \\
     \lwz & 0.51 \\
     $A_V$ & 0.17 \\
        \hline
    \end{tabular}
    \label{tab:sigma}
\end{table}

\begin{figure}
\centering
\includegraphics[width=\columnwidth]{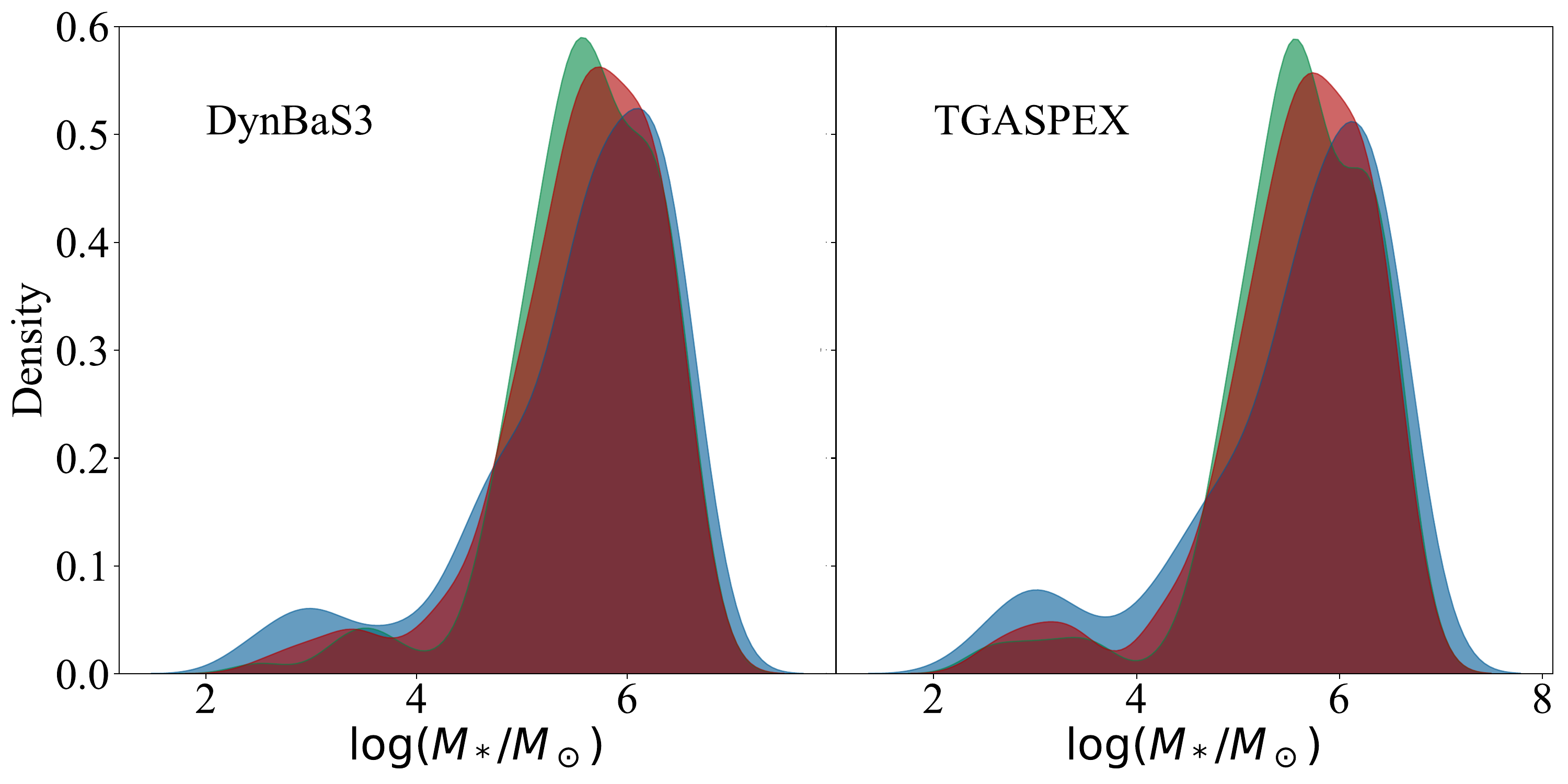}
\caption{Distribution of stellar masses obtained from SED fitting. The left panel illustrates results from \dynb\ 3, while the right panel shows results of TGASPEX. The distributions show a broad peak with log Stellar Masses between 5 and 6 M$_\odot$. Red represents results obtained using all available filters. Green represents results obtained using broad--band filters only. Blue represents results obtained using narrow--band filters only.}
\label{fig:masses}
\end{figure}

\paragraph{Stellar masses: }
Figure \ref{fig:masses} shows that the distributions of the total mass obtained with \texttt{TGASPEX} and \texttt{DynBaS3} are consistent, and range from below $10^{3}$ to above $10^{6}$ M$\odot$, 
with a peak around $10^{5.5}$ M$\odot$. Given that GC masses are known in the range $10^{4}$ to $10^{6}$ M$\odot$ \citep{Brodie2006}, we interpret the tail towards low mass to be indicative of contaminants (false positives) in our cluster-candidate catalog. Such
contaminants represent a negligible fraction, only 5 \% of our sample of GC candidates. It is unclear at this point if the more massive systems (stellar mass $>$ $10^{6}$ M$\odot$) are GCs or UCDs  \citep{phillipps2001ultracompact}.  These low-mass candidates are also intrinsically fainter, which results in lower S/N SEDs. Regarding the distributions retrieved from different filter sets, including the narrow--band filters in the fits has the effect of broadening the distribution of the {\it best} mass values.

\begin{figure}
\centering
\includegraphics[width=\columnwidth]{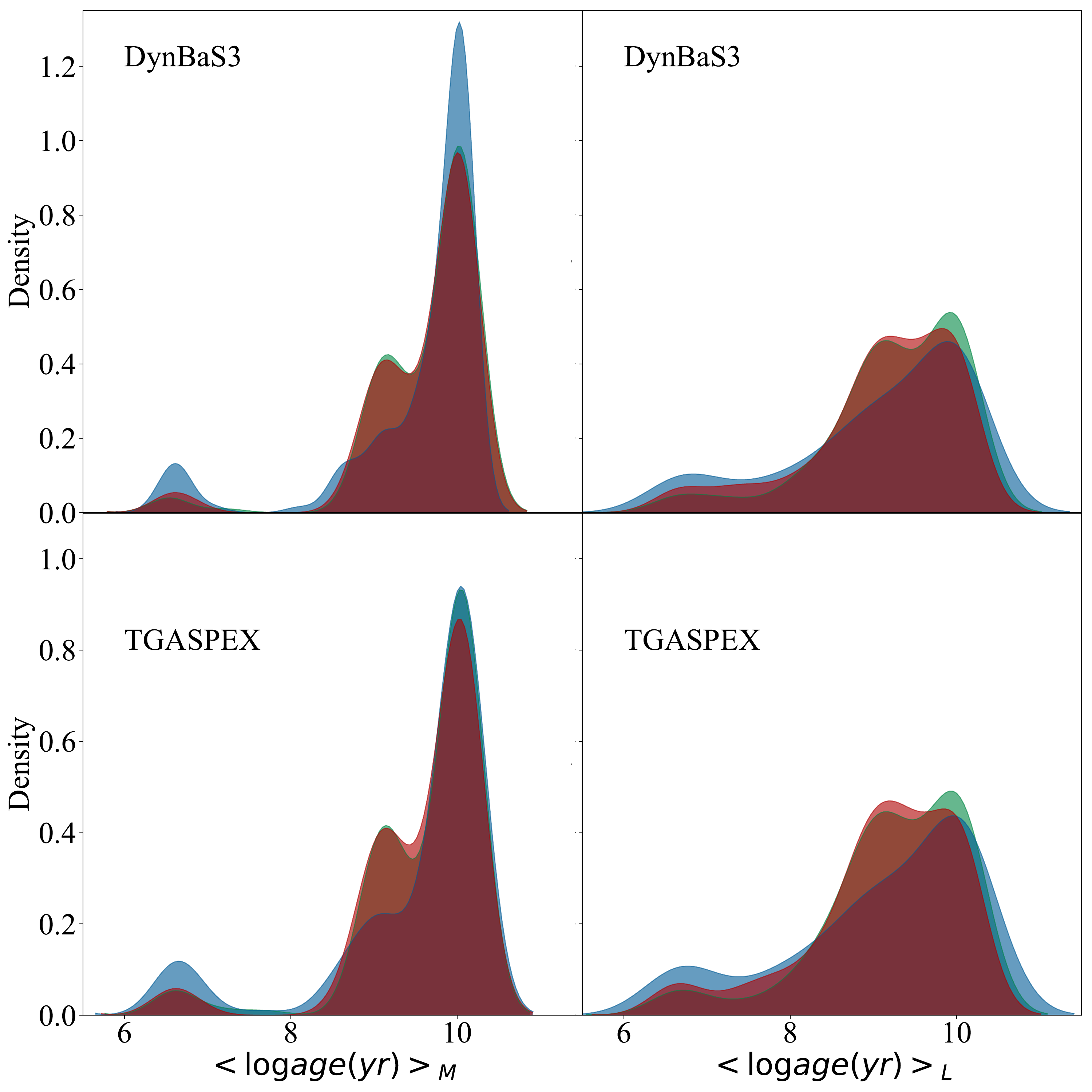}
\caption{Distribution of logarithmic ages obtained from SED fitting. Upper and lower panels illustrate results from \dynb\ 3 and TGASPEX respectively. Left and right panels show results for mass- and light-weighted ages, respectively. Red represents results obtained using all available filters. Green represents results obtained using broad--band filters only. Blue represents results obtained using narrow--band filters only.}
\label{fig:ages}
\end{figure}

\paragraph{Ages: }
Figure \ref{fig:ages} shows the distributions of our fit results for the mass-weighted and light-weighted ages. Old age clusters, with ages $\approx$ 10\,Gyr dominate the distributions for all filter combinations (narrow--band only, broad--band only, broad and narrow--bands combined), with a secondary peak occurring at ages $\approx$\,9\,Gyr. For reasons that remain unclear at this moment, this second intermediate age peak is less pronounced when only the narrow--bands are used in the fit. This may be related to the choice of sub-sample selected to be analyzed via SED fitting, where only the GC candidates with all filters measured were included. \texttt{DynBaS3} retrieves a higher fraction of old GCs than \texttt{TGASPEX}. We interpret the tail towards youngest ages ($\log age < 8$) as being produced by contaminants present in our candidate catalog. 

\begin{figure}
\centering
\includegraphics[width=\columnwidth]{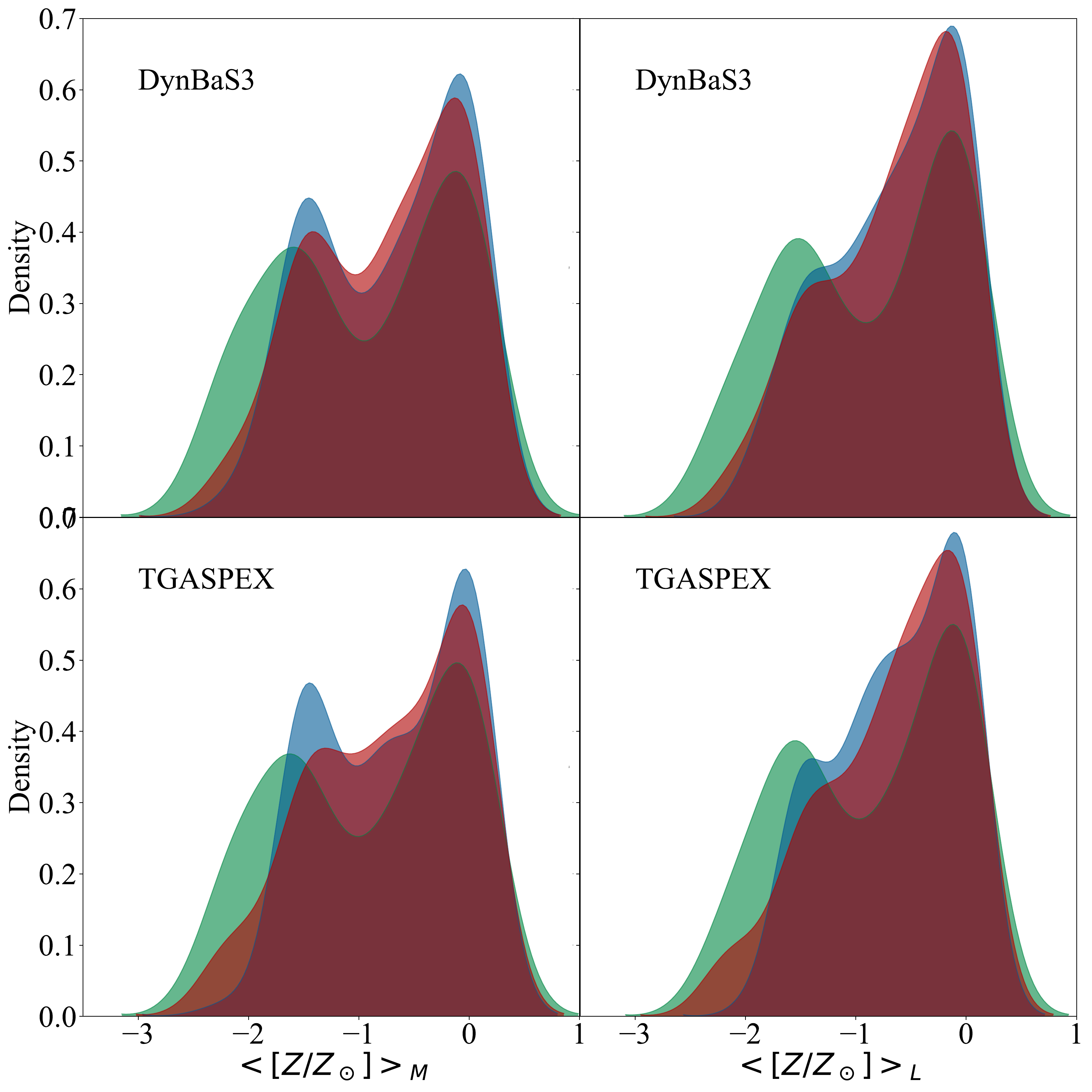}
\caption{Distribution of metallicities obtained from SED fitting. Upper and lower panels illustrate results for \dynb\ 3 and TGASPEX respectively. Left and right panels show results for mass- and light-weighted Z, respectively. Red represents results obtained using all available filters. Green represents results obtained using broad--band filters only. Blue represents results obtained using narrow--band filters only.}
\label{fig:metals}
\end{figure}

\paragraph{Metallicities: }
Figure \ref{fig:metals} shows the distributions of our fit results for the {\it best} values of mass-weighted Z and light-weighted Z.
A clear bimodal distribution of metallicities is seen in most combinations of code and filter set, while the results of light and mass-weighted Z from \tgas\ obtained only with narrow--bands show three modes, indicating three different populations. A stronger tail at low metallicities is derived when only the broad--bands are used. We interpret this result as evidence that the narrow--bands help to constrain the metallicities. 
It is unclear at this point how the metallicity distributions are bimodal when we only find evidence of color bimodality in part of the studied colors. A possible channel for this could be a non-linear color-metallicity relation, as already presented in articles such as \cite{yoon2006explaining,cantiello2007metallicity,fahrion2020fornax}.

As in other galaxies, we interpret the two families of metal-poor and metal-rich GCs to be related to two mechanisms or episodes of star formation \citep[e.g.][]{Brodie2006}, although we find evidence of three populations when using narrow--bands only and \texttt{TGASPEX}.

\begin{figure*}
\centering
\includegraphics[width=15cm]{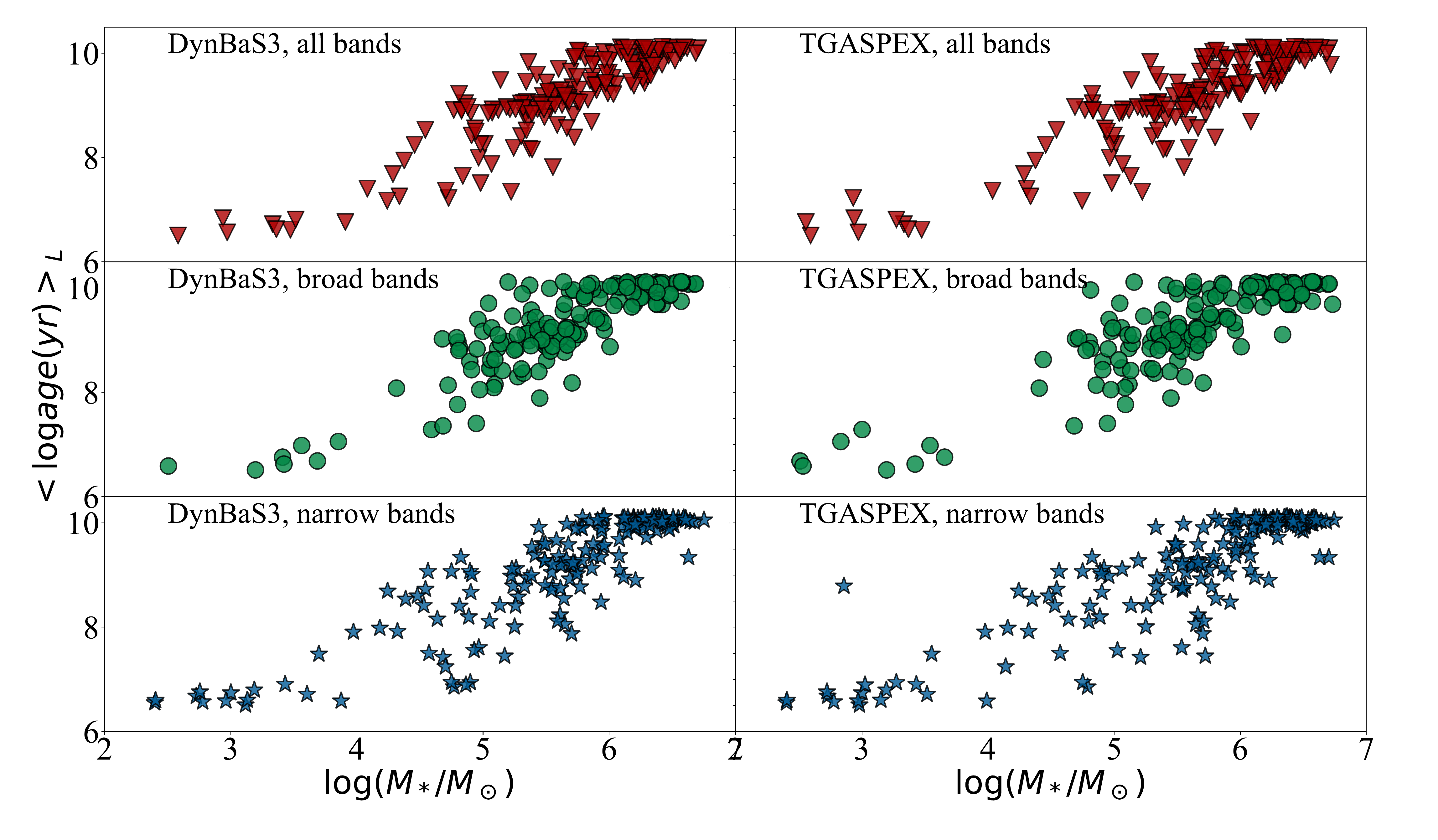}
\caption{Correlation between ages and masses derived via SED fitting. A trend where the mean ages of the cluster-candidates increase with stellar masses is seen for all combinations of code and filter set. Left and right illustrate results for \dynb\ 3 and TGASPEX respectively. Red triangles represent results obtained using all bands, green circles represent results when using only broad--bands and blue stars represent results obtained from narrow--bands only.}
\label{fig:mass_age}
\end{figure*}

\paragraph{Stellar masses \emph{vs.} ages}
We explore if there are correlations among the derived parameters. The only correlation identified was between the stellar mass and light-weighted ages, as illustrated in Figure \ref{fig:mass_age}. A correlation is also present between stellar mass and mass-weighted ages, albeit less clear.

\cite{pfeffer2018mosaics} present globular cluster models in the context of \texttt{E-MOSAICS} project. These models describe the formation, evolution, and the disruption of this class of objects.  In their work they find that based on their models and simulations most low-mass clusters were disrupted at redshift 0, therefore they conclude that clusters with higher mass are more likely to survive until the present time, which results in old GC populations having higher characteristic mass when compared with younger GCs. Therefore, we attribute the relation found in this work to be caused by the same processes found in \cite{pfeffer2018mosaics}, but we also note that we do not calculate ages for all GCs in NGC\,1023, therefore our results could be affected by selection effects that are not well characterized.

\subsection{Specific frequency}

The specific frequency (S$_N$) of the GC population of a galaxy represents the total number of GCs per unit of host galaxy luminosity. 
Following \cite{Kartha2014}, we adopt  M\textsubscript{V} = -21.07 $\pm$ 0.06 and we calculate using the GCLF a N$_{GC}$ = 553 $\pm$ 60, which was used to determine the S$_N$.
We calculate a S$_N$ = 2.1 $\pm$ 0.2, which is consistent with the S$_N$ = 1.8 $\pm$ 0.2 reported in \cite{Kartha2014} and with S$_N$ = 1.7 $\pm$ 0.3 presented in \cite{yong2012most}. 
Our S$_N$ is also consistent with estimations for lenticular galaxies (2 $\leq$ S$_N$ $\leq$ 6, \citealt{kundu1998wide,elmegreen2000specific}). 

\section{Discussion}\label{sec:discussion}

In this section we discuss the results found in Section \ref{sec:results}, to connect the observed properties of the GC system with the evolution of NGC\,1023.

\subsection{The accretion history of NGC\,1023}

The fact that we can identify bimodal distributions in metallicities from the SED fitting analysis from Section \ref{sec:color_sed_results}
could be evidence that there are at least two subpopulations of GCs in the galaxy. 

\cite{li2019formation} use a novel cluster formation model on a simulated galaxy of the same size as the Milky Way and observed that GC candidates tend to form during major merger events. The merger-induced GC formation scenario has been discussed in various recent articles \citep{li2014modeling,choksi2018formation,choksi2019formation}.

\begin{figure*}
\centering
\includegraphics[width=15cm]{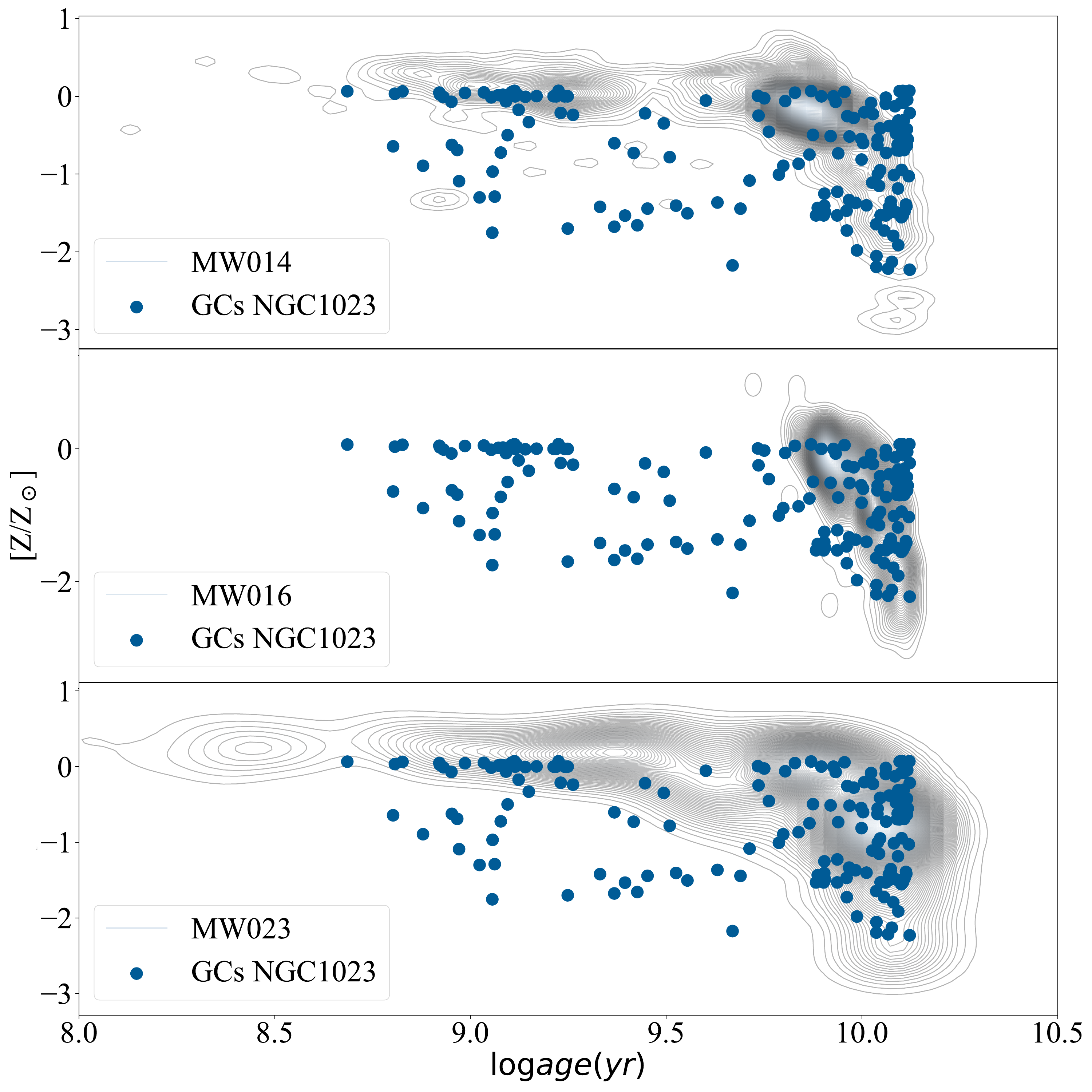}
\caption{The AMR for the  NGC\,1023 GC system  based on \dynb\ 3 (blue circles) compared to 3 E-MOSAIC simulations: MW014, MW016 and MW023 from \cite{kruijssen2019mosaics} (grey contours) that have comparable halo masses to NGC\,1023. We observe a likely broad AMR of NGC\,1023 GC system, where the objects have a broad metallicity distribution and are mainly old.}
\label{fig:agemet}
\end{figure*}

We can further investigate this scenario by studying age--metallicities relations. In Figure \ref{fig:agemet} we compare the age-metalliticy relation (hereafter, AMR) obtained from our results to 3 E-MOSAIC simulations \citep{pfeffer2018mosaics} from \cite{kruijssen2019mosaics}. 
We choose 3 simulations that have halo masses comparable to the halo mass of NGC1023 adopting the masses from \citet{alabi2016sluggs} and \citet{bilek2019study}. 

The comparison in Figure \ref{fig:agemet} is useful to get a handle on the epoch of GC assembly in the NGC\,1023. We note that the AMR of MW023 is in better agreement with the AMR of NGC\,1023 GC candidates. We also note that our AMR has outliers, similar to the ones found in MW014. Our results then favor the accretion histories of simulations such as MW014 or MW023 and rule out histories such as the one in MW016. The AMR of GC candidates when compared with the simulations indicate that NGC 1023 likely experienced an initial and rapid phase of star formation that might have formed the majority of the GC candidates, since a big amount of GC candidates were formed early in galaxy evolution considering that the age distribution has a dominating peak at $\approx$ $10^{10}$ yr, agreeing with results from \cite{kruijssen2019mosaics}. 

We also note a likely broad AMR, where the GC candidates have a wide range of metallicities. Nevertheless, there is a caveat regarding our interpretation of a broad AMR, as the selection function introduced by \texttt{GCFinder} is not well characterized. As such it remains an open question if the color cuts applied by the pipeline would introduce distortions in the age-metallicities relation.
\cite{kruijssen2019mosaics} find that a wide range of GC metallicities was related to a wide range of progenitor masses. Therefore, we believe our possible broad age--metallicity relation and our wide range of metallicities imply that NGC\,1023 experienced mergers and accretion events in the past, resulting in more than one episode of intense star formation.

This is in sync with what is been discovered about the formation of the Milky Way.
Studies about the formation of our Galaxy are motivated by many surveys created in the last years and generating huge amounts of data. In particular, Gaia survey \citep{GaiaCollaboration+2016a,GaiaCollaboration+2016b,brown2018gaia,brown2021gaia} have been revolutionising our understanding about the Milky Way. Many articles in recent years making use of Gaia were published identifying stars in the Milky Way that are claimed to be accreted from dwarf galaxies that no longer exist. In particular, stars have been claimed to be born in the progenitor galaxies Gaia-Enceladus \citep{Helmi2018,Belokurov2018,das2020ages} (which is believed to be the last major merger experienced from the Milky Way),  from the Sequoia progenitor \citep{Myeong2019}, from Thamnos 1 and Thamnos 2 \citep{Koppelman2019}, from a structure in the inner Galaxy \citep{kruijssen2019formation,kruijssen2020kraken,horta2021evidence}, among others.

We note that there is evidence from kinematic studies (e.g. \citealt{romanowsky2012ongoing,villaume2019new}) and simulations (e.g. \citealt{muratov2010modeling,choksi2018formation}) about accretion events from dwarf galaxies and that the hierarchical assembly of GC systems is well accepted. In particular, the fact that we only find evidence of color bimodality in some cases is not surprising. 
GCs trace assembly histories of galaxies and galaxies likely undergo many minor and possibly major mergers throughout their life. In this case, the lack of strong evidence of color bimodality for some colors could be an indicator that more than two subpopulations exist, but we are not able to disentangle them. \cite{puzia2002extragalactic} and \cite{blom2012sluggs,blom2012wide}, for example, found three subpopulations of GCs in the galaxy NGC\,4365.

\subsection{The ongoing interaction with NGC\,1023A}

\begin{figure}
\centering
\includegraphics[width=\columnwidth]{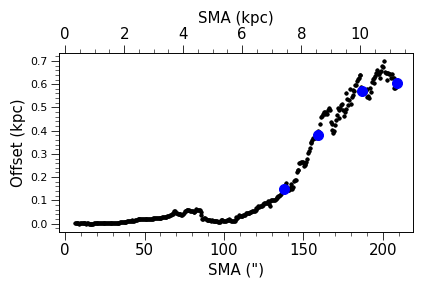}
\includegraphics[width=\columnwidth]{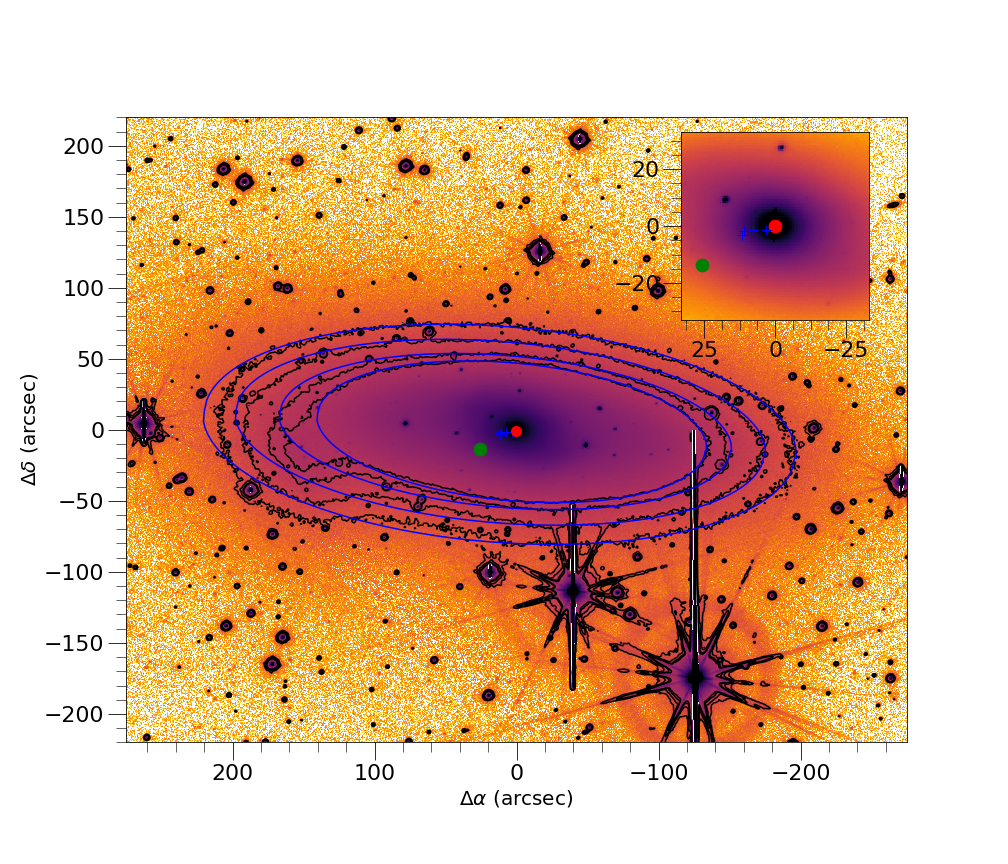}
\caption{NGC\,1023 displaced nucleus. Top panel: the offset radius profile with respect to the photometric center of the galaxy. 
Bottom panel: selected outer isophotes (in black) with their respective fitted ellipses (blue)  are overplotted on r-band image. The blue pluses are their respective centers, while the red point is the photometric center of the galaxy and the green point is the photometric center of GCs candidates. The top-right inset is a zoom-in of the inner part to see better the displacements. The blue points in the offset radius profile are the offsets for the selected isophotes.
}
\label{fig:isophote_r}
\end{figure}

NGC\,1023 is a barred galaxy in an ongoing interaction with  NGC\,1023A, a small companion at the outskirt on NGC\,1023 at East \citep{debattista02}. NGC\,1023A was recognised as an independent galaxy by \citet{barbon75}  and designated as NGC\,1023A by \citet{hart80}. \citet{capaccioli86} classified it as a Magellanic irregular or late-type dwarf galaxy, while \citet{sancisi84} from HI observation found a complex kinematic. On the other hand,
\citet{debattista02} found a faster bar pattern speed, which would be not compatible with a scenario of a  recent formation of the bar by the interaction with NGC\,1023A. Nevertheless, the ongoing interaction may have an effect on the overall NGC\,1023  structure as well as on the GCs distribution. A small fraction of the GC population of the NGC 1023 system could be associated with NGC\,1023A \citep{Cortesi2016}. 
Due to the morphology and luminosity of NGC\,1023A, a definition of its center is challenging, but we estimate that NGC\,1023A is at a projected distance of approximately 6.7 kpc from NGC\,1023.

Since NGC\,1023 is an early type galaxy the tidal effect could have two different dynamics answers depending on how longer or shorter the encounter time is compared to the galaxy's internal cross-time  \citep[e.g,][]{aguilar85,aguilar86,binney08}. For the outer parts the crossing time could be larger than the encounter time, therefore they may suffer an impulse response. On the other hand, for the inner parts, the cross-time could be smaller than the encounter time, thus they may suffer a typical tidal response.  As a consequence, the central parts can be displaced with respect to the outer ones. These offsets can be up to  $20\%$ of the observable radius of the galaxy \citep{lauer86,lauer88,davoust88,combes95, gonzales00,mora19,buzzo21}.

\cite{buzzo21} show that the nucleus of the lenticular galaxy NGC\,3115 has a displacement of 160\,arcsec with respect to the outer parts. They interpret it as the result of a recent pericenter passage with its close small companion. To probe if NGC\,1023 has a similar feature, we perform a similar photometric analysis following \citet{mora19} and \citet{buzzo21}. We model the isophote contours, in $r-$band, by using the \texttt{ELLIPSE} task from \texttt{IRAF} \citep{ellipse87}, we let free the position angle, ellipticity, and centre of the ellipses. To quantify the offset of the isophotes, we take as reference the photometric center of the galaxy. In Fig. \ref{fig:isophote_r},  we show the radial profile of the offsets at the top panel, while the outermost isophotes and their respective fitted ellipses with their centers are plotted at the bottom panel. It is clear that from $\sim$100\, arcsec the central part starts to move toward the East-South with respect to the outer parts, the maximum offset is about $\sim$700\,pc.  This nuclear displacement is strong evidence that NGC\,1023  and  NGC\,1023A had recently a pericenter passage, just a few hundred million years ago \citep{combes95,mora19}. The orientation of the offset could be used as a  strong constraint in a numerical simulation of the dynamic encounter of this pair, since the central part of NGC\,1023 must have headed into the East-South direction at the pericenter passage. This would limit the family of possible orbits to model the system. \citep[e.g,][]{combes95,mora19}.

How this interaction could have affected the distribution of the GCs of NGC1023?  To address this question, we calculate the photometric center of the GCs candidates and overlay it on the bottom panel in Fig. \ref{fig:isophote_r}. We can see that the photometric center of the GCs candidates follows the same direction as the centers of the outer isophotes. The displacement of the nucleus region with respect to the GCs is $\sim$ 1600\,pc. This behavior is expected according to impulse theory \citep{aguilar85,aguilar86,binney08} since the GCs belong to the galactic halo then they have the largest cross times of the galaxy, hence the nuclear displacement should be the largest one.   

\begin{figure*}
\centering
\includegraphics[width=\textwidth]{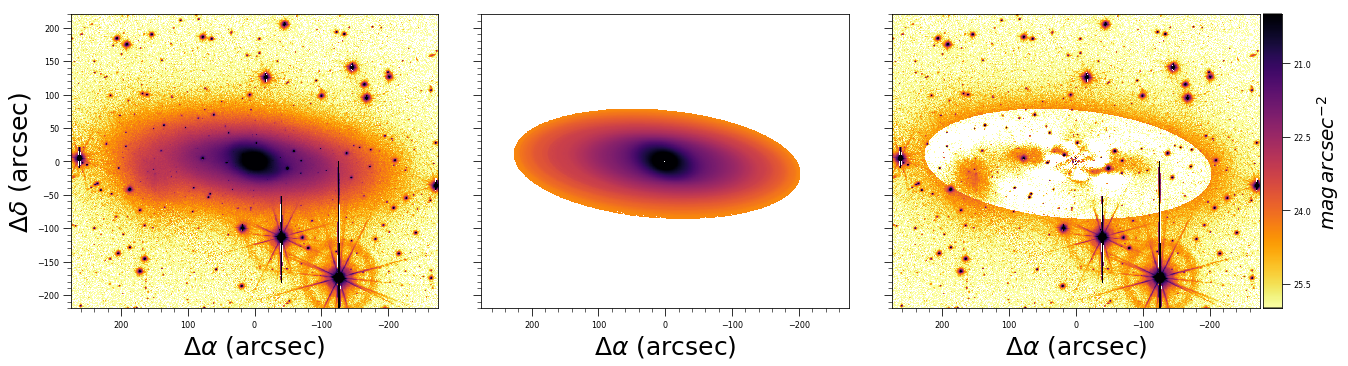}
\caption{Photometric model for NGC\,1023. Left panel:
The r-band image. Middle panel: the ellipse model. Right panel:
the residual map. NGC\,1023A is reveled in this map, together
with two spiral arms and an inner bar.}
\label{fig:model_r}
\end{figure*}

In addition to this photometric analysis, we study the residual image from the ellipse model, see Fig. \ref{fig:model_r}. The residual image unveils NGC\,1023A and the bar (oriented North-East to South-West) of NGC\,1023 \citep{mollenhoff01,debattista02}, besides two possible relic like-spiral arms. The bar has a radius of $\sim$1100\,pc. 
We note that here we are reporting the presence of these relic spiral-like arms for the first time. It is very plausible that the origin of these structures is also due to the interaction with NGC\,1023A, in this case, they would be tidal structures. However, one of the formation mechanisms of lenticular galaxies is gas removal from a spiral galaxy, then these relic spiral-like arms could be a memory of the progenitor galaxy of NGC\,1023. These features can also serve as dynamical constraints for a numerical simulation of the encounter.

\section{Summary and Conclusion}\label{sec:conclusion}

In this work, we present the first study about extragalactic globular clusters using J-PLUS data. As a test case, we detect and study the GC system in NGC\,1023 with the 12 bands of J-PLUS.
To detect the GC candidates we develop \texttt{GCFinder}, a code that can be applied to current and upcoming wide-field multi-band surveys, such as J-PAS and S-PLUS.
The pipeline presents good results within the characteristics of the survey for which it was designed. The end product is a code that 
can be adapted to other photometric surveys and other types of compact stellar systems, such as ultra-compact dwarf galaxies.

With the catalog of GC candidates provided by \texttt{GCFinder}, we perform a study of the stellar population content of a sub-sample of objects, using SED fitting techniques and photometry from broad and narrow--band filters. To calculate masses, ages, and metallicities of the GC candidates we use the codes \texttt{DynBaS3} and \texttt{TGASPEX} adapted to work with the J-PLUS filter system. We also carefully model the light of NGC\,1023 to investigate possible displacement between the outer isophotes and the distribution center of the GC candidates, which is useful to understand galaxy evolution.

In the following lines we summarize our main findings:

\begin{itemize}
\item We identify 523 GC candidates in NGC 1023 using \texttt{GCFinder}, being 335 of them not yet reported in the literature. A significant part of these new GC candidates is located in the outer regions of NGC\,1023 since we took advantage of the wide field of view of J-PLUS images. We find a specific frequency of S$_N$ = 2.1 $\pm$ 0.2, which is consistent with estimations for lenticular galaxies in the literature.
\item We investigate color distributions of the GC candidates, exploring the novel colors provided by J-PLUS. According to BIC statistics, we find evidence of color bimodality in 17 colors ($u-g$, $u-r$, $u-i$,
$g-r$, $r-z$, $i-z$, $J0378-J0430$, $J0378-J0515$, $J0378-J0660$, $J0378-J0861$, $J0410-J0515$, $J0430-J0515$, $J0430-J0660$, $J0430-J0861$, $J0515-J0660$, $J0515-J0861$, $J0660-J0861$), while according to ICL statistics we find evidence of color bimodality in 10 colors ($g-r$, $r-z$, $i-z$, $J0378-J0515$, $J0378-J0861$, $J0430-J0660$, $J0430-J0861$, $J0515-J0660$, $J0515-J0861$, $J0660-J0861$).
\item We obtain masses, ages, and metallicities for 171 GC candidates from SED fitting. We find that the peak of the mass distribution is at $10^{5.5}$ M$_\odot$. We also find a tail of GC candidates with low masses, which we interpret as likely contaminants in our list of candidates. It is unclear at this point if the more massive systems (stellar mass $>$ $10^6$ M$\odot$) are GCs or UCDs.
\item The mass-weighted and light-weighted age distributions cover a wide range of ages, with a dominant population at $\approx$ $10^{10}$ yr. The mass-weighted and light-weighted metallicity distributions of the GC candidates are bimodal in most of the cases (combining different filter sets and codes) and, in a minority of cases, we find three peaks. We note that the inclusion of narrow--band filters helps to constrain the metallicities. These results indicate that there are subpopulations of GCs in NGC\,1023 that could exist due to accretion events or different epochs/mechanisms of star formation in the galaxy.
\item We identify a correlation between light-weighted ages and stellar masses, where older GCs tend to be more massive. This suggests that massive GC candidates in NGC\,1023 are more likely to survive the turbulent history of the host galaxy than less massive objects, which is in agreement with the literature on GC systems of galaxies.
\item The age-metallicity relation is likely broad. A comparison with simulations shows evidence of a likely initial rapid phase of star formation, responsible for the formation of the majority of the GCs. Articles in the literature \citep[e.g.][]{kruijssen2019mosaics} found that a wide range of GC metallicities is related to a wide range of masses of progenitor galaxies. Therefore, the broad AMR we find is also evidence of past accretion events experienced by NGC\,1023.
\item We detect that the photometric center has a displacement of
$\sim$700\,pc and $\sim$1600\,pc with respect to the outer isophotes and the GC candidate distribution center, respectively. The offsets could be the result of ongoing interaction between NGC\,1023 and NGC\,1023A. These effects are in excellent agreement with impulse theory \citep{aguilar85,aguilar86,binney08}. 
\item The residual map obtained from the photometric model of NGC\,1023 unveils two spiral-like arms. These structures are probably due to the NGC\,1023 interaction with the NGC\,1023A satellite galaxy.
\end{itemize}

From our main findings, we conclude that it was possible to retrieve new and valuable information about the evolutionary past of NGC\,1023 as observed by J-PLUS. The multiple GC populations, relic spiral arms, and displacement between the photometric center of the GC candidates and the isophotal center that we report in this work support a formation of NGC\,1023 that involved several minor mergers and group harassment, causing a transformation from spiral to the nowadays lenticular galaxy.

\begin{acknowledgements}
We thank the referee for the valuable comments and careful revision that helped us to improve this manuscript.
We thank Diederik Kruijssen for sharing the AMRs presented in \cite{kruijssen2019mosaics}.
\newline
D.B.S. acknowledges Paula Jofr\'e for the scientific discussions, mentoring, all the support, and for the constant and invaluable availability.
D.B.S. also acknowledges Funda\c{c}\~ao de Amparo \`{a} Pesquisa do Estado de S\~ao Paulo (FAPESP) process number 2017/00204-6 for the financial support provided for the development of this project.  
\newline
PC acknowledges support from Conselho Nacional de Desenvolvimento Cient\'ifico e Tecnol\'ogico (CNPq) under grant 310041/2018-0 and from Funda\c{c}\~ao de Amparo \`{a} Pesquisa do Estado de S\~ao Paulo (FAPESP) process number 2018/05392-8.
\newline
ACS acknowledges funding from CNPq and the Rio Grande do Sul Research Foundation (FAPERGS) through grants CNPq-403580/2016-1, CNPq-11153/2018-6, PqG/FAPERGS-17/2551-0001, FAPERGS/CAPES 19/2551-0000696-9 and L'Or\'eal UNESCO ABC \emph{Para Mulheres na Ci\^encia} and the Chinese Academy of Sciences (CAS) President's International Fellowship Initiative (PIFI) through grant E085201009.
\newline
GB acknowledges financial support from the National Autonomous University of M\'exico (UNAM) through grant DGAPA/PAPIIT IG100319 and from CONACyT through grant CB2015-252364.
\newline
J. V. acknowledges the technical members of the UPAD for their invaluable work: Juan Castillo, Tamara Civera, Javier Hern\'andez, \'Angel L\'opez, Alberto Moreno, and David Muniesa.
\newline
JAHJ acknowledges Fundaç\~ao de Amparo \`{a} Pesquisa do Estado de S\~ao Paulo (FAPESP), process number 2021/08920-8.
\newline
AE acknowledges the financial support from the Spanish Ministry of Science and Innovation and the European Union - NextGenerationEU through the Recovery and Resilience Facility project ICTS-MRR-2021-03-CEFCA
and from Conselho Nacional de Desenvolvimento Cient\'ifico e Tecnol\'ogico (CNPq) under grant 313285/2020-9
\newline
DAF thanks the ARC for financial assistance via DP170102344.
\newline
Y. J-T has received funding from the European Union’s Horizon 2020 
research and innovation program under the Marie Skłodowska-Curie grant agreement No 898633. Y. J-T. also acknowledges financial support from 
the State Agency for Research of the Spanish MCIU through the “Center of Excellence Severo Ochoa” award to the Instituto de Astrofísica de 
Andalucía (SEV-2017-0709).
\newline
Based on observations made with the JAST80 telescope telescope/s at the Observatorio Astrof\'isico de Javalambre, in Teruel, owned, managed, and operated by the Centro de Estudios de F\'isica del Cosmos de Arag\'on. We thank the Centro de Estudios de F\'isica del Cosmos de Arag\'on for the allocation of the Director’s Discretionary Time to this program. We thank the OAJ Data Processing and Archiving Unit (UPAD) for reducing and calibrating the OAJ data used in this work.
\newline
Funding for the J-PLUS Project has been provided by the Governments of Spain and Arag\'on through the Fondo de Inversiones de Teruel; the Arag\'on Government through the Research Groups E96, E103, and E16\_17R; the Spanish Ministry of Science, Innovation, and Universities (MCIU/AEI/FEDER, UE) with grants PGC2018-097585-B-C21 and PGC2018-097585-B-C22; the Spanish Ministry of Economy and Competitiveness (MINECO) under AYA2015-66211-C2-1-P, AYA2015-66211-C2-2, AYA2012-30789, and ICTS-2009-14; and European FEDER funding (FCDD10-4E-867, FCDD13-4E-2685). The Brazilian agencies FINEP, FAPESP, and the National Observatory of Brazil have also contributed to this project.
\newline
This work has made use of the computing facilities of the Laboratory of Astroinformatics (IAG/USP, NAT/Unicsul), whose purchase was made possible by the Brazilian agency FAPESP (grant 2009/54006-4) and the INCT-A.
\newline
This work has made use of data from the European Space Agency (ESA) mission
{\it Gaia} (\url{https://www.cosmos.esa.int/gaia}), processed by the {\it Gaia}
Data Processing and Analysis Consortium (DPAC,
\url{https://www.cosmos.esa.int/web/gaia/dpac/consortium}). Funding for the DPAC
has been provided by national institutions, in particular, the institutions
participating in the {\it Gaia} Multilateral Agreement.
\newline
The Pan-STARRS1 Surveys (PS1) and the PS1 public science archive have been made possible through contributions by the Institute for Astronomy, the University of Hawaii, the Pan-STARRS Project Office, the Max-Planck Society, and its participating institutes, the Max Planck Institute for Astronomy, Heidelberg and the Max Planck Institute for Extraterrestrial Physics, Garching, The Johns Hopkins University, Durham University, the University of Edinburgh, the Queen's University Belfast, the Harvard-Smithsonian Center for Astrophysics, the Las Cumbres Observatory Global Telescope Network Incorporated, the National Central University of Taiwan, the Space Telescope Science Institute, the National Aeronautics and Space Administration under Grant No. NNX08AR22G was issued through the Planetary Science Division of the NASA Science Mission Directorate, the National Science Foundation Grant No. AST-1238877, the University of Maryland, Eotvos Lorand University (ELTE), the Los Alamos National Laboratory, and the Gordon and Betty Moore Foundation.
    
\end{acknowledgements}

\bibliographystyle{aa}
\bibliography{bib.bib}

\begin{thebibliography}{136}
\expandafter\ifx\csname natexlab\endcsname\relax\def\natexlab#1{#1}\fi

\bibitem[{{Aguilar} \& {White}(1985)}]{aguilar85}
{Aguilar}, L.~A. \& {White}, S.~D.~M. 1985, \apj, 295, 374

\bibitem[{{Aguilar} \& {White}(1986)}]{aguilar86}
{Aguilar}, L.~A. \& {White}, S.~D.~M. 1986, \apj, 307, 97

\bibitem[{Alabi {et~al.}(2016)Alabi, Forbes, Romanowsky, Brodie, Strader, Janz, Pota, Pastorello, Usher, Spitler, {et~al.}}]{alabi2016sluggs}
Alabi, A.~B., Forbes, D.~A., Romanowsky, A.~J., {et~al.} 2016, Monthly Notices of the Royal Astronomical Society, 460, 3838

\bibitem[{Alamo-Mart{\'\i}nez \& Blakeslee(2017)}]{alamo2017specific}
Alamo-Mart{\'\i}nez, K. \& Blakeslee, J. 2017, The Astrophysical Journal, 849, 6

\bibitem[{{Bamford} {et~al.}(2011){Bamford}, {H{\"a}u{\ss}ler}, {Rojas}, \& {Borch}}]{Bamford2011}
{Bamford}, S.~P., {H{\"a}u{\ss}ler}, B., {Rojas}, A., \& {Borch}, A. 2011, in Astronomical Society of the Pacific Conference Series, Vol. 442, Astronomical Data Analysis Software and Systems XX, ed. I.~N. {Evans}, A.~{Accomazzi}, D.~J. {Mink}, \& A.~H. {Rots}, 479

\bibitem[{{Barbon} \& {Capaccioli}(1975)}]{barbon75}
{Barbon}, R. \& {Capaccioli}, M. 1975, \aap, 42, 103

\bibitem[{{Barden} {et~al.}(2012){Barden}, {H{\"a}u{\ss}ler}, {Peng}, {McIntosh}, \& {Guo}}]{Barden2012}
{Barden}, M., {H{\"a}u{\ss}ler}, B., {Peng}, C.~Y., {McIntosh}, D.~H., \& {Guo}, Y. 2012, {GALAPAGOS: Galaxy Analysis over Large Areas: Parameter Assessment by GALFITting Objects from SExtractor}, Astrophysics Source Code Library

\bibitem[{{Beasley}(2020)}]{beasley20}
{Beasley}, M.~A. 2020, {Globular Cluster Systems and Galaxy Formation}, 245--277

\bibitem[{Beasley {et~al.}(2008)Beasley, Bridges, Peng, Harris, Harris, Forbes, \& Mackie}]{beasley20082df}
Beasley, M.~A., Bridges, T., Peng, E., {et~al.} 2008, Monthly Notices of the Royal Astronomical Society, 386, 1443

\bibitem[{Belokurov {et~al.}(2018)Belokurov, Erkal, Evans, Koposov, \& Deason}]{Belokurov2018}
Belokurov, V., Erkal, D., Evans, N., Koposov, S., \& Deason, A. 2018, Monthly Notices of the Royal Astronomical Society, 478, 611

\bibitem[{Bender {et~al.}(1988)Bender, Doebereiner, \& Moellenhoff}]{bender1988isophote}
Bender, R., Doebereiner, S., \& Moellenhoff, C. 1988, Astronomy and Astrophysics Supplement Series, 74, 385

\bibitem[{Benitez {et~al.}(2014)Benitez, Dupke, Moles, Sodre, Cenarro, Marin-Franch, Taylor, Cristobal, Fernandez-Soto, de~Oliveira, {et~al.}}]{benitez2014j}
Benitez, N., Dupke, R., Moles, M., {et~al.} 2014, arXiv preprint arXiv:1403.5237

\bibitem[{{Berriman} {et~al.}(2004){Berriman}, {Deelman}, {Good}, {Jacob}, {Katz}, {Kesselman}, {Laity}, {Prince}, {Singh}, \& {Su}}]{Berriman2004}
{Berriman}, G.~B., {Deelman}, E., {Good}, J.~C., {et~al.} 2004, in Proceedings of the International Society for Optical Engineering, Vol. 5493, Optimizing Scientific Return for Astronomy through Information Technologies, ed. P.~J. {Quinn} \& A.~{Bridger}, 221--232

\bibitem[{{Bertin} \& {Arnouts}(1996)}]{Bertin1996}
{Bertin}, E. \& {Arnouts}, S. 1996, \aaps, 117, 393

\bibitem[{B{\'\i}lek {et~al.}(2019)B{\'\i}lek, Samurovi{\'c}, \& Renaud}]{bilek2019study}
B{\'\i}lek, M., Samurovi{\'c}, S., \& Renaud, F. 2019, Astronomy \& Astrophysics, 625, A32

\bibitem[{{Binney} \& {Tremaine}(2008)}]{binney08}
{Binney}, J. \& {Tremaine}, S. 2008, {Galactic Dynamics: Second Edition}

\bibitem[{{Blakeslee} {et~al.}(2012){Blakeslee}, {Cho}, {Peng}, {Ferrarese}, {Jord{\'a}n}, \& {Martel}}]{Blakeslee2012}
{Blakeslee}, J.~P., {Cho}, H., {Peng}, E.~W., {et~al.} 2012, \apj, 746, 88

\bibitem[{Blom {et~al.}(2012{\natexlab{a}})Blom, Forbes, Brodie, Foster, Romanowsky, Spitler, \& Strader}]{blom2012sluggs}
Blom, C., Forbes, D.~A., Brodie, J.~P., {et~al.} 2012{\natexlab{a}}, Monthly Notices of the Royal Astronomical Society, 426, 1959

\bibitem[{Blom {et~al.}(2012{\natexlab{b}})Blom, Spitler, \& Forbes}]{blom2012wide}
Blom, C., Spitler, L.~R., \& Forbes, D.~A. 2012{\natexlab{b}}, Monthly Notices of the Royal Astronomical Society, 420, 37

\bibitem[{Brodie {et~al.}(2014)Brodie, Romanowsky, Strader, Forbes, Foster, Jennings, Pastorello, Pota, Usher, Blom, {et~al.}}]{brodie2014sages}
Brodie, J.~P., Romanowsky, A.~J., Strader, J., {et~al.} 2014, The Astrophysical Journal, 796, 52

\bibitem[{{Brodie} \& {Strader}(2006)}]{Brodie2006}
{Brodie}, J.~P. \& {Strader}, J. 2006, \araa, 44, 193

\bibitem[{{Brodie} {et~al.}(2012){Brodie}, {Usher}, {Conroy}, {Strader}, {Arnold}, {Forbes}, \& {Romanowsky}}]{brodie2012}
{Brodie}, J.~P., {Usher}, C., {Conroy}, C., {et~al.} 2012, \apjl, 759, L33

\bibitem[{Brown {et~al.}(2018)Brown, Vallenari, Prusti, De~Bruijne, Babusiaux, Bailer-Jones, Biermann, Evans, Eyer, Jansen, {et~al.}}]{brown2018gaia}
Brown, A., Vallenari, A., Prusti, T., {et~al.} 2018, Astronomy \& astrophysics, 616, A1

\bibitem[{Brown {et~al.}(2021)Brown, Vallenari, Prusti, De~Bruijne, Babusiaux, Biermann, Creevey, Evans, Eyer, Hutton, {et~al.}}]{brown2021gaia}
Brown, A.~G., Vallenari, A., Prusti, T., {et~al.} 2021, Astronomy \& Astrophysics, 649, A1

\bibitem[{{Bruzual} \& {Charlot}(2003)}]{Bruzual2003}
{Bruzual}, G. \& {Charlot}, S. 2003, \mnras, 344, 1000

\bibitem[{Buzzo {et~al.}(2022)Buzzo, Cortesi, Forbes, Brodie, Couch, Eduardo~Barbosa, de~Brito~Silva, Coelho, Chies-Santos, Escudero, {et~al.}}]{buzzo2022new}
Buzzo, M.~L., Cortesi, A., Forbes, D.~A., {et~al.} 2022, Monthly Notices of the Royal Astronomical Society, 510, 1383

\bibitem[{{Buzzo} {et~al.}(2021){Buzzo}, {Cortesi}, {Hernandez-Jimenez}, {Coccato}, {Werle}, {Beraldo e Silva}, {Grossi}, {Vika}, {Barbosa}, {Lucatelli}, {Santana-Silva}, {Bamford}, {Debattista}, {Forbes}, {Overzier}, {Romanowsky}, {Ferrari}, {Brodie}, \& {Mendes de Oliveira}}]{buzzo21}
{Buzzo}, M.~L., {Cortesi}, A., {Hernandez-Jimenez}, J.~A., {et~al.} 2021, \mnras, 504, 2146

\bibitem[{Cantiello \& Blakeslee(2007)}]{cantiello2007metallicity}
Cantiello, M. \& Blakeslee, J.~P. 2007, The Astrophysical Journal, 669, 982

\bibitem[{{Cantiello} {et~al.}(2014){Cantiello}, {Blakeslee}, {Raimondo}, {Chies-Santos}, {Jennings}, {Norris}, \& {Kuntschner}}]{Cantiello2014}
{Cantiello}, M., {Blakeslee}, J.~P., {Raimondo}, G., {et~al.} 2014, \aap, 564, L3

\bibitem[{{Capaccioli} {et~al.}(1986){Capaccioli}, {Lorenz}, \& {Afanasjev}}]{capaccioli86}
{Capaccioli}, M., {Lorenz}, H., \& {Afanasjev}, V.~L. 1986, \aap, 169, 54

\bibitem[{Cenarro {et~al.}(2019)Cenarro, Moles, Crist{\'o}bal-Hornillos, Mar{\'\i}n-Franch, Ederoclite, Varela, L{\'o}pez-Sanjuan, Hern{\'a}ndez-Monteagudo, Angulo, Rami{\'o}, {et~al.}}]{cenarro2019j}
Cenarro, A.~J., Moles, M., Crist{\'o}bal-Hornillos, D., {et~al.} 2019, Astronomy \& Astrophysics, 622, A176

\bibitem[{{Chabrier}(2003)}]{chabrier2003}
{Chabrier}, G. 2003, \pasp, 115, 763

\bibitem[{Chambers {et~al.}(2016)Chambers, Magnier, Metcalfe, Flewelling, Huber, Waters, Denneau, Draper, Farrow, Finkbeiner, {et~al.}}]{chambers2016pan}
Chambers, K.~C., Magnier, E., Metcalfe, N., {et~al.} 2016, arXiv preprint arXiv:1612.05560

\bibitem[{{Chen} {et~al.}(2015){Chen}, {Bressan}, {Girardi}, {Marigo}, {Kong}, \& {Lanza}}]{chen2015}
{Chen}, Y., {Bressan}, A., {Girardi}, L., {et~al.} 2015, \mnras, 452, 1068

\bibitem[{Chies-Santos {et~al.}(2011)Chies-Santos, Larsen, Kuntschner, Anders, Wehner, Strader, Brodie, \& Santos}]{chies2011optical}
Chies-Santos, A., Larsen, S., Kuntschner, H., {et~al.} 2011, Astronomy \& Astrophysics, 525, A20

\bibitem[{{Chies-Santos} {et~al.}(2013){Chies-Santos}, {Cortesi}, {Fantin}, {Merrifield}, {Bamford}, \& {Serra}}]{Chies2013}
{Chies-Santos}, A.~L., {Cortesi}, A., {Fantin}, D.~S.~M., {et~al.} 2013, \aap, 559, A67

\bibitem[{Chies-Santos {et~al.}(2022)Chies-Santos, de~Souza, Caso, Ennis, de~Souza, Barbosa, Chen, Cenarro, Ederoclite, Crist{\'o}bal-Hornillos, {et~al.}}]{chies2022j}
Chies-Santos, A.~L., de~Souza, R.~S., Caso, J.~P., {et~al.} 2022, arXiv preprint arXiv:2202.11472

\bibitem[{{Cho} {et~al.}(2016){Cho}, {Blakeslee}, {Chies-Santos}, {Jee}, {Jensen}, {Peng}, \& {Lee}}]{Cho2016}
{Cho}, H., {Blakeslee}, J.~P., {Chies-Santos}, A.~L., {et~al.} 2016, \apj, 822, 95

\bibitem[{Choksi \& Gnedin(2019)}]{choksi2019formation}
Choksi, N. \& Gnedin, O.~Y. 2019, Monthly Notices of the Royal Astronomical Society, 486, 331

\bibitem[{Choksi {et~al.}(2018)Choksi, Gnedin, \& Li}]{choksi2018formation}
Choksi, N., Gnedin, O.~Y., \& Li, H. 2018, Monthly Notices of the Royal Astronomical Society, 480, 2343

\bibitem[{Chung {et~al.}(2016)Chung, Yoon, Lee, \& Lee}]{chung2016nonlinear}
Chung, C., Yoon, S.-J., Lee, S.-Y., \& Lee, Y.-W. 2016, The Astrophysical Journal, 818, 201

\bibitem[{{Ciambur}(2015)}]{Ciambur2015}
{Ciambur}, B.~C. 2015, \apj, 810, 120

\bibitem[{{Cid Fernandes} {et~al.}(2013){Cid Fernandes}, {P{\'e}rez}, {Garc{\'{\i}}a Benito}, {Gonz{\'a}lez Delgado}, {de Amorim}, {S{\'a}nchez}, {Husemann}, {Falc{\'o}n Barroso}, {S{\'a}nchez-Bl{\'a}zquez}, {Walcher}, \& {Mast}}]{cid-fernandes2013}
{Cid Fernandes}, R., {P{\'e}rez}, E., {Garc{\'{\i}}a Benito}, R., {et~al.} 2013, \aap, 557, A86

\bibitem[{{Coe} {et~al.}(2012){Coe}, {Umetsu}, {Zitrin}, {Donahue}, {Medezinski}, {Postman}, {Carrasco}, {Anguita}, {Geller}, {Rines}, {Diaferio}, {Kurtz}, {Bradley}, {Koekemoer}, {Zheng}, {Nonino}, {Molino}, {Mahdavi}, {Lemze}, {Infante}, {Ogaz}, {Melchior}, {Host}, {Ford}, {Grillo}, {Rosati}, {Jim{\'e}nez-Teja}, {Moustakas}, {Broadhurst}, {Ascaso}, {Lahav}, {Bartelmann}, {Ben{\'{\i}}tez}, {Bouwens}, {Graur}, {Graves}, {Jha}, {Jouvel}, {Kelson}, {Moustakas}, {Maoz}, {Meneghetti}, {Merten}, {Riess}, {Rodney}, \& {Seitz}}]{Coe2012}
{Coe}, D., {Umetsu}, K., {Zitrin}, A., {et~al.} 2012, \apj, 757, 22

\bibitem[{{Combes} {et~al.}(1995){Combes}, {Rampazzo}, {Bonfanti}, {Prugniel}, \& {Sulentic}}]{combes95}
{Combes}, F., {Rampazzo}, R., {Bonfanti}, P.~P., {Prugniel}, P., \& {Sulentic}, J.~W. 1995, \aap, 297, 37

\bibitem[{{Cortesi} {et~al.}(2016){Cortesi}, {Chies-Santos}, {Pota}, {Foster}, {Coccato}, {Mendes de Oliveira}, {Forbes}, {Merrifield}, {Bamford}, {Romanowsky}, {Brodie}, {Kartha}, {Alabi}, {Proctor}, \& {Almeida}}]{Cortesi2016}
{Cortesi}, A., {Chies-Santos}, A.~L., {Pota}, V., {et~al.} 2016, \mnras, 456, 2611

\bibitem[{{Cortesi} {et~al.}(2011){Cortesi}, {Merrifield}, \& {Arnaboldi}}]{Cortesi2011}
{Cortesi}, A., {Merrifield}, M., \& {Arnaboldi}, M. 2011, in Astrophysics and Space Science Proceedings, Vol.~27, Environment and the Formation of Galaxies: 30 years later, 109

\bibitem[{Costa-Duarte {et~al.}(2019)Costa-Duarte, Sampedro, Molino, Xavier, Herpich, Chies-Santos, Barbosa, Cortesi, Schoenell, Kanaan, {et~al.}}]{costa2019s}
Costa-Duarte, M., Sampedro, L., Molino, A., {et~al.} 2019, arXiv preprint arXiv:1909.08626

\bibitem[{{Crist{\'o}bal-Hornillos} {et~al.}(2014){Crist{\'o}bal-Hornillos}, {Varela}, {Ederoclite}, {V{\'a}zquez Rami{\'o}}, {L{\'o}pez-Sainz}, {Hern{\'a}ndez-Fuertes}, {Civera}, {Muniesa}, {Moles}, {Cenarro}, {Mar{\'{\i}}n-Franch}, \& {Yanes-D{\'{\i}}az}}]{cristobal+14}
{Crist{\'o}bal-Hornillos}, D., {Varela}, J., {Ederoclite}, A., {et~al.} 2014, in \procspie, Vol. 9152, Software and Cyberinfrastructure for Astronomy III, 91520O

\bibitem[{Das {et~al.}(2020)Das, Hawkins, \& Jofr{\'e}}]{das2020ages}
Das, P., Hawkins, K., \& Jofr{\'e}, P. 2020, Monthly Notices of the Royal Astronomical Society, 493, 5195

\bibitem[{{Davoust} \& {Prugniel}(1988)}]{davoust88}
{Davoust}, E. \& {Prugniel}, P. 1988, \aap, 201, L30

\bibitem[{De~Souza {et~al.}(2017)De~Souza, Dantas, Costa-Duarte, Feigelson, Killedar, Lablanche, Vilalta, Krone-Martins, Beck, \& Gieseke}]{de2017probabilistic}
De~Souza, R., Dantas, M., Costa-Duarte, M., {et~al.} 2017, Monthly Notices of the Royal Astronomical Society, 472, 2808

\bibitem[{{Debattista} {et~al.}(2002){Debattista}, {Corsini}, \& {Aguerri}}]{debattista02}
{Debattista}, V.~P., {Corsini}, E.~M., \& {Aguerri}, J.~A.~L. 2002, \mnras, 332, 65

\bibitem[{Dolfi {et~al.}(2021)Dolfi, Forbes, Couch, Bekki, Ferr{\'e}-Mateu, Romanowsky, \& Brodie}]{dolfi2021sluggs}
Dolfi, A., Forbes, D.~A., Couch, W.~J., {et~al.} 2021, Monthly Notices of the Royal Astronomical Society, 504, 4923

\bibitem[{Eadie {et~al.}(2022)Eadie, Harris, \& Springford}]{eadie2022clearing}
Eadie, G.~M., Harris, W.~E., \& Springford, A. 2022, The Astrophysical Journal, 926, 162

\bibitem[{Elmegreen(2000)}]{elmegreen2000specific}
Elmegreen, B.~G. 2000, in Toward a New Millennium in Galaxy Morphology (Springer), 469--484

\bibitem[{Fahrion {et~al.}(2020)Fahrion, Lyubenova, Hilker, van~de Ven, Falc{\'o}n-Barroso, Leaman, Mart{\'\i}n-Navarro, Bittner, Coccato, Corsini, {et~al.}}]{fahrion2020fornax}
Fahrion, K., Lyubenova, M., Hilker, M., {et~al.} 2020, Astronomy \& Astrophysics, 637, A27

\bibitem[{{Falc{\'o}n-Barroso} {et~al.}(2011){Falc{\'o}n-Barroso}, {S{\'a}nchez-Bl{\'a}zquez}, {Vazdekis}, {Ricciardelli}, {Cardiel}, {Cenarro}, {Gorgas}, \& {Peletier}}]{falcon-barroso2011}
{Falc{\'o}n-Barroso}, J., {S{\'a}nchez-Bl{\'a}zquez}, P., {Vazdekis}, A., {et~al.} 2011, \aap, 532, A95

\bibitem[{Finlator {et~al.}(2000)Finlator, Ivezi{\'c}, Fan, Strauss, Knapp, Lupton, Gunn, Rockosi, Anderson, Csabai, {et~al.}}]{finlator2000optical}
Finlator, K., Ivezi{\'c}, {\v{Z}}., Fan, X., {et~al.} 2000, The Astronomical Journal, 120, 2615

\bibitem[{{Forbes} {et~al.}(2014){Forbes}, {Almeida}, {Spitler}, \& {Pota}}]{Forbes2014}
{Forbes}, D.~A., {Almeida}, A., {Spitler}, L.~R., \& {Pota}, V. 2014, \mnras, 442, 1049

\bibitem[{{Forbes} {et~al.}(2016){Forbes}, {Romanowsky}, {Pastorello}, {Foster}, {Brodie}, {Strader}, {Usher}, \& {Pota}}]{forbes2016}
{Forbes}, D.~A., {Romanowsky}, A.~J., {Pastorello}, N., {et~al.} 2016, \mnras, 457, 1242

\bibitem[{{Gaia Collaboration} {et~al.}(2016{\natexlab{a}}){Gaia Collaboration}, {Brown}, {Vallenari}, {Prusti}, {de Bruijne}, {Mignard}, {Drimmel}, {Babusiaux}, {Bailer-Jones}, {Bastian}, \& et~al.}]{GaiaCollaboration+2016a}
{Gaia Collaboration}, {Brown}, A.~G.~A., {Vallenari}, A., {et~al.} 2016{\natexlab{a}}, \aap, 595, A2

\bibitem[{{Gaia Collaboration} {et~al.}(2016{\natexlab{b}}){Gaia Collaboration}, {Prusti}, {de Bruijne}, {Brown}, {Vallenari}, {Babusiaux}, {Bailer-Jones}, {Bastian}, {Biermann}, {Evans}, \& et~al.}]{GaiaCollaboration+2016b}
{Gaia Collaboration}, {Prusti}, T., {de Bruijne}, J.~H.~J., {et~al.} 2016{\natexlab{b}}, \aap, 595, A1

\bibitem[{{Gonz{\'a}lez Delgado} {et~al.}(2021){Gonz{\'a}lez Delgado}, {D{\'\i}az-Garc{\'\i}a}, {de Amorim}, {Bruzual}, {Cid Fernandes}, {P{\'e}rez}, {Bonoli}, {Cenarro}, {Coelho}, {Cortesi}, {Garc{\'\i}a-Benito}, {L{\'o}pez Fern{\'a}ndez}, {Mart{\'\i}nez-Solaeche}, {Rodr{\'\i}guez-Mart{\'\i}n}, {Magris}, {Mej{\'\i}a-Narvaez}, {Brito-Silva}, {Abramo}, {Diego}, {Dupke}, {Hern{\'a}n-Caballero}, {Hern{\'a}ndez-Monteagudo}, {L{\'o}pez-Sanjuan}, {Mar{\'\i}n-Franch}, {Marra}, {Moles}, {Montero-Dorta}, {Queiroz}, {Sodr{\'e}}, {Varela}, {V{\'a}zquez Rami{\'o}}, {V{\'\i}lchez}, {Baqui}, {Ben{\'\i}tez}, {Crist{\'o}bal-Hornillos}, {Ederoclite}, {Mendes de Oliveira}, {Civera}, {Muniesa}, {Taylor}, {Tempel}, \& {J-PAS Collaboration}}]{gd+21}
{Gonz{\'a}lez Delgado}, R.~M., {D{\'\i}az-Garc{\'\i}a}, L.~A., {de Amorim}, A., {et~al.} 2021, \aap, 649, A79

\bibitem[{{Gonz{\'a}lez-Serrano} \& {Carballo}(2000)}]{gonzales00}
{Gonz{\'a}lez-Serrano}, J.~I. \& {Carballo}, R. 2000, \aaps, 142, 353

\bibitem[{Goto {et~al.}(2002)Goto, Sekiguchi, Nichol, Bahcall, Kim, Annis, Ivezi{\'c}, Brinkmann, Hennessy, Szokoly, {et~al.}}]{goto2002cut}
Goto, T., Sekiguchi, M., Nichol, R.~C., {et~al.} 2002, The Astronomical Journal, 123, 1807

\bibitem[{Harris(2001)}]{harris2001globular}
Harris, W.~E. 2001, in Star clusters (Springer), 223--408

\bibitem[{{Hart} {et~al.}(1980){Hart}, {Davies}, \& {Johnson}}]{hart80}
{Hart}, L., {Davies}, R.~D., \& {Johnson}, S.~C. 1980, \mnras, 191, 269

\bibitem[{H{\"a}u{\ss}ler {et~al.}(2013)H{\"a}u{\ss}ler, Bamford, Vika, Rojas, Barden, Kelvin, Alpaslan, Robotham, Driver, Baldry, {et~al.}}]{haussler2013megamorph}
H{\"a}u{\ss}ler, B., Bamford, S.~P., Vika, M., {et~al.} 2013, Monthly Notices of the Royal Astronomical Society, 430, 330

\bibitem[{Helmi {et~al.}(2018)Helmi, Babusiaux, Koppelman, Massari, Veljanoski, \& Brown}]{Helmi2018}
Helmi, A., Babusiaux, C., Koppelman, H.~H., {et~al.} 2018, Nature, 563, 85

\bibitem[{Horta {et~al.}(2021)Horta, Schiavon, Mackereth, Pfeffer, Mason, Kisku, Fragkoudi, Allende~Prieto, Cunha, Hasselquist, {et~al.}}]{horta2021evidence}
Horta, D., Schiavon, R.~P., Mackereth, J.~T., {et~al.} 2021, Monthly Notices of the Royal Astronomical Society, 500, 1385

\bibitem[{Ivezi{\'c} {et~al.}(2014)Ivezi{\'c}, Connolly, VanderPlas, \& Gray}]{ivezic2014statistics}
Ivezi{\'c}, {\v{Z}}., Connolly, A.~J., VanderPlas, J.~T., \& Gray, A. 2014, Statistics, data mining, and machine learning in astronomy (Princeton University Press)

\bibitem[{{Jedrzejewski}(1987)}]{ellipse87}
{Jedrzejewski}, R.~I. 1987, \mnras, 226, 747

\bibitem[{Jedrzejewski(1987)}]{jedrzejewski1987ccd}
Jedrzejewski, R.~I. 1987, Monthly Notices of the Royal Astronomical Society, 226, 747

\bibitem[{{Kartha} {et~al.}(2014){Kartha}, {Forbes}, {Spitler}, {Romanowsky}, {Arnold}, \& {Brodie}}]{Kartha2014}
{Kartha}, S.~S., {Forbes}, D.~A., {Spitler}, L.~R., {et~al.} 2014, \mnras, 437, 273

\bibitem[{Kim {et~al.}(2021)Kim, Yoon, Lee, Chung, \& Sohn}]{kim2021nonlinear}
Kim, S., Yoon, S.-J., Lee, S.-Y., Chung, C., \& Sohn, S.~T. 2021, The Astrophysical Journal Supplement Series, 256, 29

\bibitem[{Koppelman {et~al.}(2019)Koppelman, Helmi, Massari, Price-Whelan, \& Starkenburg}]{Koppelman2019}
Koppelman, H.~H., Helmi, A., Massari, D., Price-Whelan, A.~M., \& Starkenburg, T.~K. 2019, Astronomy \& Astrophysics, 631, L9

\bibitem[{Kruijssen {et~al.}(2020)Kruijssen, Pfeffer, Chevance, Bonaca, Trujillo-Gomez, Bastian, Reina-Campos, Crain, \& Hughes}]{kruijssen2020kraken}
Kruijssen, J.~D., Pfeffer, J.~L., Chevance, M., {et~al.} 2020, Monthly Notices of the Royal Astronomical Society, 498, 2472

\bibitem[{Kruijssen {et~al.}(2019{\natexlab{a}})Kruijssen, Pfeffer, Crain, \& Bastian}]{kruijssen2019mosaics}
Kruijssen, J.~D., Pfeffer, J.~L., Crain, R.~A., \& Bastian, N. 2019{\natexlab{a}}, Monthly Notices of the Royal Astronomical Society, 486, 3134

\bibitem[{Kruijssen {et~al.}(2019{\natexlab{b}})Kruijssen, Pfeffer, Reina-Campos, Crain, \& Bastian}]{kruijssen2019formation}
Kruijssen, J.~D., Pfeffer, J.~L., Reina-Campos, M., Crain, R.~A., \& Bastian, N. 2019{\natexlab{b}}, Monthly Notices of the Royal Astronomical Society, 486, 3180

\bibitem[{Kundu \& Whitmore(1998)}]{kundu1998wide}
Kundu, A. \& Whitmore, B.~C. 1998, The Astronomical Journal, 116, 2841

\bibitem[{Kundu \& Zepf(2007)}]{kundu2007bimodal}
Kundu, A. \& Zepf, S.~E. 2007, The Astrophysical Journal Letters, 660, L109

\bibitem[{{Larsen}(2001)}]{Larsen2001}
{Larsen}, S.~S. 2001, \aj, 122, 1782

\bibitem[{Larsen {et~al.}(2001)Larsen, Brodie, Huchra, Forbes, \& Grillmair}]{larsen2001properties}
Larsen, S.~S., Brodie, J.~P., Huchra, J.~P., Forbes, D.~A., \& Grillmair, C.~J. 2001, The Astronomical Journal, 121, 2974

\bibitem[{{Lauer}(1986)}]{lauer86}
{Lauer}, T.~R. 1986, \apj, 311, 34

\bibitem[{{Lauer}(1988)}]{lauer88}
{Lauer}, T.~R. 1988, \apj, 325, 49

\bibitem[{{Lawson} \& {Hanson}(1974)}]{lawson1974}
{Lawson}, C.~L. \& {Hanson}, R.~J. 1974, {Solving least squares problems} (Prentice-Hall, Inc., Englewood Cliffs, New Jersey)

\bibitem[{Lee {et~al.}(2018)Lee, Chung, \& Yoon}]{lee2018nonlinear}
Lee, S.-Y., Chung, C., \& Yoon, S.-J. 2018, The Astrophysical Journal Supplement Series, 240, 2

\bibitem[{Lee {et~al.}(2020)Lee, Chung, \& Yoon}]{lee2020nonlinear}
Lee, S.-Y., Chung, C., \& Yoon, S.-J. 2020, The Astrophysical Journal, 905, 124

\bibitem[{Lenzen {et~al.}(2003)Lenzen, Hartung, Brandner, Finger, Hubin, Lacombe, Lagrange, Lehnert, Moorwood, \& Mouillet}]{lenzen2003naos}
Lenzen, R., Hartung, M., Brandner, W., {et~al.} 2003, in Instrument Design and Performance for Optical/Infrared Ground-based Telescopes, Vol. 4841, International Society for Optics and Photonics, 944--952

\bibitem[{Li \& Gnedin(2019)}]{li2019formation}
Li, H. \& Gnedin, O. 2019, Proceedings of the International Astronomical Union, 14, 34

\bibitem[{Li \& Gnedin(2014)}]{li2014modeling}
Li, H. \& Gnedin, O.~Y. 2014, The Astrophysical Journal, 796, 10

\bibitem[{L{\'o}pez-Sanjuan {et~al.}(2019)L{\'o}pez-Sanjuan, Rami{\'o}, Varela, Spinoso, Angulo, Muniesa, Viironen, Crist{\'o}bal-Hornillos, Cenarro, Ederoclite, {et~al.}}]{lopez2019j}
L{\'o}pez-Sanjuan, C., Rami{\'o}, H.~V., Varela, J., {et~al.} 2019, Astronomy \& Astrophysics, 622, A177

\bibitem[{Lotz {et~al.}(2004)Lotz, Miller, \& Ferguson}]{lotz2004colors}
Lotz, J.~M., Miller, B.~W., \& Ferguson, H.~C. 2004, The Astrophysical Journal, 613, 262

\bibitem[{{Magris C.} {et~al.}(2015){Magris C.}, {Mateu P.}, {Mateu}, {Bruzual A.}, {Cabrera-Ziri}, \& {Mej{\'{\i}}a-Narv{\'a}ez}}]{magris+15}
{Magris C.}, G., {Mateu P.}, J., {Mateu}, C., {et~al.} 2015, \pasp, 127, 16

\bibitem[{{Marigo} {et~al.}(2013){Marigo}, {Bressan}, {Nanni}, {Girardi}, \& {Pumo}}]{marigo2013}
{Marigo}, P., {Bressan}, A., {Nanni}, A., {Girardi}, L., \& {Pumo}, M.~L. 2013, \mnras, 434, 488

\bibitem[{{Mar{\'{\i}}n-Franch} {et~al.}(2015){Mar{\'{\i}}n-Franch}, {Taylor}, {Cenarro}, {Cristobal-Hornillos}, \& {Moles}}]{t80cam}
{Mar{\'{\i}}n-Franch}, A., {Taylor}, K., {Cenarro}, J., {Cristobal-Hornillos}, D., \& {Moles}, M. 2015, in IAU General Assembly, Vol.~29, 2257381

\bibitem[{{Mej{\'{\i}}a-Narv{\'a}ez} {et~al.}(2017){Mej{\'{\i}}a-Narv{\'a}ez}, {Bruzual}, {Magris}, {Alcaniz}, {Ben{\'{\i}}tez}, {Carneiro}, {Cenarro}, {Crist{\'o}bal-Hornillos}, {Dupke}, {Ederoclite}, {Mar{\'{\i}}n-Franch}, {de Oliveira}, {Moles}, {Sodre}, {Taylor}, {Varela}, \& {Rami{\'o}}}]{mejia+17}
{Mej{\'{\i}}a-Narv{\'a}ez}, A., {Bruzual}, G., {Magris}, C.~G., {et~al.} 2017, \mnras, 471, 4722

\bibitem[{Mendes~de Oliveira {et~al.}(2019)Mendes~de Oliveira, Ribeiro, Schoenell, Kanaan, Overzier, Molino, Sampedro, Coelho, Barbosa, Cortesi, {et~al.}}]{mendes2019southern}
Mendes~de Oliveira, C., Ribeiro, T., Schoenell, W., {et~al.} 2019, Monthly Notices of the Royal Astronomical Society, 489, 241

\bibitem[{{M{\"o}llenhoff} \& {Heidt}(2001)}]{mollenhoff01}
{M{\"o}llenhoff}, C. \& {Heidt}, J. 2001, \aap, 368, 16

\bibitem[{{Mora} {et~al.}(2019){Mora}, {Torres-Flores}, {Firpo}, {Hernandez-Jimenez}, {Urrutia-Viscarra}, \& {Mendes de Oliveira}}]{mora19}
{Mora}, M.~D., {Torres-Flores}, S., {Firpo}, V., {et~al.} 2019, \mnras, 488, 830

\bibitem[{{Morganti} {et~al.}(2006){Morganti}, {de Zeeuw}, {Oosterloo}, {McDermid}, {Krajnovi{\'c}}, {Cappellari}, {Kenn}, {Weijmans}, \& {Sarzi}}]{Morganti2006}
{Morganti}, R., {de Zeeuw}, P.~T., {Oosterloo}, T.~A., {et~al.} 2006, \mnras, 371, 157

\bibitem[{Muratov \& Gnedin(2010)}]{muratov2010modeling}
Muratov, A.~L. \& Gnedin, O.~Y. 2010, The Astrophysical Journal, 718, 1266

\bibitem[{Myeong {et~al.}(2019)Myeong, Vasiliev, Iorio, Evans, \& Belokurov}]{Myeong2019}
Myeong, G., Vasiliev, E., Iorio, G., Evans, N., \& Belokurov, V. 2019, arXiv preprint arXiv:1904.03185

\bibitem[{Nakazono {et~al.}(2021)Nakazono, de~Oliveira, Hirata, Jeram, Queiroz, Eikenberry, Gonzalez, Abramo, Overzier, Espadoto, {et~al.}}]{nakazono2021discovery}
Nakazono, L., de~Oliveira, C.~M., Hirata, N., {et~al.} 2021, arXiv preprint arXiv:2106.11986

\bibitem[{{Noordermeer} {et~al.}(2008){Noordermeer}, {Merrifield}, {Coccato}, {Arnaboldi}, {Capaccioli}, {Douglas}, {Freeman}, {Gerhard}, {Kuijken}, {de Lorenzi}, {Napolitano}, \& {Romanowsky}}]{Noordermeer2008}
{Noordermeer}, E., {Merrifield}, M.~R., {Coccato}, L., {et~al.} 2008, \mnras, 384, 943

\bibitem[{Pedregosa {et~al.}(2011)Pedregosa, Varoquaux, Gramfort, Michel, Thirion, Grisel, Blondel, Prettenhofer, Weiss, Dubourg, Vanderplas, Passos, Cournapeau, Brucher, Perrot, \& Duchesnay}]{scikit-learn}
Pedregosa, F., Varoquaux, G., Gramfort, A., {et~al.} 2011, Journal of Machine Learning Research, 12, 2825

\bibitem[{{Peng} {et~al.}(2002){Peng}, {Ho}, {Impey}, \& {Rix}}]{Peng2002}
{Peng}, C.~Y., {Ho}, L.~C., {Impey}, C.~D., \& {Rix}, H.-W. 2002, \aj, 124, 266

\bibitem[{Peng {et~al.}(2006)Peng, Jord{\'a}n, C{\^o}t{\'e}, Blakeslee, Ferrarese, Mei, West, Merritt, Milosavljevi{\'c}, \& Tonry}]{peng2006acs}
Peng, E.~W., Jord{\'a}n, A., C{\^o}t{\'e}, P., {et~al.} 2006, The Astrophysical Journal, 639, 95

\bibitem[{Pfeffer {et~al.}(2018)Pfeffer, Kruijssen, Crain, \& Bastian}]{pfeffer2018mosaics}
Pfeffer, J., Kruijssen, J.~D., Crain, R.~A., \& Bastian, N. 2018, Monthly Notices of the Royal Astronomical Society, 475, 4309

\bibitem[{Phillipps {et~al.}(2001)Phillipps, Drinkwater, Gregg, \& Jones}]{phillipps2001ultracompact}
Phillipps, S., Drinkwater, M., Gregg, M., \& Jones, J. 2001, The Astrophysical Journal, 560, 201

\bibitem[{{Plat} {et~al.}(2019){Plat}, {Charlot}, {Bruzual}, {Feltre}, {Vidal-Garc{\'\i}a}, {Morisset}, {Chevallard}, \& {Todt}}]{plat2019}
{Plat}, A., {Charlot}, S., {Bruzual}, G., {et~al.} 2019, \mnras, 490, 978

\bibitem[{Prakash {et~al.}(2015)Prakash, Licquia, Newman, \& Rao}]{prakash2015luminous}
Prakash, A., Licquia, T.~C., Newman, J.~A., \& Rao, S.~M. 2015, The Astrophysical Journal, 803, 105

\bibitem[{{Prugniel} {et~al.}(2011){Prugniel}, {Vauglin}, \& {Koleva}}]{prugniel2011}
{Prugniel}, P., {Vauglin}, I., \& {Koleva}, M. 2011, \aap, 531, A165

\bibitem[{Puzia {et~al.}(2002)Puzia, Zepf, Kissler-Patig, Hilker, Minniti, \& Goudfrooij}]{puzia2002extragalactic}
Puzia, T.~H., Zepf, S.~E., Kissler-Patig, M., {et~al.} 2002, Astronomy \& Astrophysics, 391, 453

\bibitem[{Richtler(2005)}]{richtler2005some}
Richtler, T. 2005, arXiv preprint astro-ph/0512545

\bibitem[{Romanowsky {et~al.}(2012)Romanowsky, Strader, Brodie, Mihos, Spitler, Forbes, Foster, \& Arnold}]{romanowsky2012ongoing}
Romanowsky, A.~J., Strader, J., Brodie, J.~P., {et~al.} 2012, The astrophysical journal, 748, 29

\bibitem[{{S{\'a}nchez-Bl{\'a}zquez} {et~al.}(2006){S{\'a}nchez-Bl{\'a}zquez}, {Peletier}, {Jim{\'e}nez-Vicente}, {Cardiel}, {Cenarro}, {Falc{\'o}n-Barroso}, {Gorgas}, {Selam}, \& {Vazdekis}}]{sanchez-blazquez2006}
{S{\'a}nchez-Bl{\'a}zquez}, P., {Peletier}, R.~F., {Jim{\'e}nez-Vicente}, J., {et~al.} 2006, \mnras, 371, 703

\bibitem[{{Sancisi} {et~al.}(1984{\natexlab{a}}){Sancisi}, {van Woerden}, {Davies}, \& {Hart}}]{Sancisi1984}
{Sancisi}, R., {van Woerden}, H., {Davies}, R.~D., \& {Hart}, L. 1984{\natexlab{a}}, \mnras, 210, 497

\bibitem[{{Sancisi} {et~al.}(1984{\natexlab{b}}){Sancisi}, {van Woerden}, {Davies}, \& {Hart}}]{sancisi84}
{Sancisi}, R., {van Woerden}, H., {Davies}, R.~D., \& {Hart}, L. 1984{\natexlab{b}}, \mnras, 210, 497

\bibitem[{{Schlafly} \& {Finkbeiner}(2011)}]{schlafly_finkbeiner11}
{Schlafly}, E.~F. \& {Finkbeiner}, D.~P. 2011, \apj, 737, 103

\bibitem[{{Sharma} {et~al.}(2016){Sharma}, {Prugniel}, \& {Singh}}]{sharma2016}
{Sharma}, K., {Prugniel}, P., \& {Singh}, H.~P. 2016, \aap, 585, A64

\bibitem[{Strader {et~al.}(2005)Strader, Brodie, Cenarro, Beasley, \& Forbes}]{strader2005extragalactic}
Strader, J., Brodie, J.~P., Cenarro, A., Beasley, M.~A., \& Forbes, D.~A. 2005, The Astronomical Journal, 130, 1315

\bibitem[{Strauss {et~al.}(2002)Strauss, Weinberg, Lupton, Narayanan, Annis, Bernardi, Blanton, Burles, Connolly, Dalcanton, {et~al.}}]{strauss2002spectroscopic}
Strauss, M.~A., Weinberg, D.~H., Lupton, R.~H., {et~al.} 2002, The Astronomical Journal, 124, 1810

\bibitem[{{Taylor}(2006)}]{Taylor2006A}
{Taylor}, M.~B. 2006, in Astronomical Society of the Pacific Conference Series, Vol. 351, Astronomical Data Analysis Software and Systems XV, ed. C.~{Gabriel}, C.~{Arviset}, D.~{Ponz}, \& S.~{Enrique}, 666

\bibitem[{{Tody}(1993)}]{iraf1993}
{Tody}, D. 1993, in Astronomical Society of the Pacific Conference Series, Vol.~52, Astronomical Data Analysis Software and Systems II, ed. R.~J. {Hanisch}, R.~J.~V. {Brissenden}, \& J.~{Barnes}, 173

\bibitem[{{Tully}(1980)}]{Tully1980}
{Tully}, R.~B. 1980, \apj, 237, 390

\bibitem[{Usher {et~al.}(2012)Usher, Forbes, Brodie, Foster, Spitler, Arnold, Romanowsky, Strader, \& Pota}]{usher2012sluggs}
Usher, C., Forbes, D.~A., Brodie, J.~P., {et~al.} 2012, Monthly Notices of the Royal Astronomical Society, 426, 1475

\bibitem[{{Valdes} {et~al.}(2004){Valdes}, {Gupta}, {Rose}, {Singh}, \& {Bell}}]{valdes2004}
{Valdes}, F., {Gupta}, R., {Rose}, J.~A., {Singh}, H.~P., \& {Bell}, D.~J. 2004, \apjs, 152, 251

\bibitem[{Varela {et~al.}(2009)Varela, d'Onofrio, Marmo, Fasano, Bettoni, Cava, Couch, Dressler, Kj{\ae}rgaard, Moles, {et~al.}}]{varela2009wings}
Varela, J., d'Onofrio, M., Marmo, C., {et~al.} 2009, Astronomy \& Astrophysics, 497, 667

\bibitem[{Villaume {et~al.}(2019)Villaume, Romanowsky, Brodie, \& Strader}]{villaume2019new}
Villaume, A., Romanowsky, A.~J., Brodie, J., \& Strader, J. 2019, The Astrophysical Journal, 879, 45

\bibitem[{Wang {et~al.}(2021)Wang, Bai, Yuan, Wang, \& Liu}]{wang2021machine}
Wang, C., Bai, Y., Yuan, H., Wang, S., \& Liu, J. 2021, arXiv preprint arXiv:2106.12787

\bibitem[{Yong {et~al.}(2012)Yong, Norris, Bessell, Christlieb, Asplund, Beers, Barklem, Frebel, \& Ryan}]{yong2012most}
Yong, D., Norris, J.~E., Bessell, M.~S., {et~al.} 2012, The Astrophysical Journal, 762, 26

\bibitem[{Yoon {et~al.}(2011)Yoon, Lee, Blakeslee, Peng, Sohn, Cho, Kim, Chung, Kim, \& Lee}]{yoon2011nonlinearb}
Yoon, S.-J., Lee, S.-Y., Blakeslee, J.~P., {et~al.} 2011, The Astrophysical Journal, 743, 150

\bibitem[{Yoon {et~al.}(2006)Yoon, Yi, \& Lee}]{yoon2006explaining}
Yoon, S.-J., Yi, S.~K., \& Lee, Y.-W. 2006, Science, 311, 1129

\bibitem[{York {et~al.}(2000)York, Adelman, Anderson~Jr, Anderson, Annis, Bahcall, Bakken, Barkhouser, Bastian, Berman, {et~al.}}]{york2000sloan}
York, D.~G., Adelman, J., Anderson~Jr, J.~E., {et~al.} 2000, The Astronomical Journal, 120, 1579

\end{thebibliography}

\begin{appendix}

\section{Identifying GC candidates: the \texttt{GCFinder} pipeline}\label{sec:methods_gcfinder_appendix}

\subsection{Handling the host galaxy}\label{subsec:modeling}

Historically, the first step when investigating extragalactic globular clusters is modeling the surface brightness of the host galaxy (e.g. \citealt{Forbes2014,Kartha2014,Cho2016}).
This is done to enhance the detection of point-like objects inlaid in the extended galaxy halo light. Following this approach, we first attempt to remove the smooth galaxy light profiles from the individual images, and to perform that step we carry out numerous tests with different software, named ELLIPSE \citep{iraf1993}, ISOFIT \citep{Ciambur2015} and GALFITM \citep{Bamford2011,haussler2013megamorph}, 
as well as median smoothing technique. A challenge found was that NGC\,1023 has a very large image size of approximately 700 X 260 pixels ($\approx$ 0.004 square degrees),
which makes the modeling very time-consuming as well as computational power consuming. The other main challenge found was that NGC\,1023 has a companion that overlaps with it in the image. As a consequence, when we subtract the model from the observed image, the residual image did not have the desired quality. More details about the different methods tested as well as about intermediate results are presented in Appendix \ref{different_methods}.

From the several tests we carry out, we learn that we are not able to retrieve GCs projected over the central brightest regions of the galaxy, even when modeling and subtracting the galaxy's two-dimensional light profile. We, therefore, explore alternative methods to retrieve GCs in J-PLUS.

\subsection{GCFinder}\label{subsec:GCFinder}

To detect and select GC candidates in J-PLUS-like images we develop a pipeline named \texttt{GCFinder}, that consists of an approach that does not require modeling the host galaxy and is based on a careful detection of GC candidates using  Source Extractor \citep{Bertin1996} and criteria based on the data quality, morphology, color and magnitude of the objects.
GCs are not detected \textit{a priori} by the data reduction pipeline of J-PLUS (JYPE, \citealt{cristobal+14}), therefore developing a straightforward way to detect and select these objects 
is fundamental to perform GC studies for a large sample of galaxies. 

\paragraph{White image:}
The detection image is 
a "white" image, that is, an image originated from the sum of frames of 4 broad--bands ($g$, $r$, $i$, and $z$), while the photometry was performed in each band independently.
We do not use the u filter because it has a low response (see \citealt{cenarro2019j}) and it could include noise to the white image. The use of a white image increases the chances of detecting faint sources, that are harder to detect in separate bands. 
To construct the white images, Montage \citep{Berriman2004} is 
used and included in the pipeline. With the use of Montage, it 
is possible to align the images before combining them, to build white images without the displacement of frames of different bands and it is  
also possible to perform background correction on the studied images. It is important to avoid displacement of frames because it could introduce an effect of expanding the objects as well as it could produce fake detections since light could be detected in false positions. This methodology was adopted to increase the signal-to-noise ratio of the sources and thus enhance object detection. To prevent possible noise associated with the narrow--band images from being introduced in the detection image, only the broad--bands were adopted in the construction of the white images.

\paragraph{Detection of point-like sources as GC candidates: }
To perform detection of GCs that also includes objects close to the center of NGC\,1023, we perform
extensive testing of the different input parameters of Source Extractor to optimize our detection. We identify 3 key input parameters of Source Extractor to perform the detection of GCs under these conditions: BACK$\_$SIZE, BACK$\_$FILTERSIZE, and PHOT$\_$AUTOPARAMS. BACK$\_$SIZE determines the pixel size of the area used to estimate the background and is
one of the most important parameters. If the BACK$\_$SIZE is too small, the background estimate can be affected by the presence of objects and noise and it is also possible that part of the surrounding galaxy light
is absorbed in the background map. If the BACK$\_$SIZE is too large, it does not consider small variations in the background. BACK$\_$FILTERSIZE is the parameter that controls the size of the filter used to estimate the background. Finally, PHOT$\_$AUTOPARAMS is the parameter that controls the elliptical opening used for object detection.

\begin{figure}
\centering
\includegraphics[width=9cm]{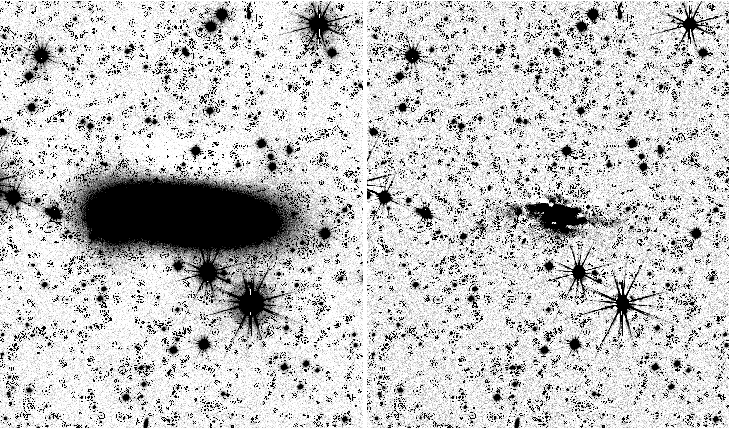}
\caption{Left panel: zoom in the NGC\,1023 image. 
Right panel: Illustration of how Source Extractor interprets the image with the chosen input parameters.}  
\label{fig:metodosex}
\end{figure}

We note  
that when these key parameters are included as a function of the FWHM of each image, Source Extractor does not consider the extended light profile of the galactic halo in its detection, making it possible to recover the GCs in this inner region of NGC\,1023 as shown in Figure \ref{fig:metodosex}. The functions of the adopted BACK$\_$SIZE, BACK$\_$FILTERSIZE and PHOT$\_$AUTOPARAMS can be seen in Equations \ref{eq1}, \ref{eq2} and \ref{eq3}, respectively. For more details about the input parameters of \texttt{GCFinder}, please see Appendix \ref{sec:append}.
 
\begin{equation}
    {\rm BACK\_SIZE} = 4 \cdot {\rm FWHM} \cdot 1.05
\label{eq1}
\end{equation}
\begin{equation}
    {\rm BACK\_FILTERSIZE}= 2 \cdot {\rm FWHM} \cdot 1.05
\label{eq2}
\end{equation}
\begin{equation}
    \text{Minimum radius for } {\rm PHOT\_AUTOPARAMS} = 3 \cdot {\rm FWHM} \cdot 1.05
\label{eq3}
\end{equation}

The factor of 1.05 appears to increase the FWHM value by 5\% to compensate for variations through the field of the images since the Point Spread Function (PSF) in the images used in the work is not homogenized. In this work, we always use magnitude MAG$\_$AUTO.

After detecting all sources using Source Extractor in dual mode, the pipeline performs the selection of GC candidates. We adopt criteria based on the shape, magnitude, color, and data quality of the objects.

\begin{figure}
\centering
\includegraphics[width=8cm]{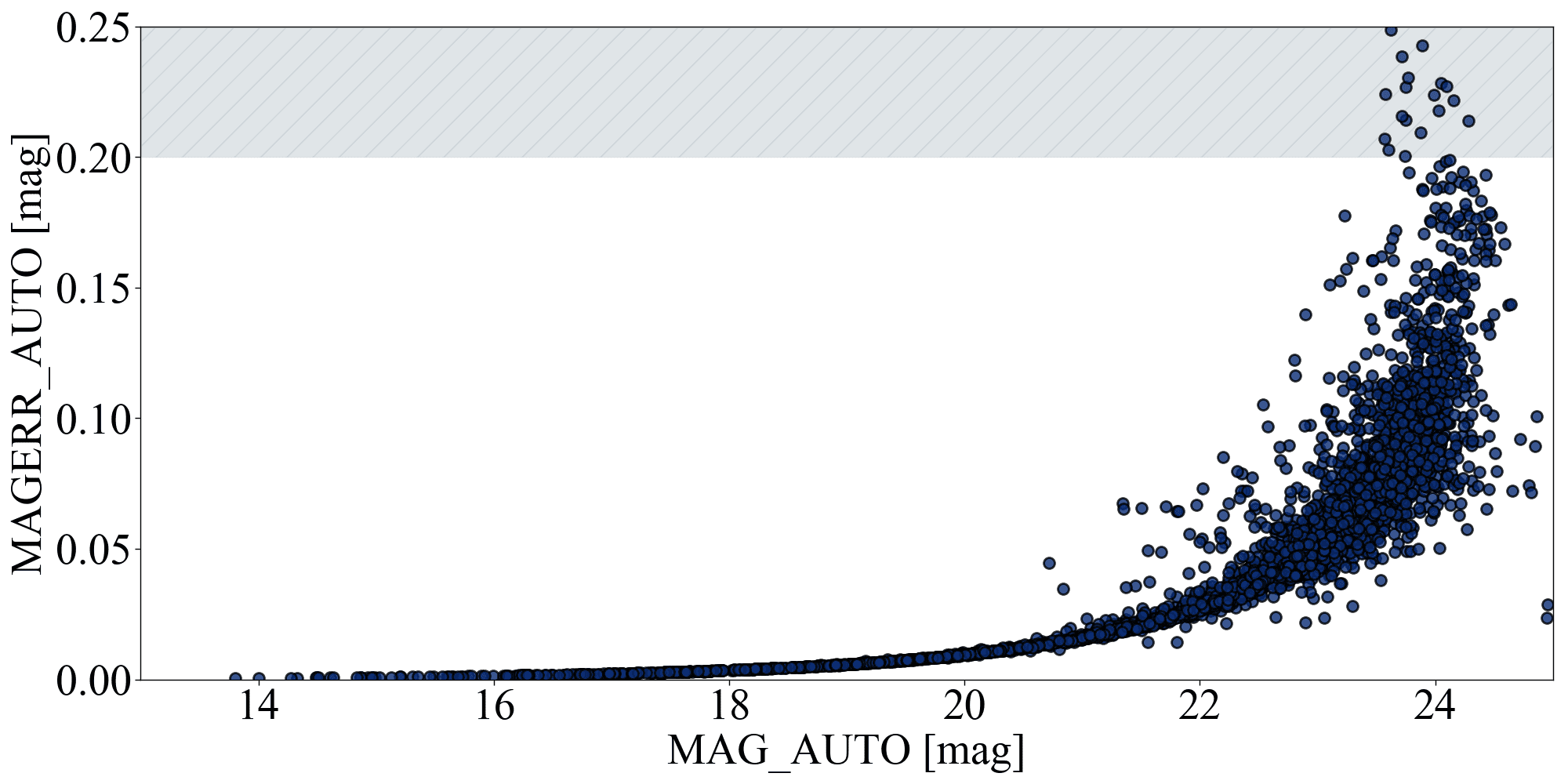}
\includegraphics[width=8cm]{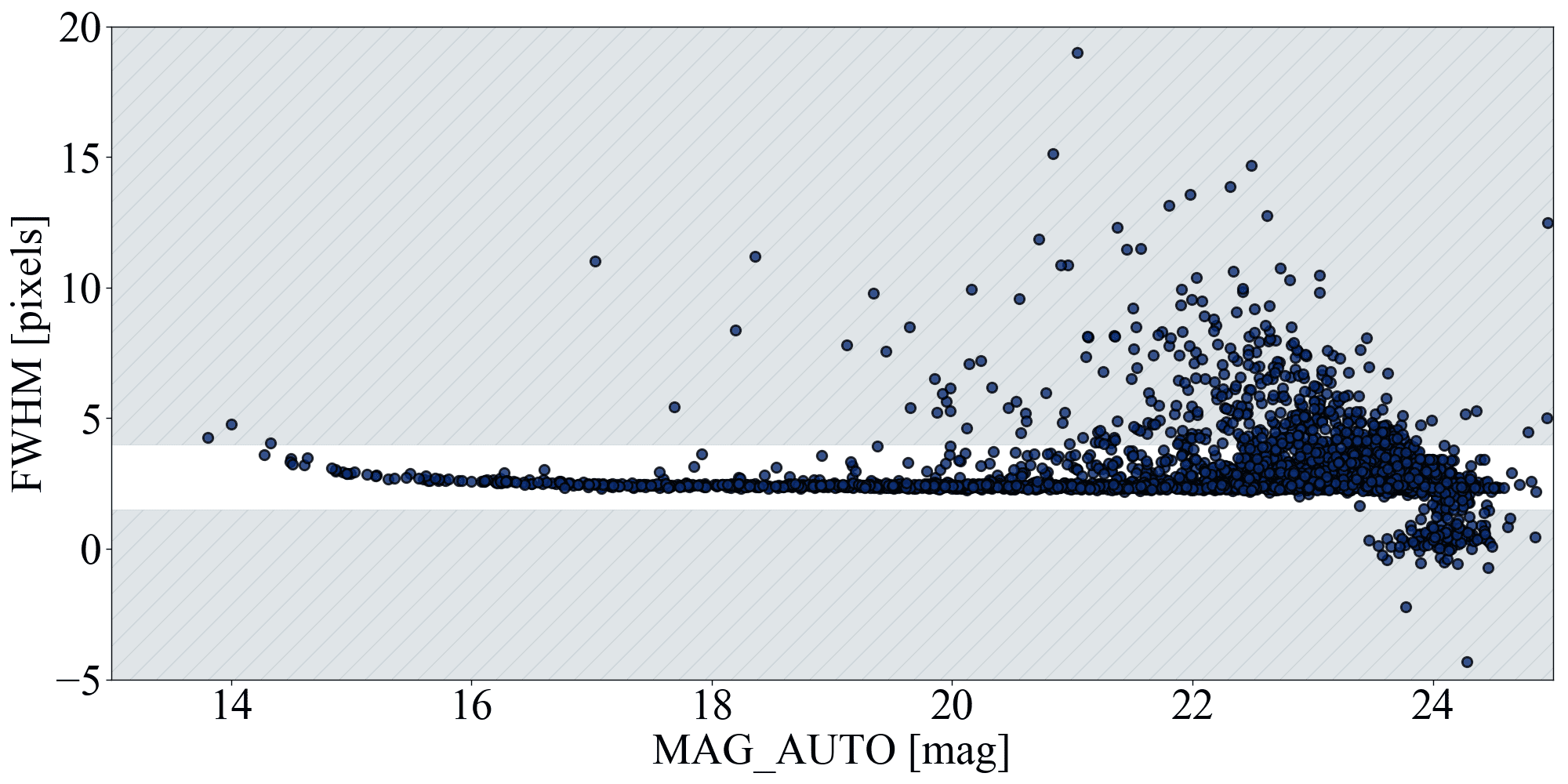}
\caption{Top panel: Magnitude error as function of the source magnitude. Bottom panel: FWHM of the sources as a function of the source magnitude. Objects in the gray area have been discarded, following the selection criteria described in the text. 
}
\label{fig:phase1}
\end{figure}

\paragraph{Phase 1 - Selection by quality and shape: }
The first selection done by \texttt{GCFinder} refers to data quality and shape of objects (hereafter Phase 1), adapted from \cite{Cho2016}. Phase 1 is done using the white image only. In the case of detections done in the white image, we adopt as Source Extractor input values those associated with the band with the worst PSF.
The catalogs generated in this Phase 1 are used to select data quality and object format.
We set white source magnitude error (MAGERR$\_$AUTO) < 0.2 to have an S/N > 5 on the selected data. 
With the creation of the white image, we observe that few objects are excluded at this stage since the adopted methodology improves the S/N ratio of the data, as can be seen in Figure \ref{fig:phase1}. We make one more data quality selection to exclude objects that were saturated or that were too close to the edge of the images. This type of object has compromised photometry, which can affect the magnitude and color selection that is performed in the following phases of the pipeline. For this, we adopt the Source Extractor FLAGS output parameter < 4, in agreement with \cite{Cho2016}.
To select only compact objects, we visually set limits for the FWHM, as can be seen in Figure \ref{fig:phase1}. The region identified in this Figure corresponds to objects that are point-like sources. Such selection makes it possible to exclude detections that are possibly galaxies. An example of selection using such criteria is shown in Figure \ref{fig:phase1}.

\begin{figure}
\centering
\includegraphics[width=8cm]{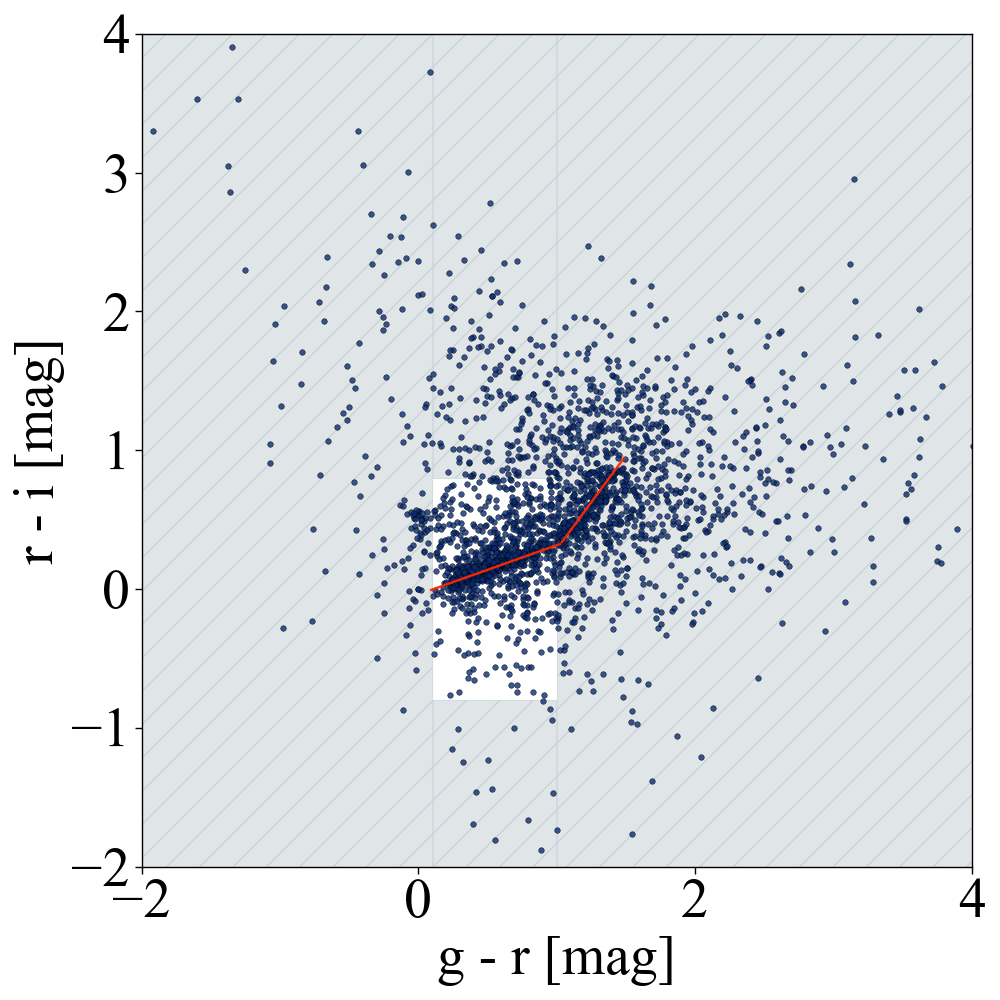}
\includegraphics[width=8cm]{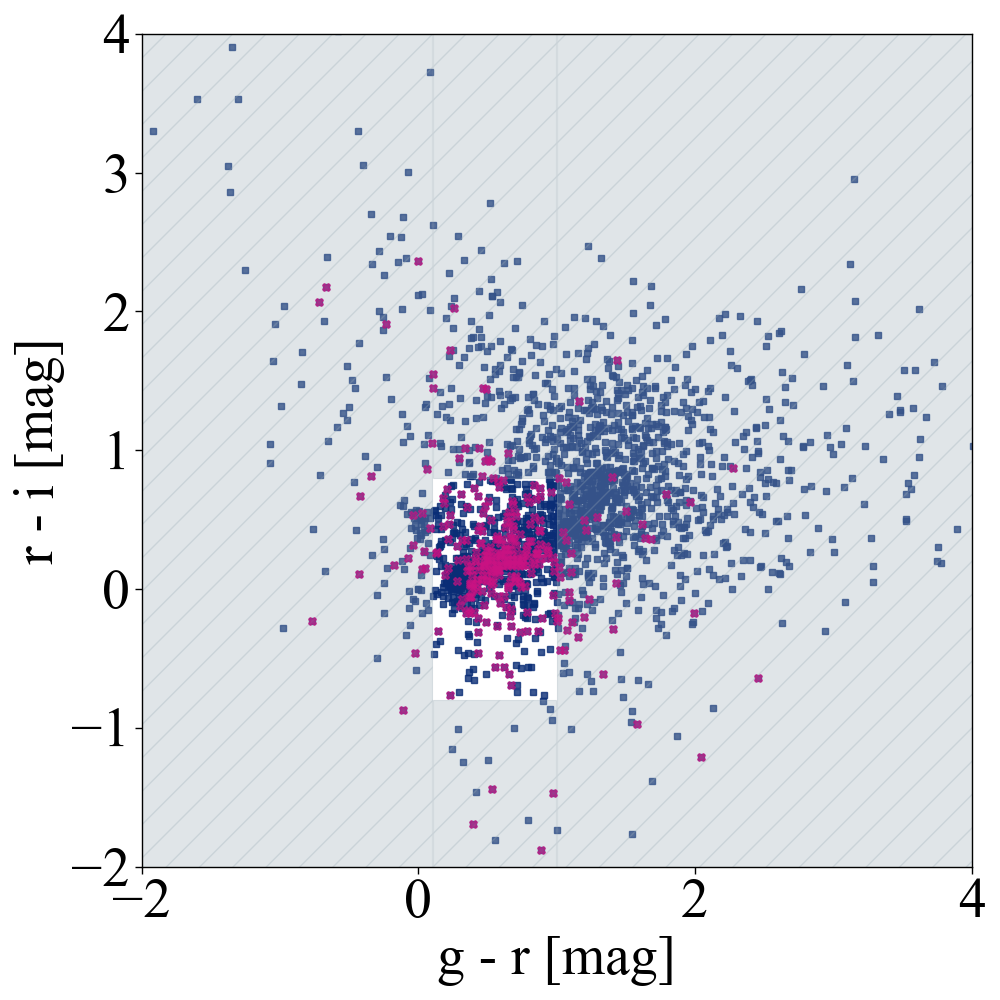}
\caption{Top panel: Selection of the GC candidates on the color-color diagram of the detections. Objects in the gray area have been discarded. A red line was included to guide the eye. Bottom panel: same as top panel with GC candidates also presented in \cite{Kartha2014} in magenta.}
\label{fig:phase2}
\end{figure}

\paragraph{Phase 2 - Selection by color:} 
The next selection done was related to the color of objects (hereafter Phase 2). Phase 2 is done using the individual images of $g$, $r$, and $i$ bands. At this point, we establish threshold values for the colors of the selected objects, to separate possible GCs from other objects, such as passive galaxies and low-mass stars.
To do this, we make a selection on a color-color diagram of $g - r$ versus $r - i$. We choose to use these colors since the number of detections is high in each band. In this diagram, a concentration of objects appears in a well-defined region of the image, which we refer to as the main branch. To select objects in color $r - i$, we exclude sources that were far from the main branch, and to select objects in color $g - r$, we exclude objects in the region where there is a more accentuated growth in the value of color $r - i$. An example of the cut established for NGC\,1023 can be seen in Figure \ref{fig:phase2}.
Objects in the region where the more accentuated growth of the color $r - i$ begins are possible low-mass stars \citep{finlator2000optical} and high redshift galaxies \citep{goto2002cut,prakash2015luminous}. The chosen region also corresponds to the same color interval from the majority of GC candidates reported in \cite{Kartha2014}, which we also show in Figure \ref{fig:phase2}.

\begin{figure}
\centering
\includegraphics[width=9cm]{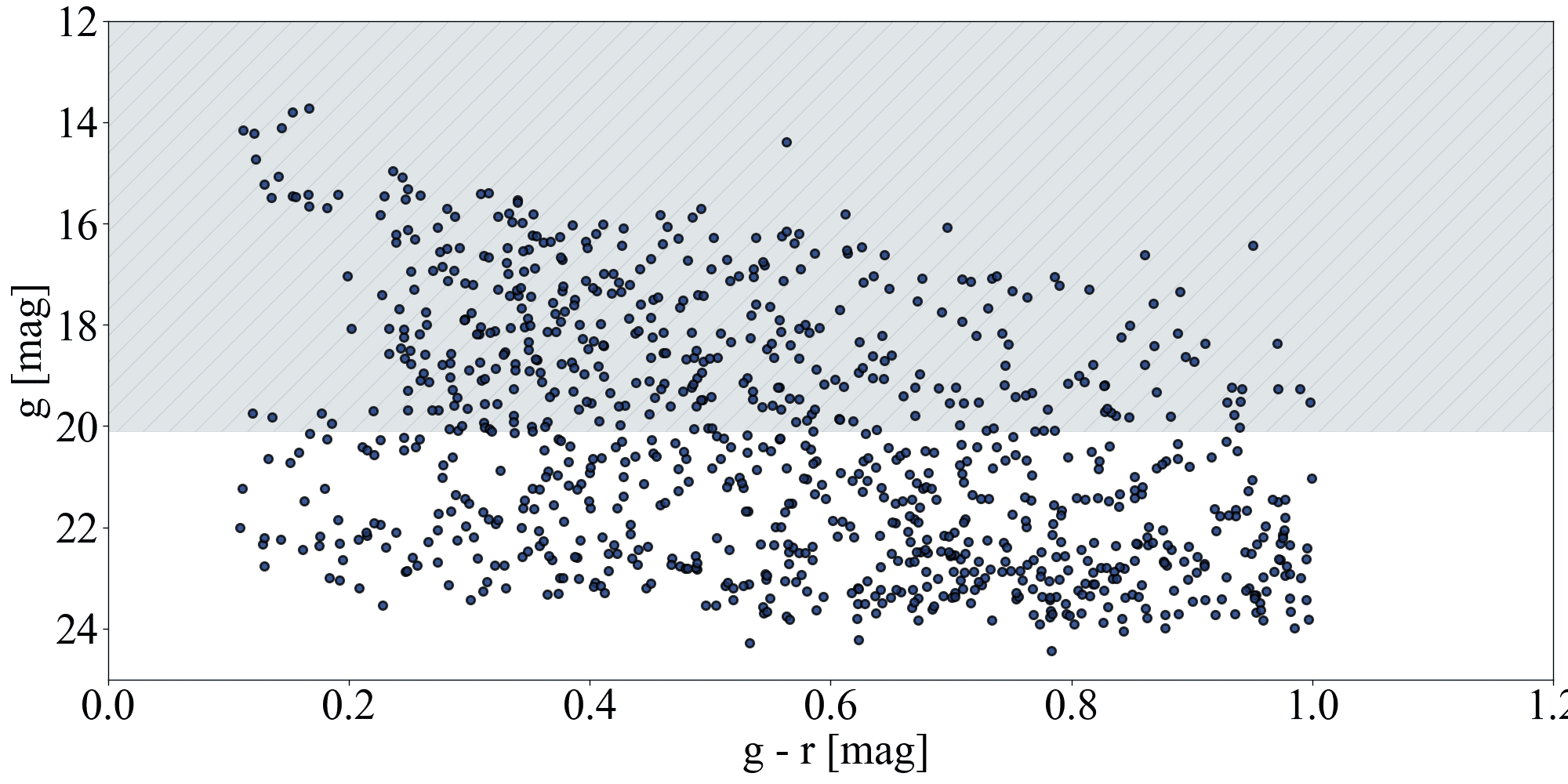}
\caption{Upper limit in magnitude adopted. Objects in the gray area have been discarded.}
\label{fig:phase3}
\end{figure}

\paragraph{Phase 3 - Magnitude limit cut: }
The third selection (hereafter Phase 3) is performed 
according to the magnitude of the objects in $g$-band. In this last step in the selection of GC candidates, very bright objects were excluded to clean our sample of Galactic stars, objects between the Milky Way and NGC\,1023, as well as possible Ultra-Compact Dwarf Galaxies (UCDs, \citealt{phillipps2001ultracompact}) for example. For this, the magnitude of one of the largest GCs from \cite{forbes2016} (a reference catalog  containing only spectroscopically confirmed globular clusters and other compact  stellar systems) was used as a reference: its absolute magnitude in the $g$-band was calculated from the distance of the galaxy, and adopted as the typical magnitude of the brightest GCs (see Figure \ref{fig:phase3}).

\paragraph{Phase 4 - Matching the GC candidates in all bands: }
At the end of all these selection steps, a final catalog with GC candidates is created (hereafter Phase 4). Afterward, matches are made from the final selection catalog with the detection catalogs of each band, using STILTS (integrated into the pipeline, \citealt{Taylor2006A}), to obtain the GC candidates in each band. This procedure is necessary since the \texttt{Source Extractor} input parameters of the detection image and photometric images are different, therefore the same objects do not necessarily have the same ID in all bands.

\paragraph{The pipeline -}
The current version of the pipeline \texttt{GCFinder} consists of a code in python that performs the process described in the previous paragraphs in a semiautomatic way. The code run inside a support folder prepared with the necessary directories for the correct functioning of Montage and the files for the correct functioning of the Source Extractor. For more details, please see Appendix \ref{sec:append}.  

\subsection{GCFinder performance}\label{sec:detection_results}

\subsubsection{Comparison with Gaia EDR3 data} \label{sec:gaiaed3_gcfinder}

In terms of the possible contamination by field stars, the contaminants have shape, color, and magnitude in the $g$-band equivalent to GCs and we are not able to exclude such objects using the techniques presented in this work even if the selection criteria adopted in \texttt{GCFinder}
encompass the main photometric selection techniques adopted in the literature.  

A sanity check done to evaluate the foreground stars in our sample was to inspect the parallax of the GCs selected by \texttt{GCFinder}. First, we cross-match our catalog of GC candidates with Gaia EDR3 \citep{GaiaCollaboration+2016b,brown2021gaia} data to acquire parallax measurements. From the 523 objects selected by \texttt{GCFinder}, we find only 153 in Gaia EDR3 considering a searching radius of 1 arcsec. Then we verify if the parallax values were compatible with zero within 3 sigmas (which indicates that the GC candidates are at a distance that is compatible with extragalactic objects). In general, all the GC candidates found in Gaia EDR3 have large parallax uncertainties (always larger than 0.4 mas), meaning all their parallax values are compatible with zero. 

\begin{figure}
\centering
\includegraphics[width=\columnwidth]{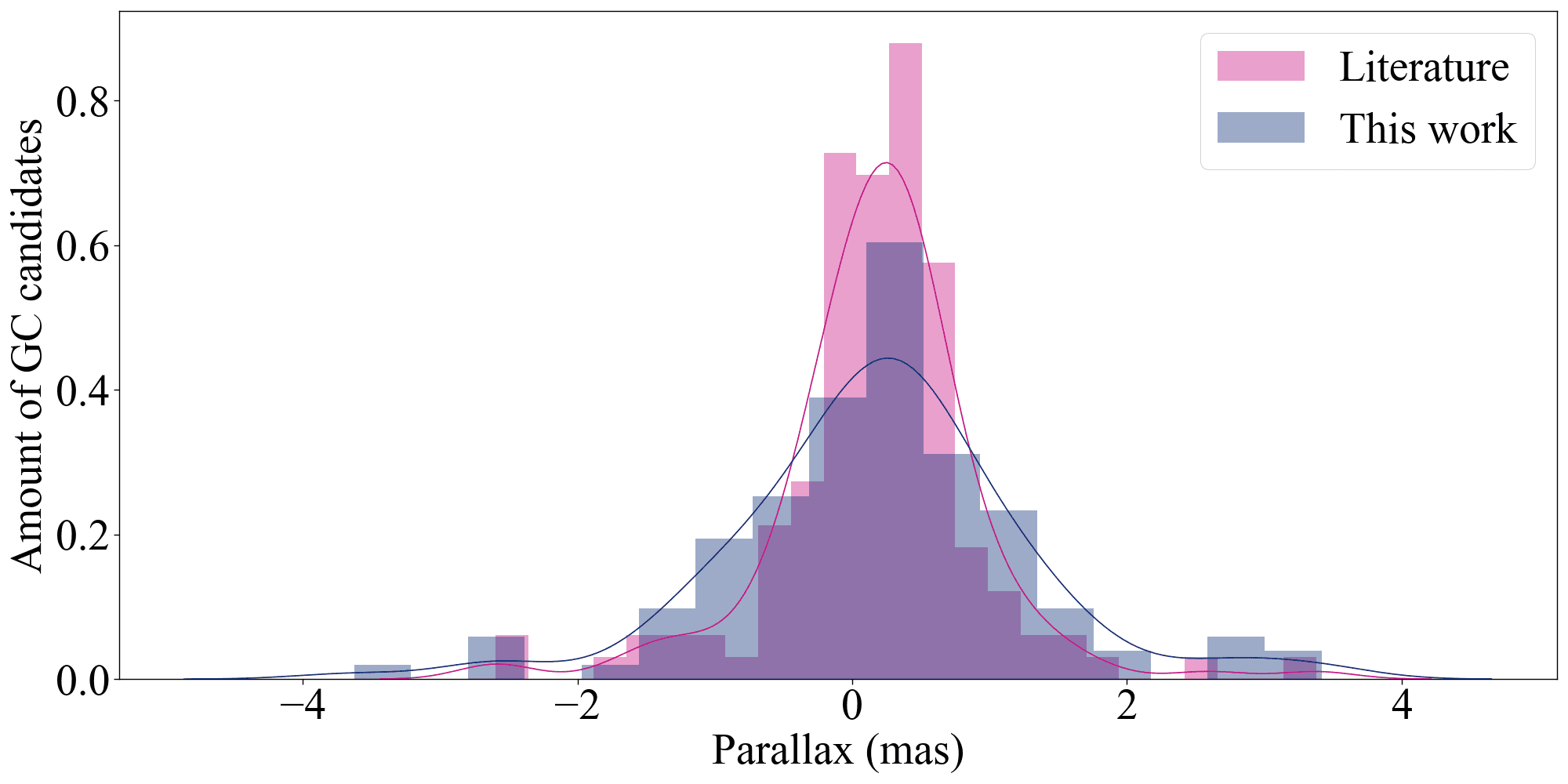}
\caption{Distribution of parallaxes for the GC candidates as obtained from Gaia EDR3. The blue distribution corresponds to the candidates of the present wok and in magenta the distribution of GC candidates identified by \cite{Kartha2014} is shown.}
\label{parallax}
\end{figure}

In Figure \ref{parallax} we compare the distribution of parallaxes from our catalog with the distribution of GC candidates in \cite{Kartha2014}. From the 627 GC candidates presented in \cite{Kartha2014}, we find only 164 in Gaia EDR3 considering a search radius of 1 arcsec. The parallax distribution of objects from the literature is narrower than the distribution of GC candidates from \texttt{GCFinder}, but the range covered by the two data sets is similar. 

Therefore, using our methodology, we obtain a catalog of GC candidates that is consistent with previous articles, with the advantage of not requiring modeling to remove the host galaxy light, either through modeling of the host galaxies' structural components or through median filtering.

This shows that the pipeline could be easily applied automatically in surveys such as J-PLUS, J-PAS, and S-PLUS, which could potentially generate large catalogs of GC candidates, especially in the outer halo regions and for spectroscopic follow-up.

\subsubsection{Comparison using colors}\label{comparison_using_colors_appendix}

To explore the nature of the new GC candidates found in this work, we analyze their colors and compared them with GCs detected by \texttt{GCFinder} that were also reported in \cite{Kartha2014}. 

\begin{figure*}
\centering
\includegraphics[width=4.5cm]{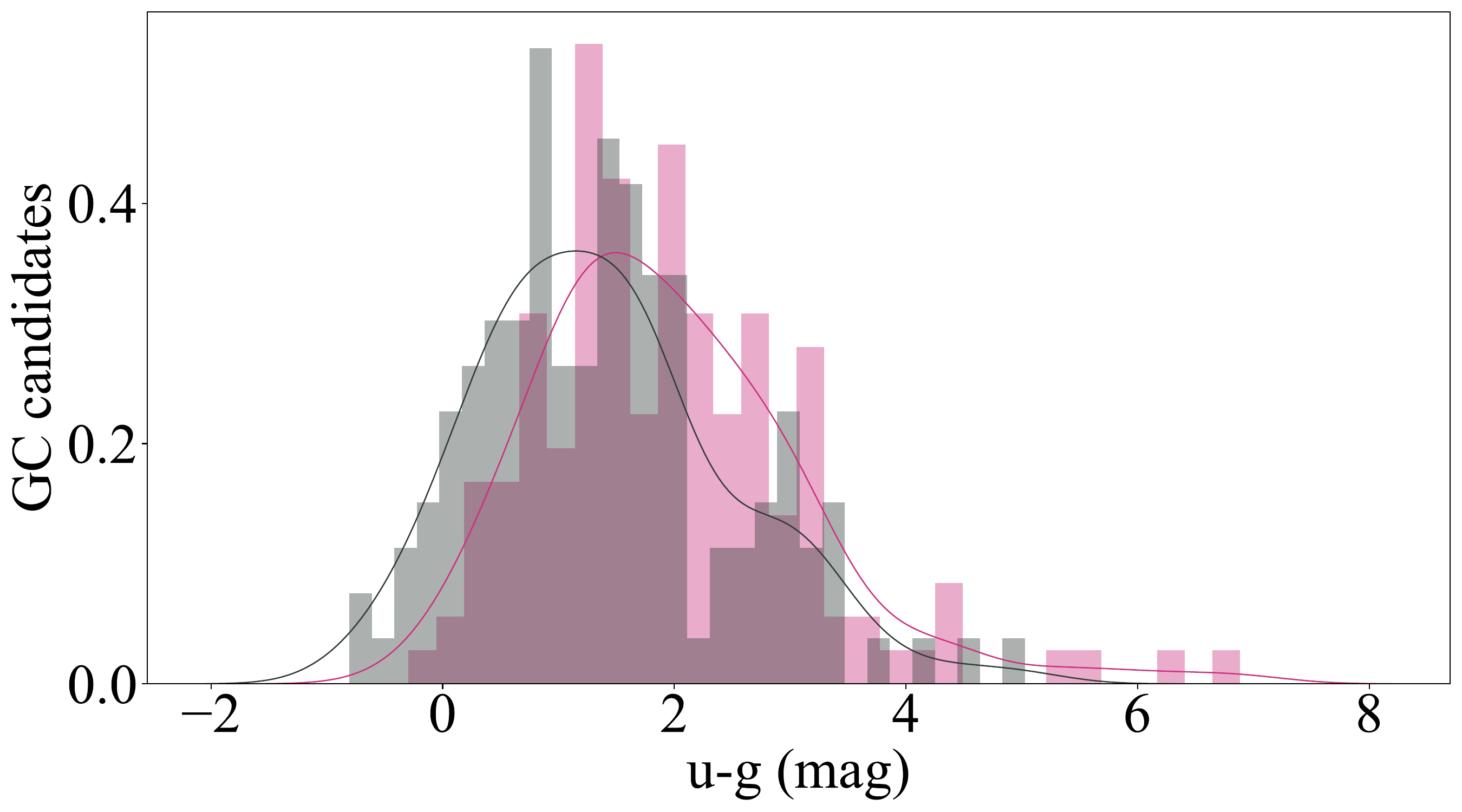}
\includegraphics[width=4.5cm]{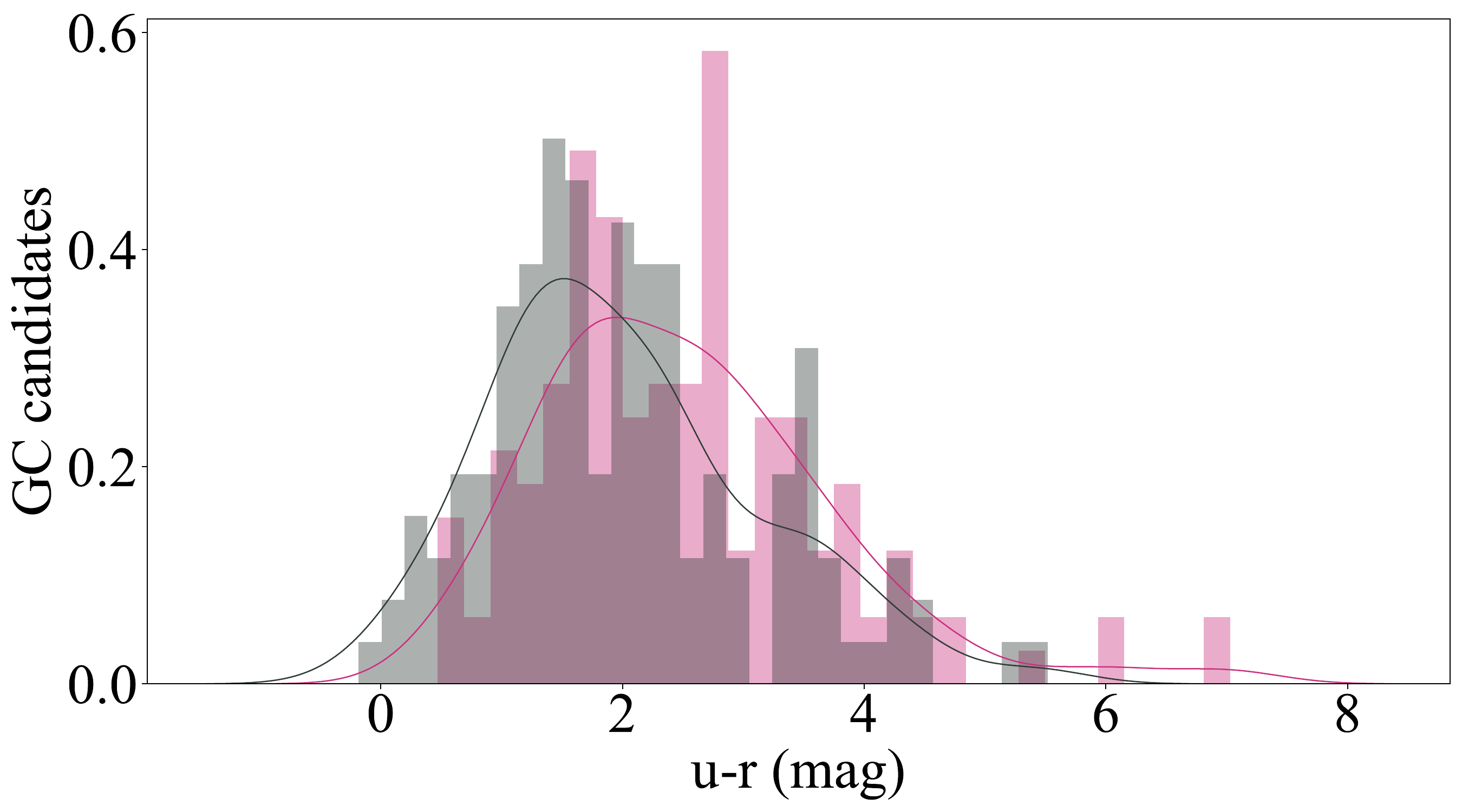}
\includegraphics[width=4.5cm]{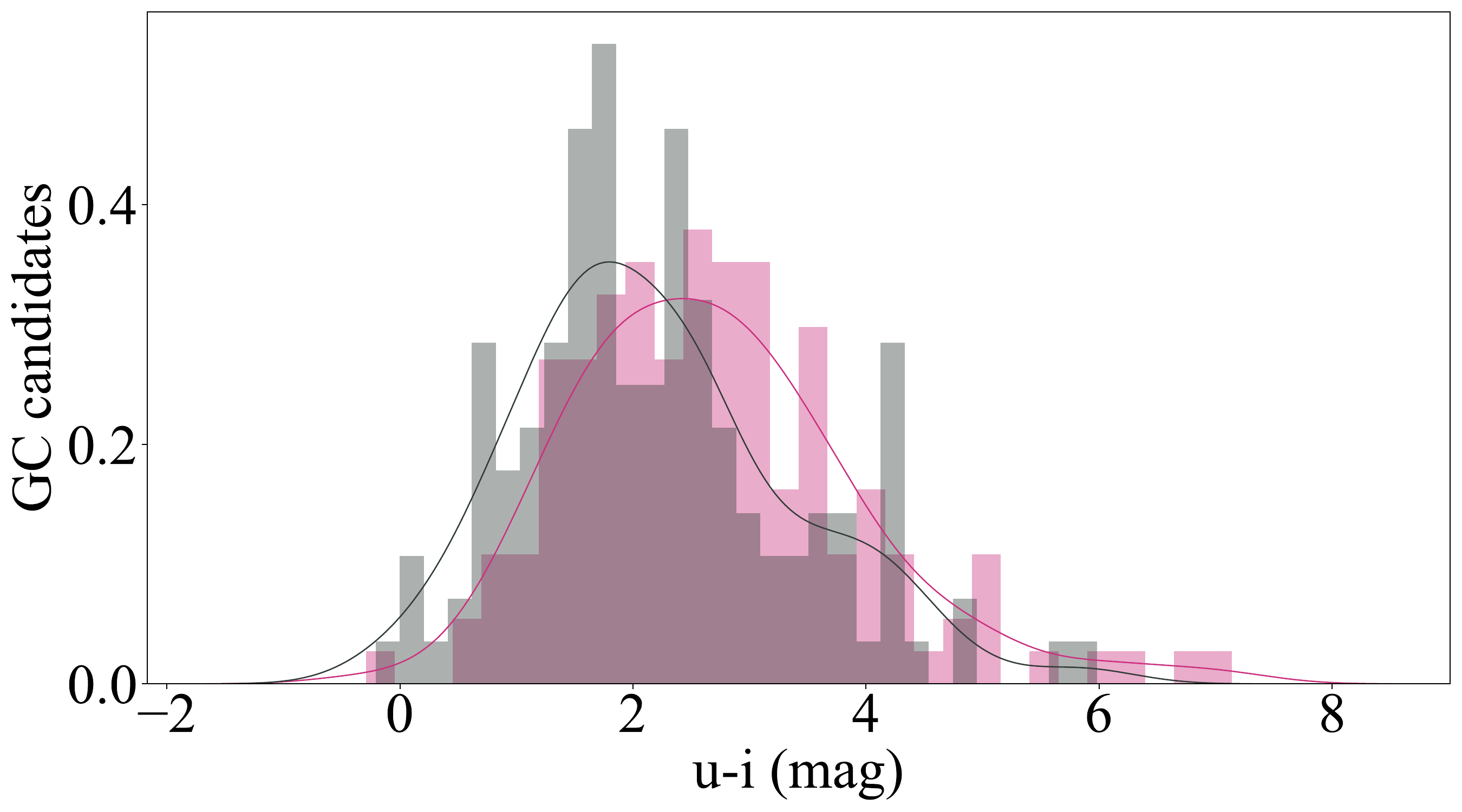}
\includegraphics[width=4.5cm]{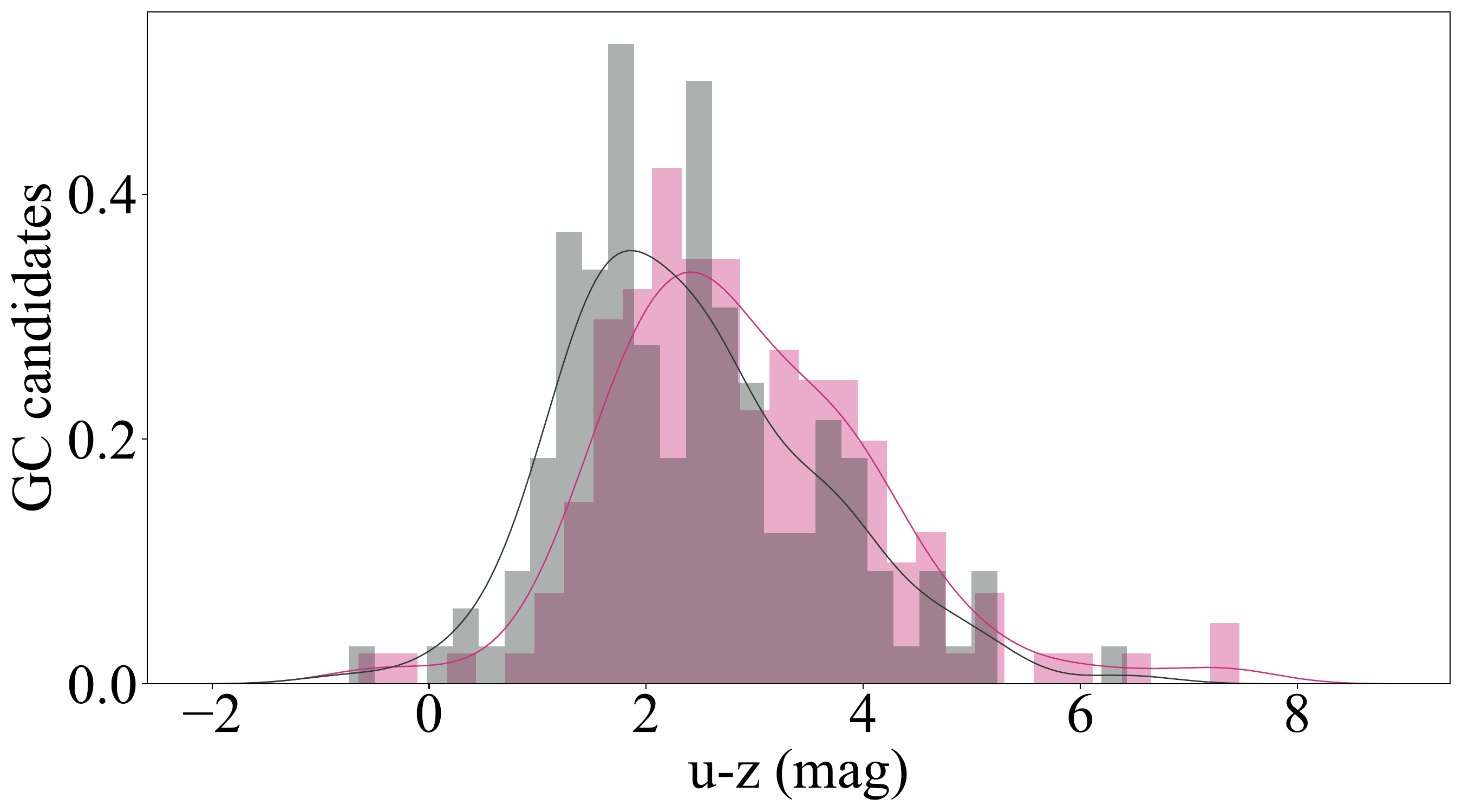}\\
\includegraphics[width=4.5cm]{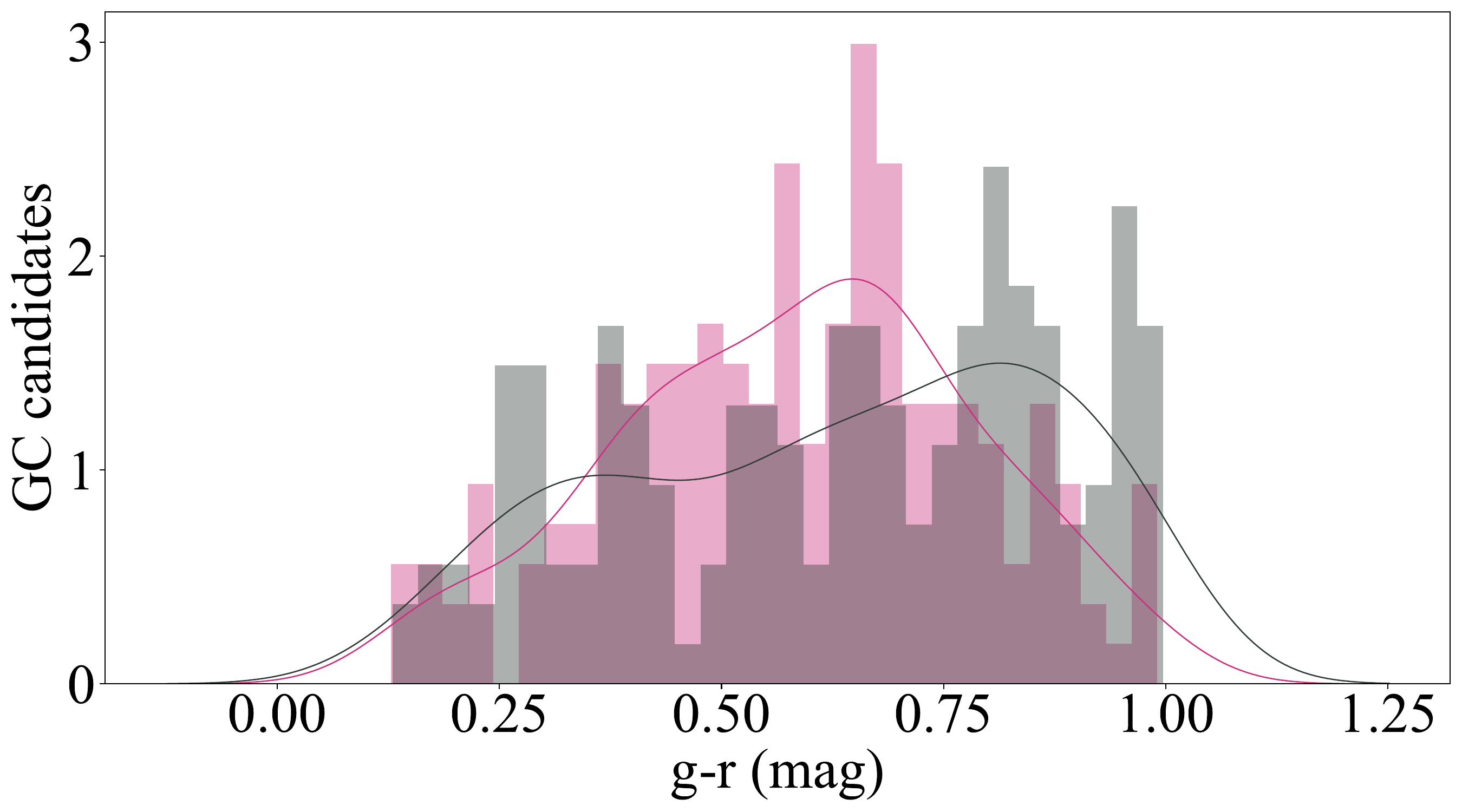}
\includegraphics[width=4.5cm]{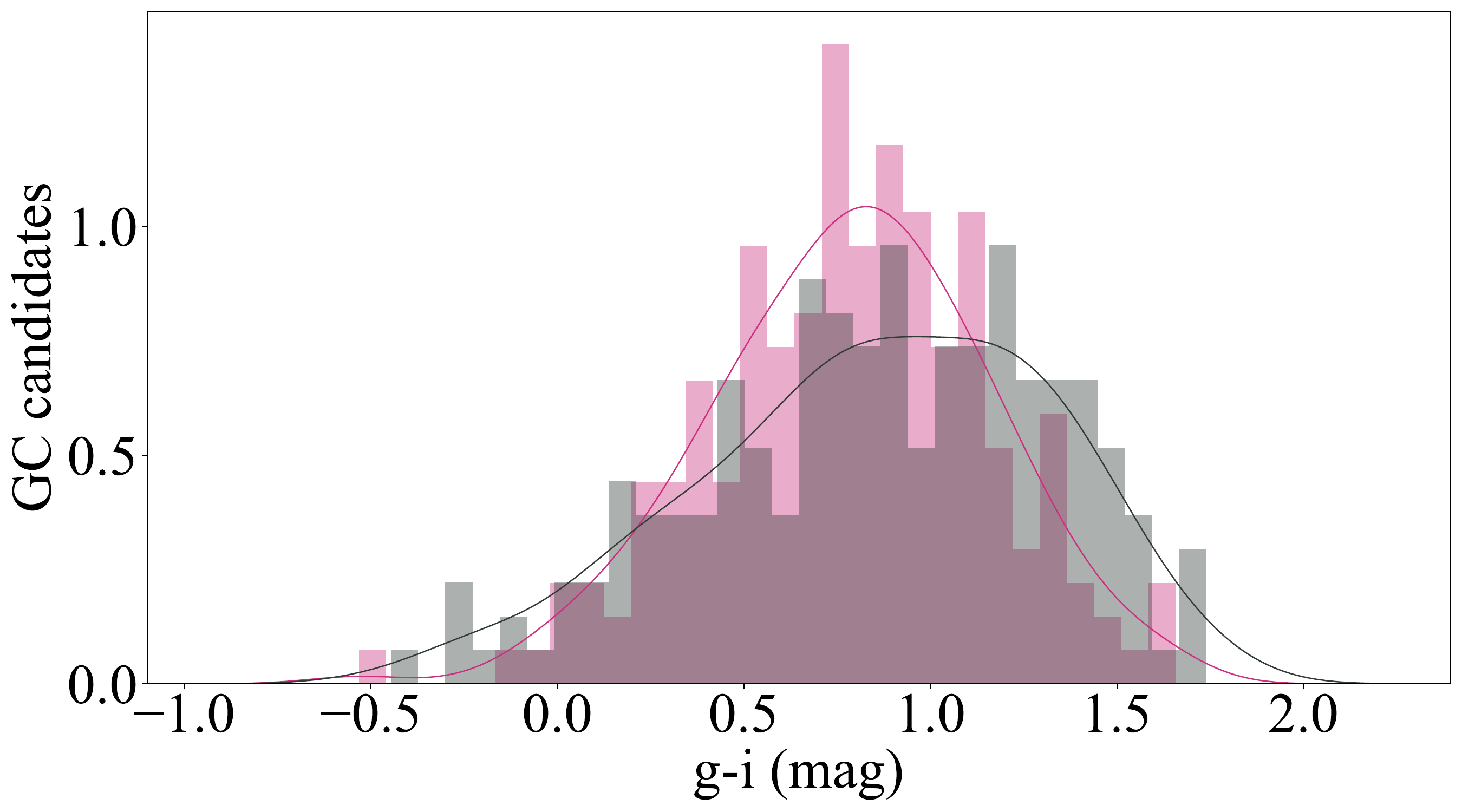}
\includegraphics[width=4.5cm]{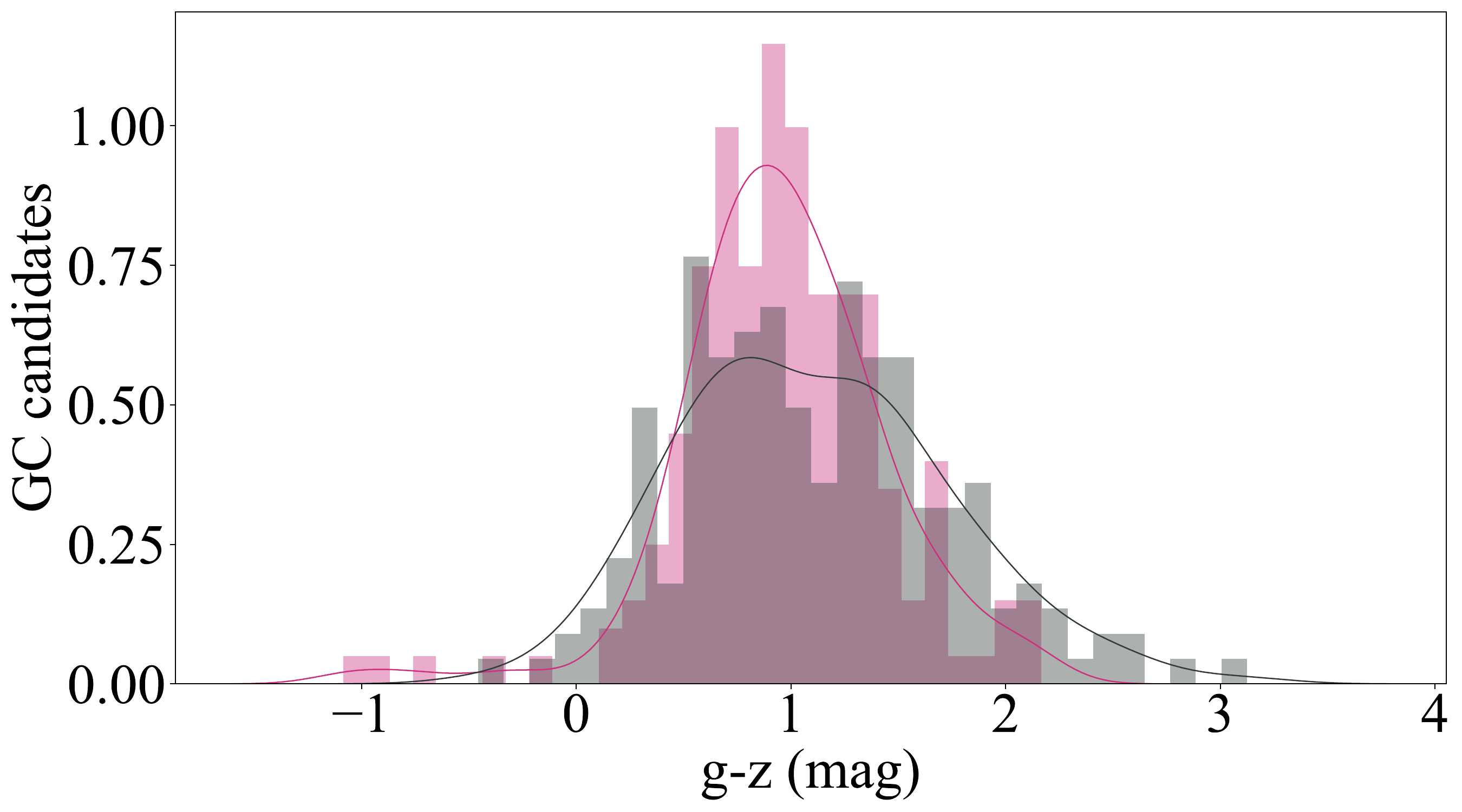}
\includegraphics[width=4.5cm]{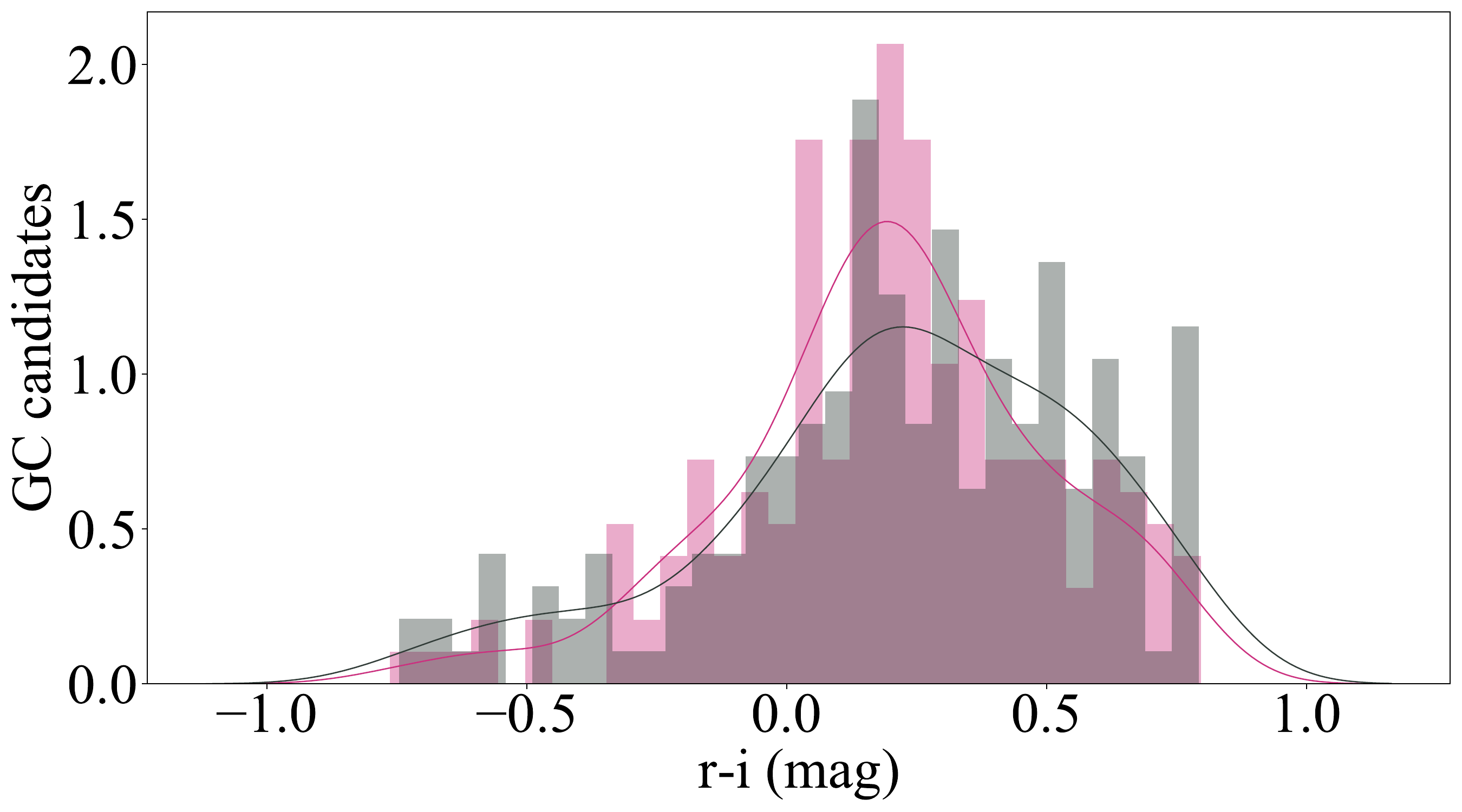}\\
\includegraphics[width=4.5cm]{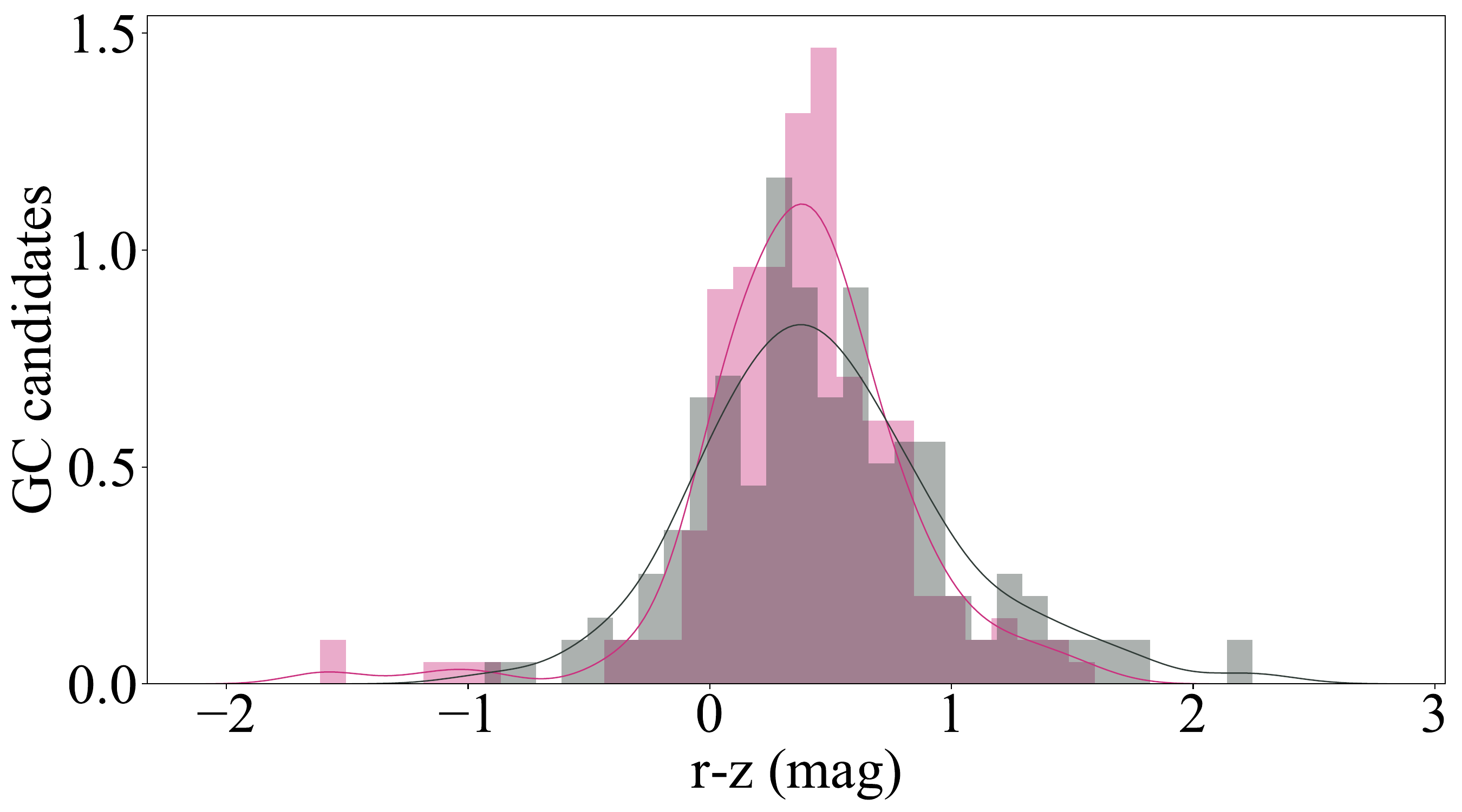}
\includegraphics[width=4.5cm]{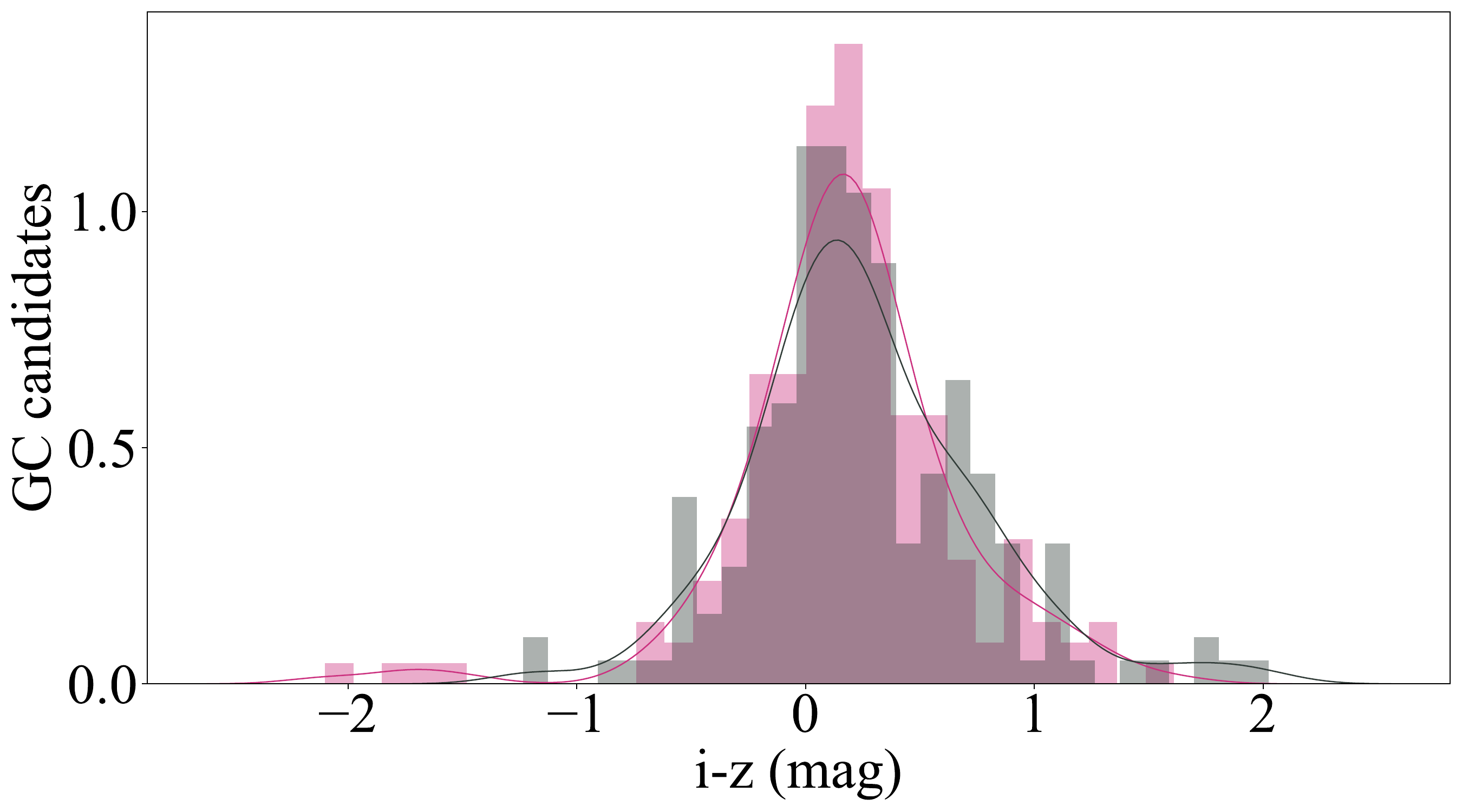}
\includegraphics[width=4.5cm]{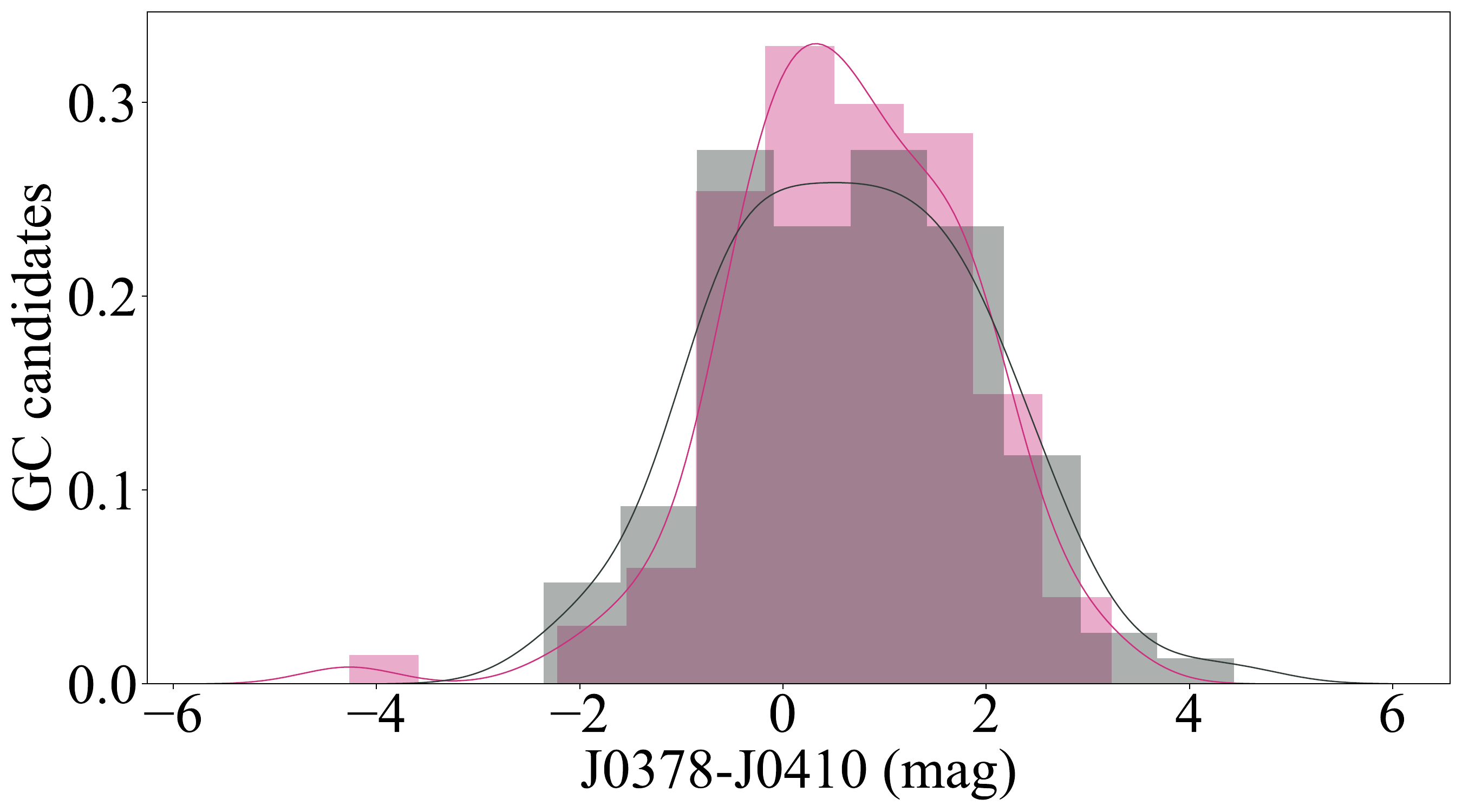}
\includegraphics[width=4.5cm]{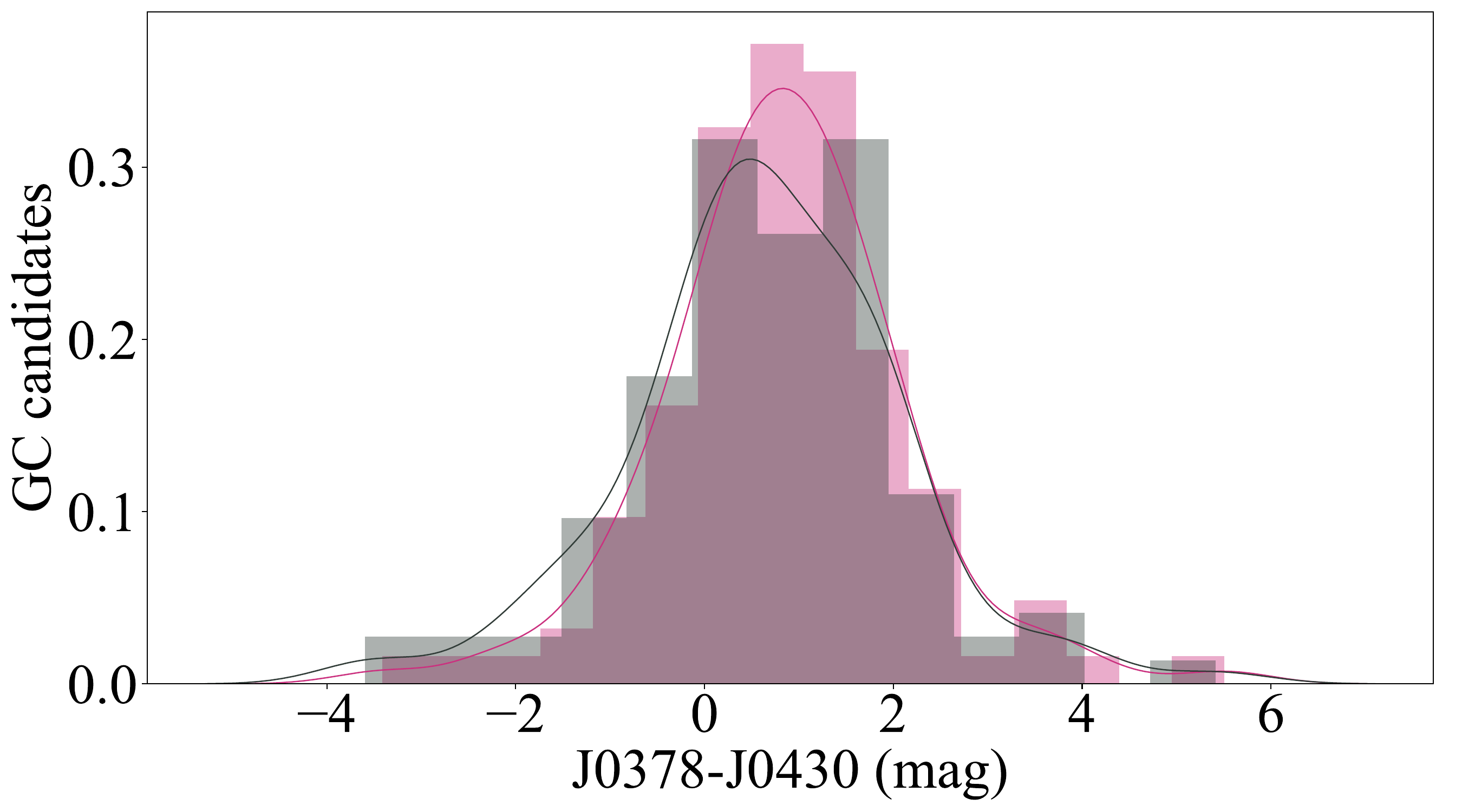}\\
\includegraphics[width=4.5cm]{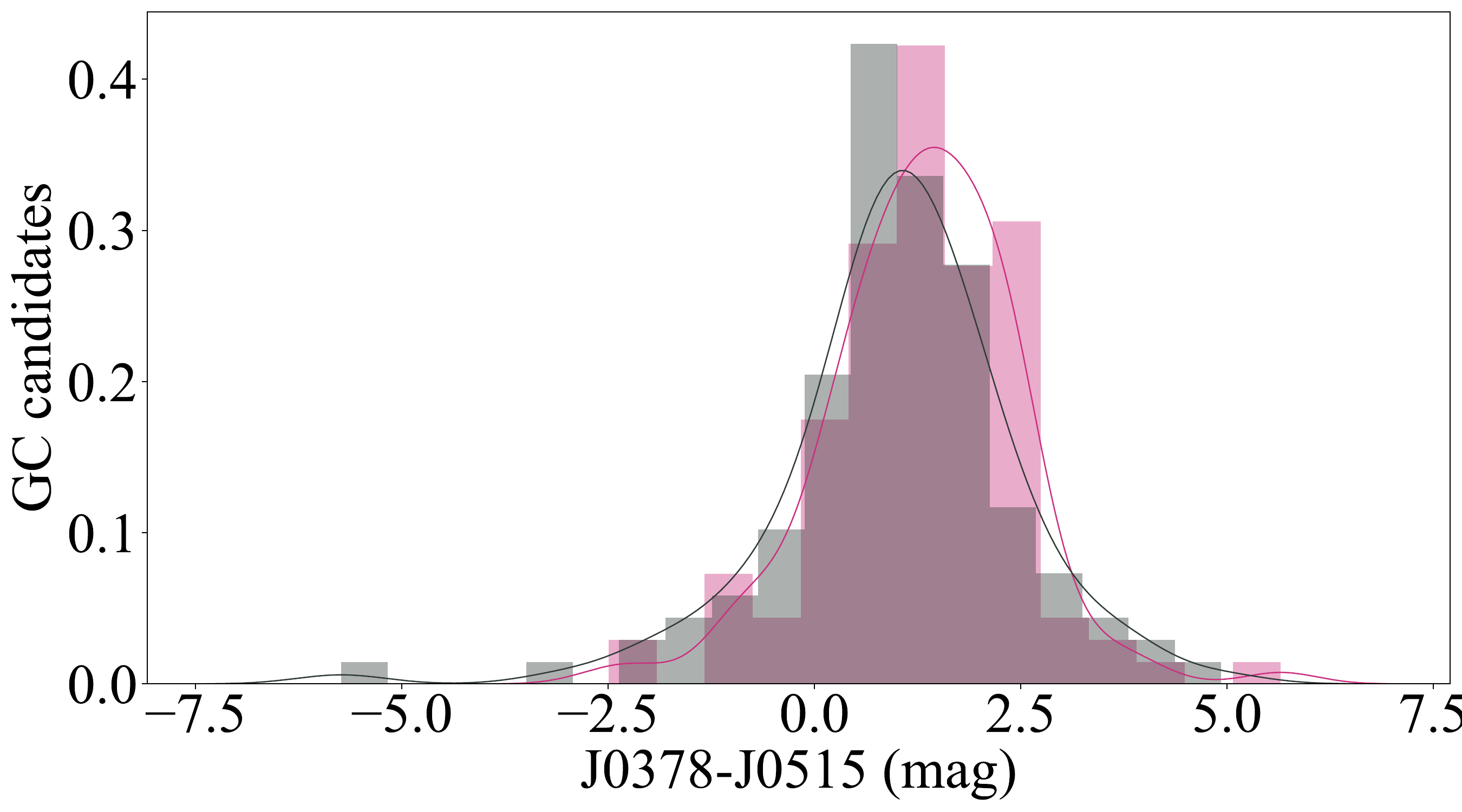}
\includegraphics[width=4.5cm]{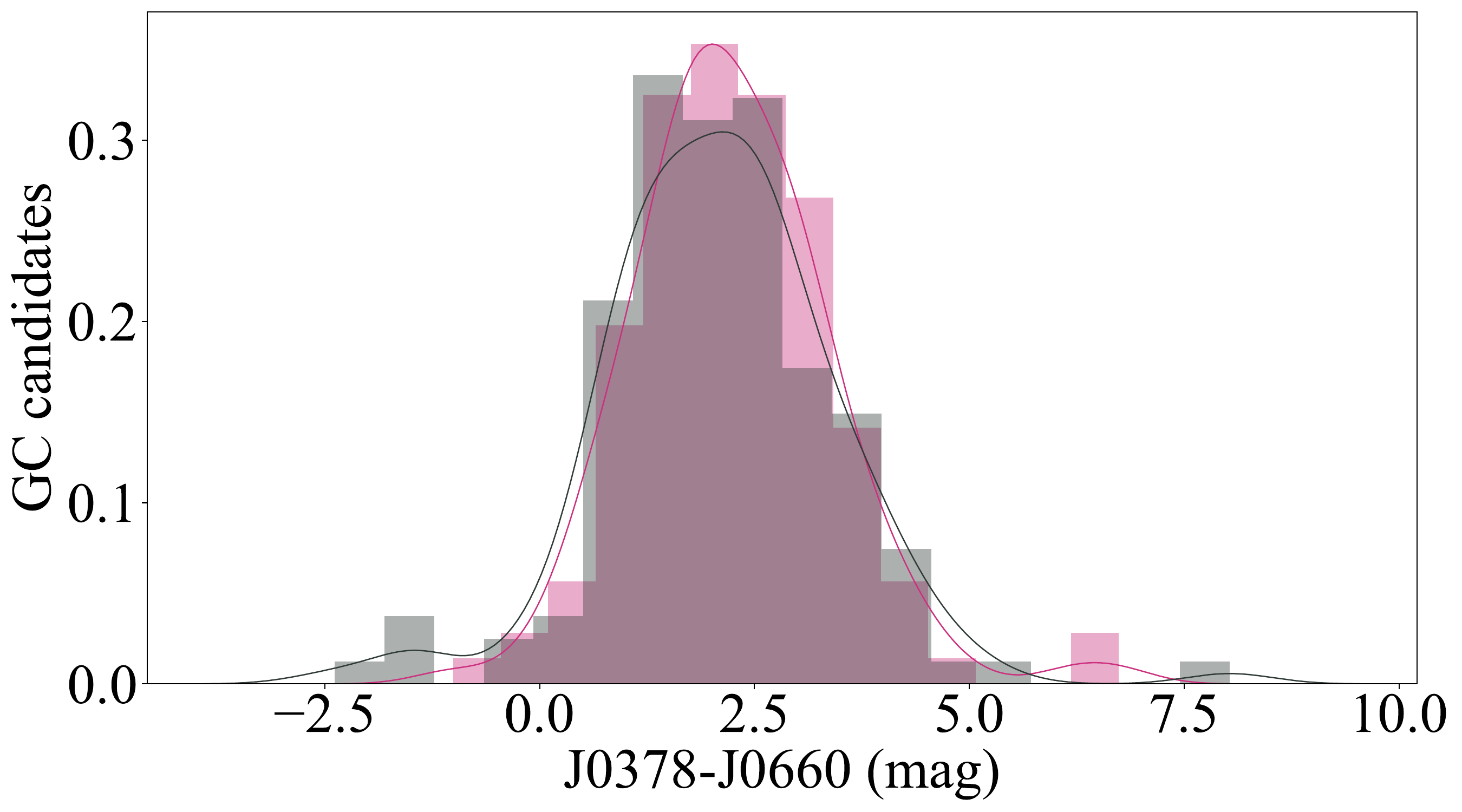}
\includegraphics[width=4.5cm]{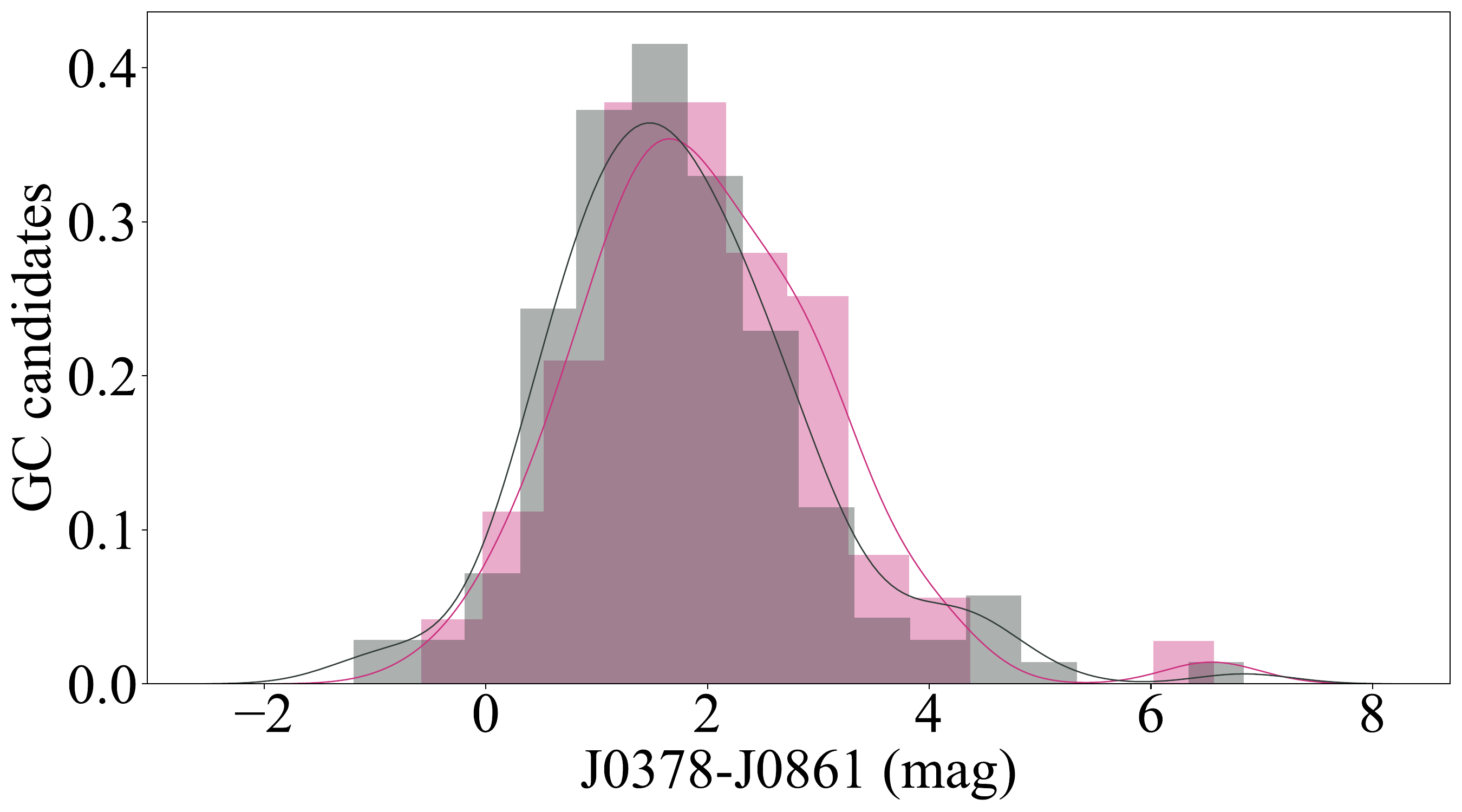}
\includegraphics[width=4.5cm]{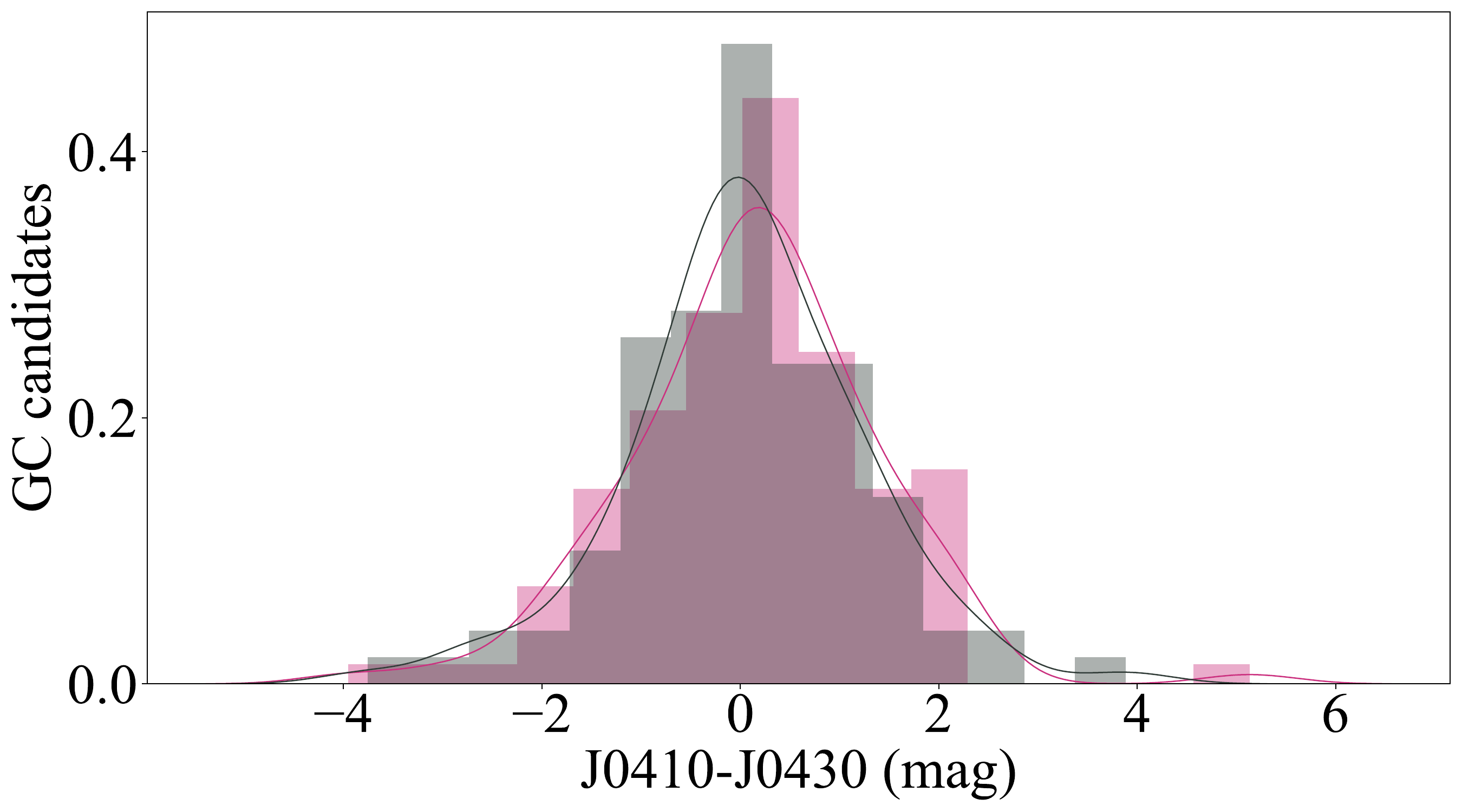}\\
\includegraphics[width=4.5cm]{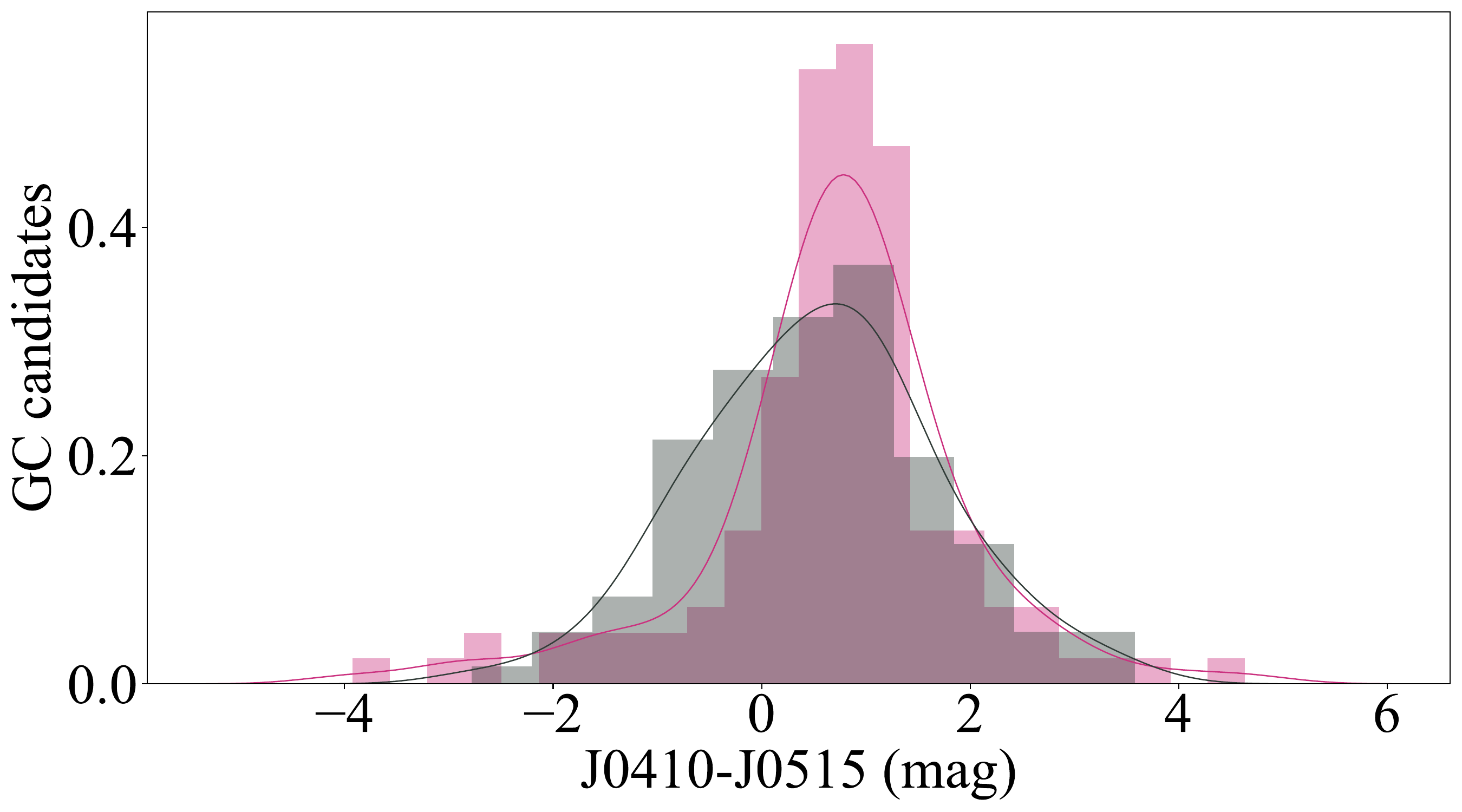}
\includegraphics[width=4.5cm]{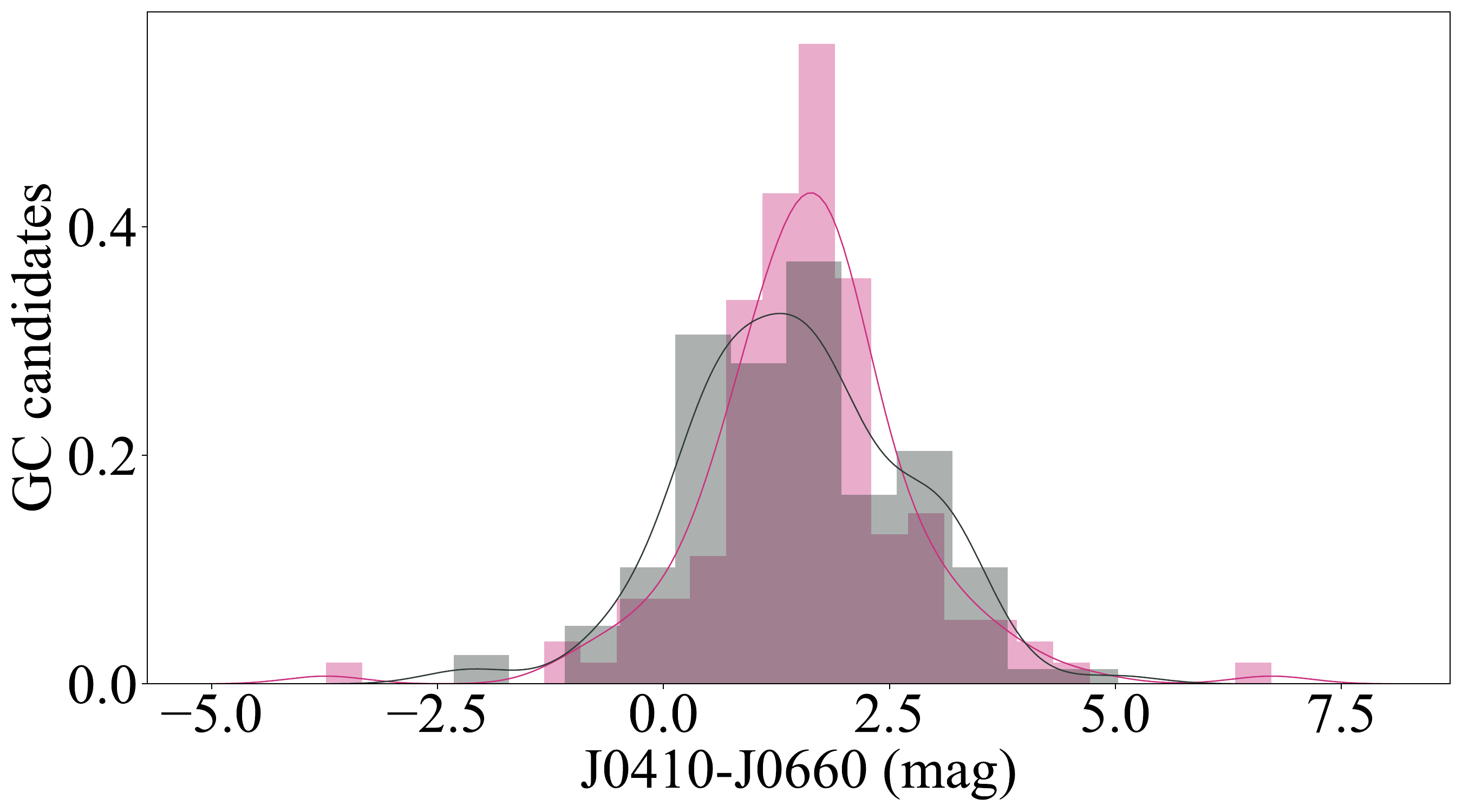}
\includegraphics[width=4.5cm]{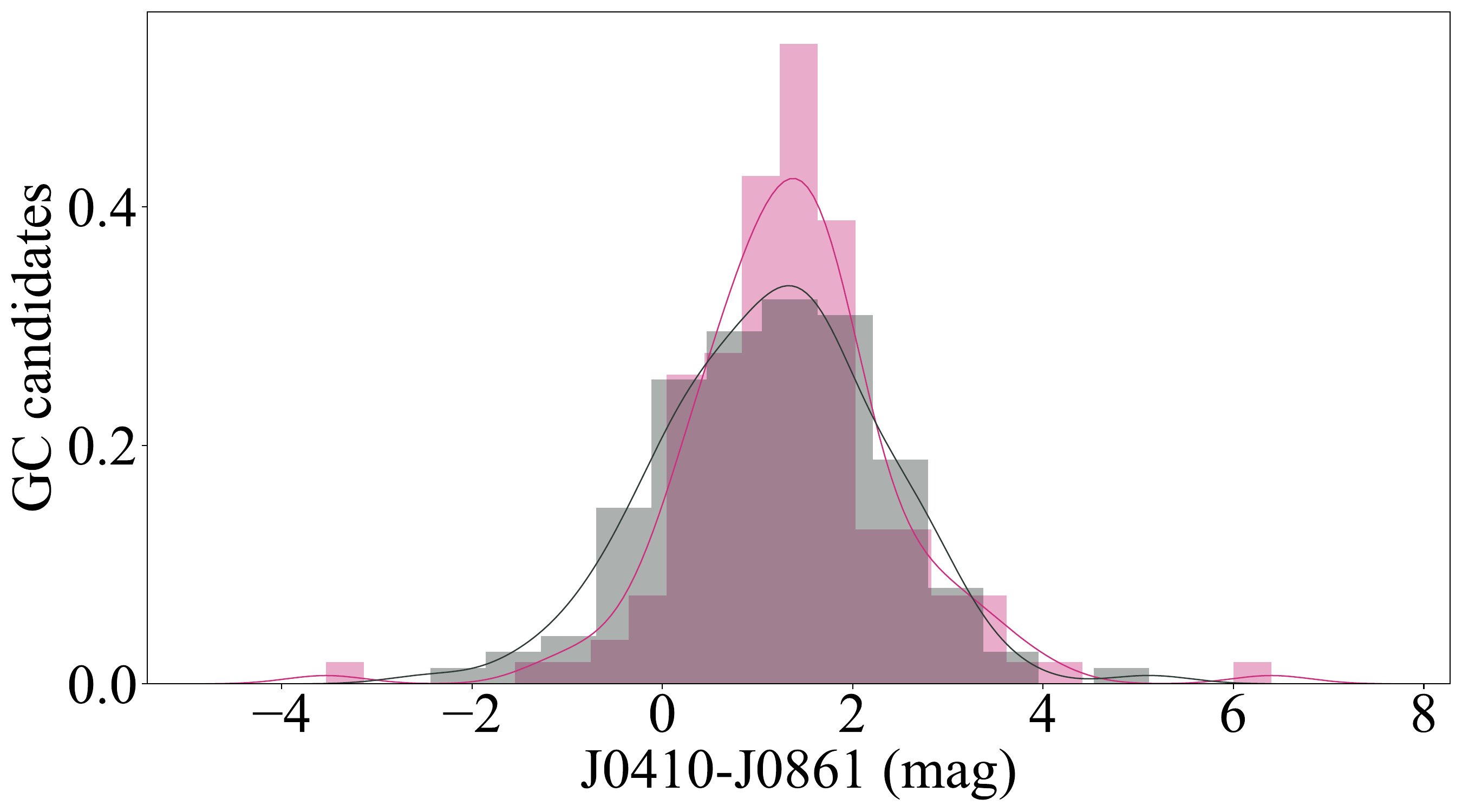}
\includegraphics[width=4.5cm]{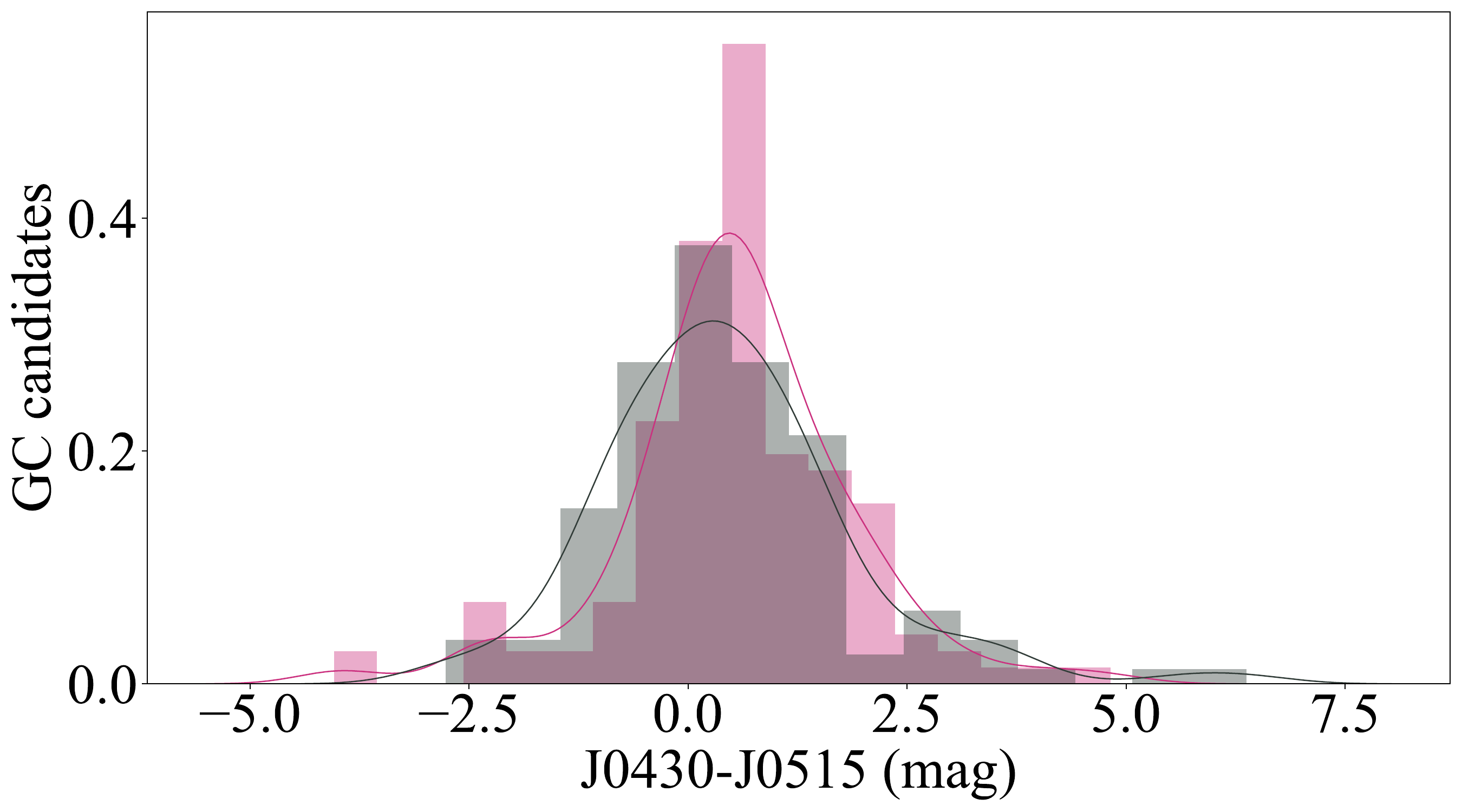}\\
\includegraphics[width=4.5cm]{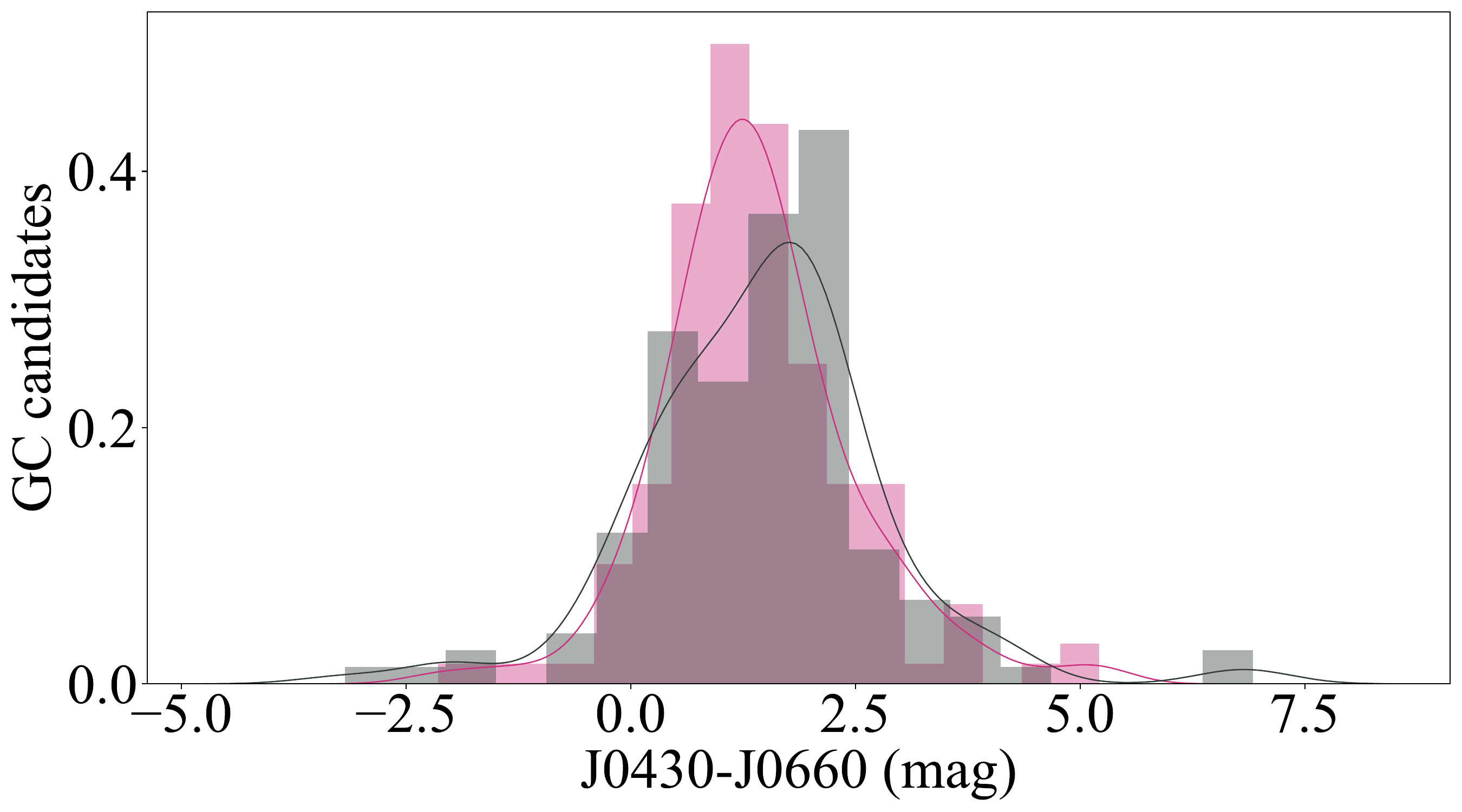}
\includegraphics[width=4.5cm]{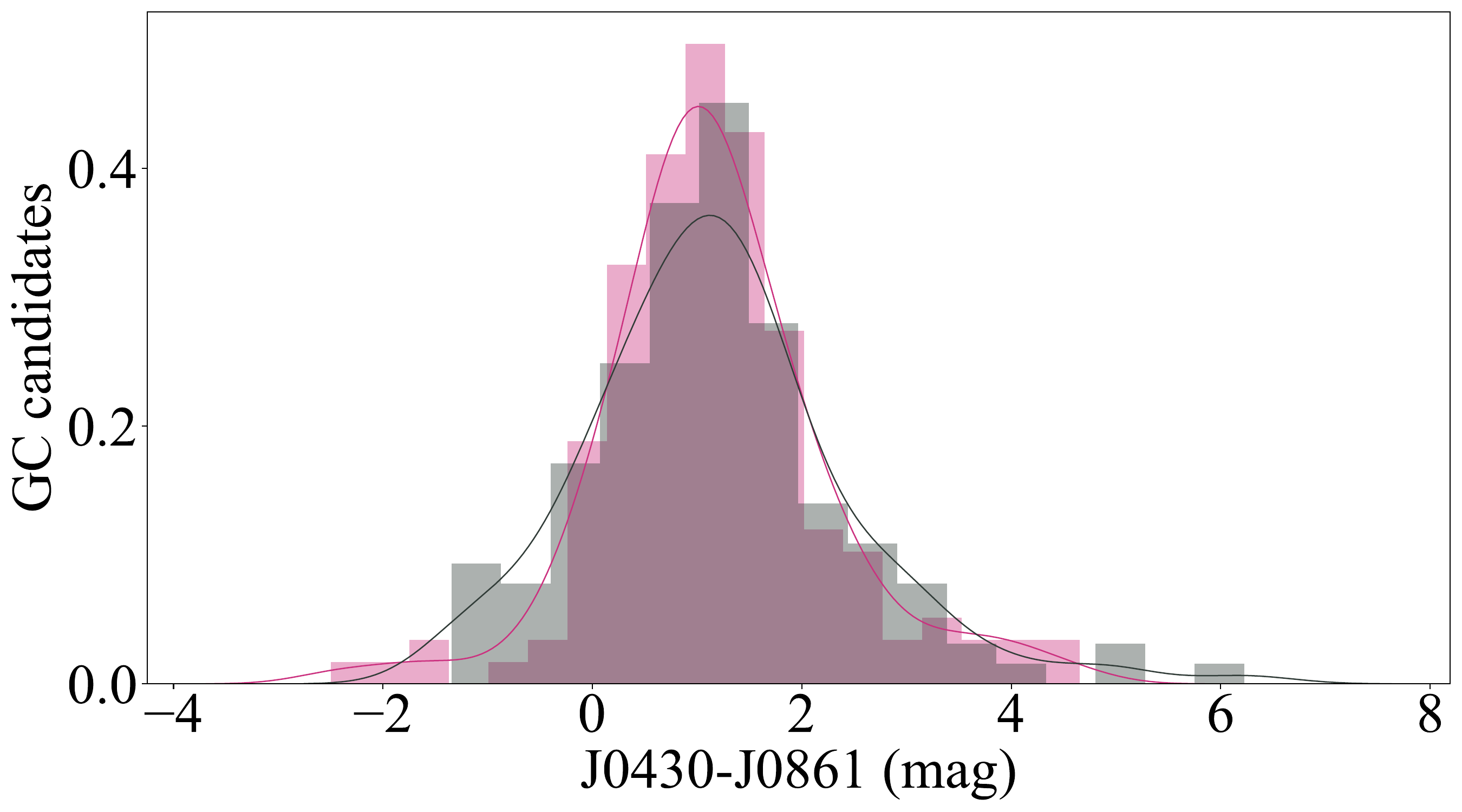}
\includegraphics[width=4.5cm]{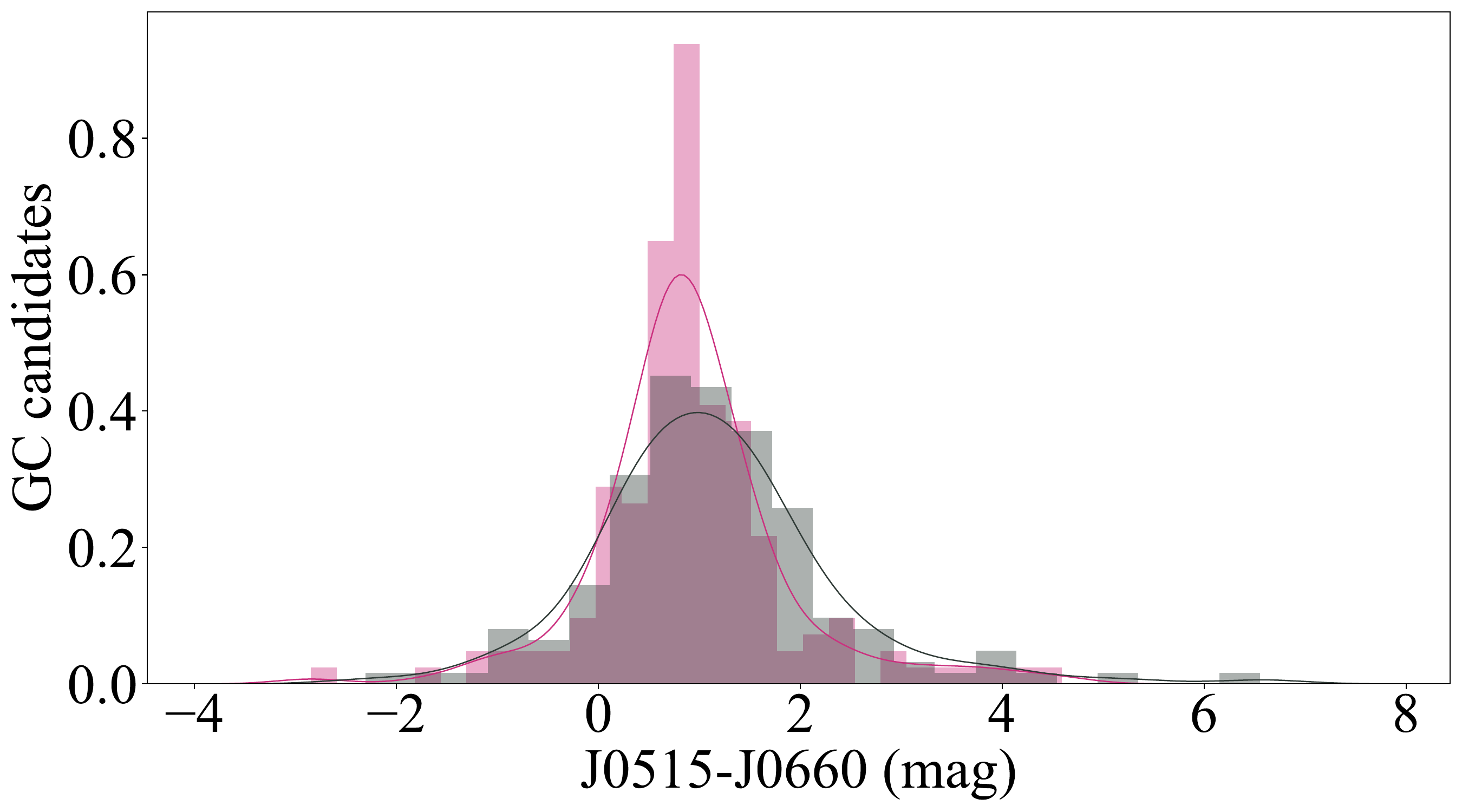}
\includegraphics[width=4.5cm]{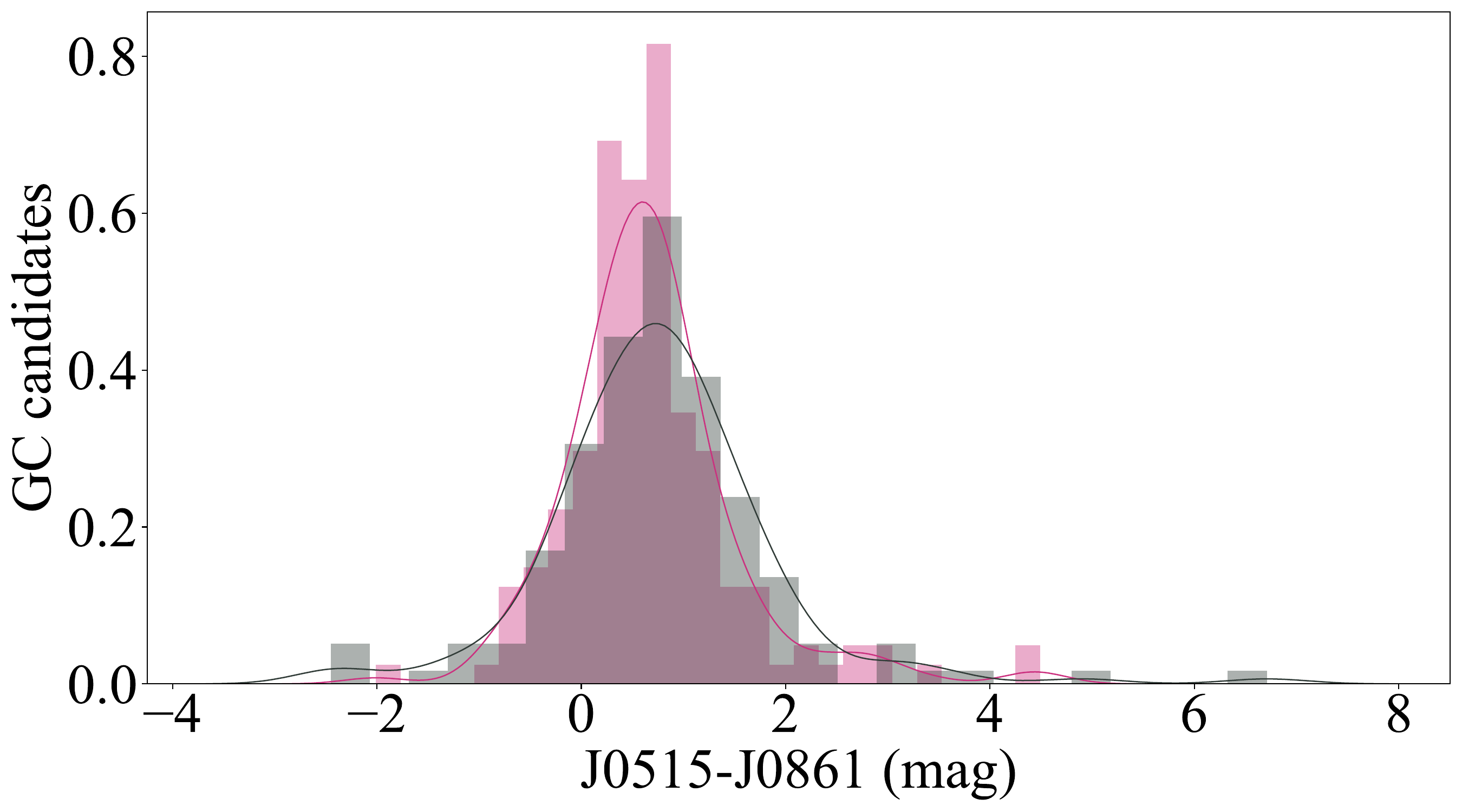}\\
\includegraphics[width=4.5cm]{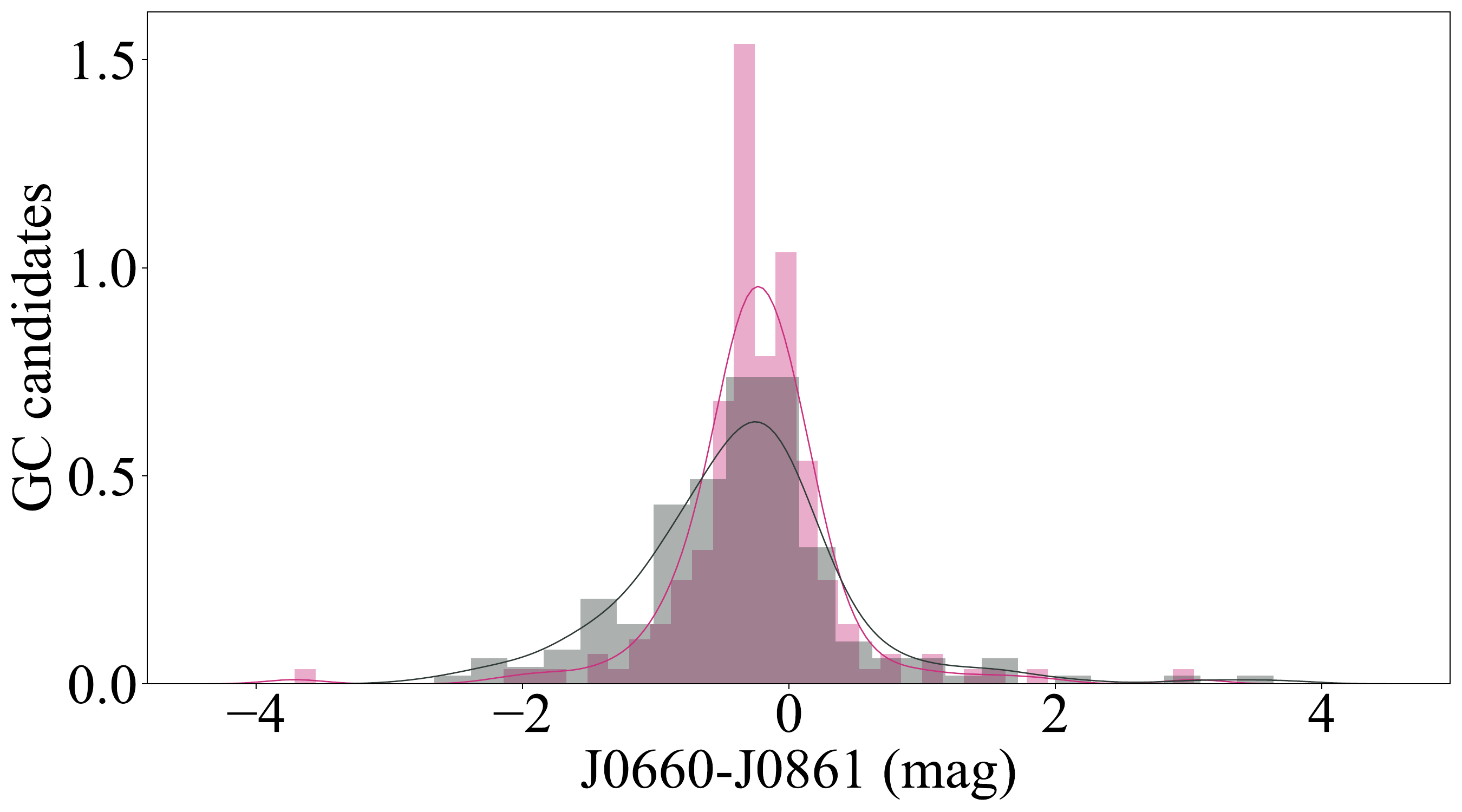}
\caption{Density distributions of colors based on J-PLUS filters. Magenta: Color distribution of GC candidates identified by \texttt{GCFinder} that are also present in the reference catalog \citep{Kartha2014}. Grey:  Color distribution of GC candidates identified by \texttt{GCFinder} which are not present in the reference catalog.}
\label{fig:cor_cor_lit_gcfinder}
\end{figure*}

Figure \ref{fig:cor_cor_lit_gcfinder} compares the distributions of colors of GC candidates selected by \texttt{GCFinder} and those of \cite{Kartha2014}.
We observe that there are color shifts (e.g. $u-g$, $u-r$, $u-z$). The GC candidates that are not found in \cite{Kartha2014} are bluer, which is consistent with metal-poor halo GCs. Although the distribution profile is not the same among the two groups, there is no clear separation between them. 

We note that many papers in the literature use machine learning to classify objects. In particular, \cite{lopez2019j} study the star galaxy separation of objects in J-PLUS data considering their morphology, \cite{wang2021machine} build a supervised machine learning algorithm to classify objects (stars, galaxies, and quasars) in J-PLUS, \cite{costa2019s} use machine learning to perform star galaxy separation in S-PLUS data while \cite{nakazono2021discovery} train a random forest classifier and provided catalogs of stars, galaxies, and quasars also in S-PLUS survey. 
We believe that such approaches would be complementary 
in the case of identification of GCs, but it was not the objective of this work. Having a pipeline based only on astrophysical selection is useful to characterize the properties of this class of objects in new surveys and is also efficient, as shown. We hope our work could also serve as a training base for future pipelines based on machine learning techniques.

\section{The structure of the pipeline \texttt{GCFinder}}\label{sec:append}

\subsection{Technical requirements}\label{sec:technical_requirements}

For the \texttt{GCFinder} pipeline to work, the following resources must be installed on your computer:

\begin{itemize}
\item \textit{Python} 2.7
\item \texttt{Montage}
\item \texttt{Source Extractor}
\item \texttt{STILTS}
\end{itemize}

The pipeline was developed and tested only on the Unix system, more precisely on Ubuntu 16.04.

The pipeline needs a processing time of approximately 7 minutes, 
with 70 \% of this time being consumed in the construction of the white image. This information comes from results obtained with a computer with 4 GB of RAM and an Intel core I5 processor. The pipeline has no special requirements for RAM or processing capacity of the machine used. However, the original J-PLUS images have a large field (9500 pixels x 9500 pixels, $\approx$ 2 deg$^{2}$) and we only work with images cropped in the region of the galaxy, so the studied images have smaller fields. As Montage has been integrated into the pipeline, working with the original images makes the necessary processing time longer and computers with little RAM face difficulties in the white image construction stage. Considering what was studied in this work, this particular result is very satisfactory, given that the main difficulty 
we face was the fact that many packages for modeling and removing the light profile of the galaxy need hours to run and machines with large processing power.

\subsection{Inputs required for pipeline operation}\label{sec:required_inputs}

Before the \texttt{GCFinder} pipeline starts working, the user must include the images of all bands in a specific directory inside the support folder where the code is inserted and provide a file with the zero point values for each band. When the pipeline starts working, the user is asked to provide the path to the pipeline directory, so that the code can perform the necessary operations between files and folders.
During code execution, 3 more pieces of information are requested. The first one is the FWHM cutoff that must be used, the second one is the limits in the color-color diagram, and finally, the distance from the galaxy so that the calculation of the magnitude cutoff is carried out. The FWHM and color cuts remained interactive, as the distribution of such quantities in the graphs might be particular to each galaxy. Keeping these steps interactive ensures better results and greater user control.

\subsection{Pipeline outputs}\label{sec:pipeline_outputs}

The final product of the \texttt{GCFinder} pipeline consists of catalogs of GC candidates for each band, with information on coordinates and magnitudes. In addition, a file is generated with the number of GC candidates in each band, to facilitate the visualization of the results.
The pipeline also provides intermediate catalogs at the end of each execution step and provides figures like those presented in the previous sections with the criteria adopted in each selection. This allows the user to have control of what is done during the code execution and have access to partial results.

\section{Other methods tested for dealing with the host galaxies} \label{different_methods}

Here we present more details about the different methods tested to detect globular clusters in J-PLUS images.

\subsection{ELLIPSE and BMODEL method}\label{ellipse_and_bmodal_1}

ELLIPSE and BMODEL \citep{iraf1993} are packages widely used with the objective of removing extended light profiles of galaxies so that globular clusters can be detected. ELLIPSE is an IRAF task that adjusts elliptical isophotes in images of galaxies, having as input a set of parameters based on the geometry of the object that is modeled
and as output a table with information about the fit. BMODEL creates a noise-free two-dimensional photometric model of the galaxy built from the data table generated by ELLIPSE. 
After making the model of the galaxy, the Imarith task present in IRAF \citep{iraf1993} was used to subtract from the original image of the galaxy the constructed light model to obtain a residual image from which GC candidates would be selected. All input parameters were determined through several tests with the images and visually evaluating the quality of the residual image obtained. 

As mentioned in the literature \citep{Ciambur2015}, we observe that the simpler the galaxy, the better the residual image formed from this process. This is because the galaxy modeling potential of ELLIPSE and BMODEL is greater for simple galaxies. Since the objective of this work was to create a pipeline as automated as possible, ELLIPSE was not used interactively in this work.

\subsection{ISOFIT and CMODEL method}\label{isofit_and_cmodel_1}

ISOFIT and CMODEL \citep{Ciambur2015} are new versions of ELLIPSE and BMODEL respectively. Thus, the functioning of these IRAF tasks is analogous to what was presented in the previous section. The modeling of more complex galaxies carried out with ELLIPSE and BMODEL has limitations when the object of study is a galaxy with a certain degree of complexity (such as arms or bars). 

However, the updated versions of ISOFIT and CMODEL generate smaller residues for the cases of galaxies that present a more complex structure \citep{Ciambur2015}. ISOFIT is more efficient than ELLIPSE for modeling more complex galaxies because there is the possibility of working with higher harmonics \citep{jedrzejewski1987ccd,bender1988isophote,Ciambur2015}. The number of maximum possible harmonics can be influenced by the quality of the images studied \footnote{https://github.com/BogdanCiambur/ISOFIT}.

Input parameters were obtained in the same way as with ELLIPSE, just selecting input parameters without using the interactive mode and performing several tests to determine the parameters that would generate the best possible residual images. We observe in this work that ISOFIT and CMODEL create better residual images, which improve our detections of GC candidates.

\subsection{GALFITM method}\label{galfitm_1}

\begin{figure*}
\centering
\includegraphics[width=7cm]{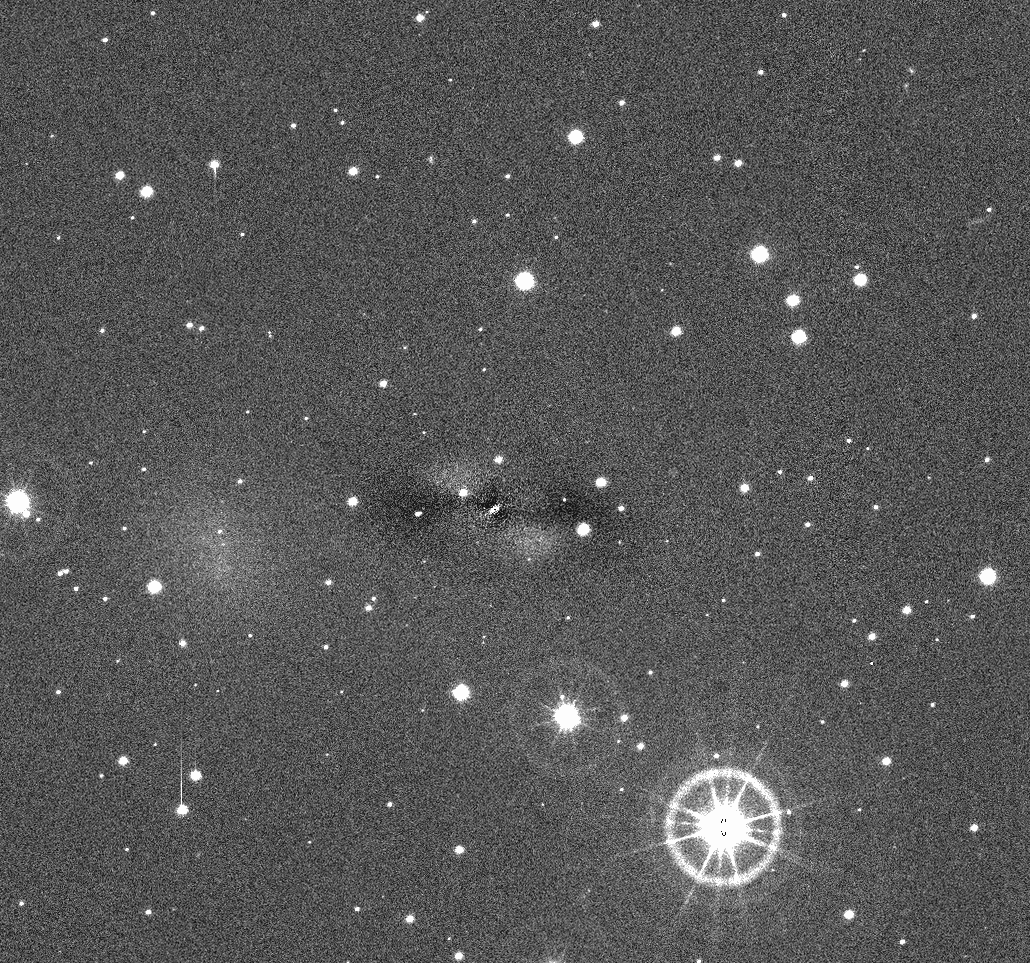}
\caption{Residual image produced by the \texttt{GALFITM} of the galaxy NGC 1023 observed in the $u$ band. The FoV is $\approx$ 0.02 $\deg^2$.}.
\label{fig:galfitm_ex}
\end{figure*}

GALFITM \citep{Bamford2011} is a software that models and removes the light profile of galaxies. It was developed within the context of the Megamorph project \citep{Bamford2011}, which created new versions of the GALFIT \citep{Peng2002} and GALAPAGOS \citep{Barden2012} tools, which are capable of modeling galaxies in multiple bands. Thus, the Megamorph project software builds galaxy models depending on the wavelengths of the filters used. Furthermore, GALFITM encompasses the morphology of galaxies in the modeling process. Several tests were performed with different input parameters and the quality of the residual images obtained was visually assessed. An example of a residual image obtained using this method is presented in Figure \ref{fig:galfitm_ex}.

GALFITM is much more automatic than ELLIPSE
and ISOFIT, so that among all the software studied to model and remove the light profile of the galaxy, it proved to be the most suitable for this work. Due to the processing power required to create the models, we use the uv100 machine from the Astroinformatics Laboratory at IAG-USP \footnote{https://lai.iag.usp.br/} for the Brazilian astronomical community (LAi). Using the large processing capacity at LAi, we see that GALFITM required at least 2 hours to process a dataset if an exponential disk profile was used. 
After performing tests with GALFITM, we conclude that non-parametric alternatives should be tested for this work, since using classic interactive software was unfeasible for our purpose. Using classic automatic software also requires a lot of time and processing power from the machine used, adding an unwanted level of complexity to the pipeline. As additional methods, we test median smoothing and direct detection of objects in the image using only Source Extractor.

\subsection{Median Smoothing method}\label{sec:median_smoothing_1}

The median smoothing technique consists of applying a filter to an astronomical image. In this technique, windows are created on the studied images and the central pixel is replaced by the median of all pixels in the window. To perform median smoothing, the Median feature of the IRAF package was used \citep{iraf1993}. The window size (xwindow and ywindow parameters) chosen was 25 for each axis.

\begin{figure*}
\centering
\includegraphics[width=18cm]{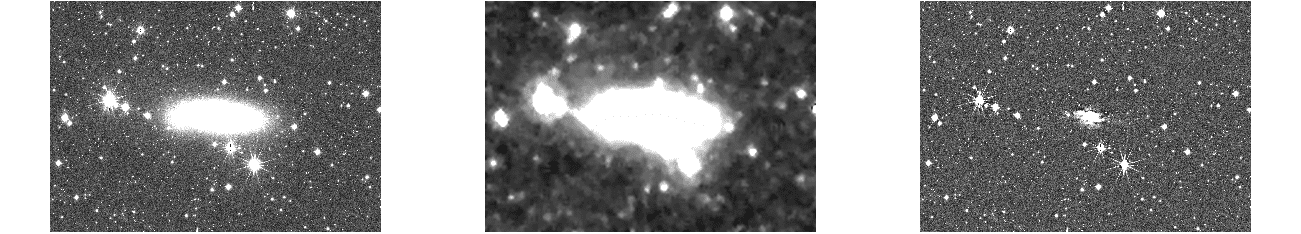}
\caption{Left panel: original image. Central panel: median of the original image produced with the \texttt{MEDIAN} tool from the \texttt{IRAF} package. Right panel: residual image after median subtraction. The FoV is $\approx$ 0.1 $\deg^2$.}.
\label{fig:median_smoothing}
\end{figure*}

Once the filtered image was created, it was subtracted from the original image using the Imarith resource, also from the IRAF package. As a result, an image without the extended light profile of the galactic halo was obtained. In such an image, however, the central part of the galaxy is not removed. An example of a residual image obtained using this method is presented in Figure \ref{fig:median_smoothing}.

\subsection{The Source Extractor method}\label{sec:source_extractor_method_1}

The Source Extractor \citep{Bertin1996} performs photometry of astronomical objects (mainly extragalactic ones) and is used in the JYPE pipeline \citep{cristobal+14}, responsible for the J-PLUS data reduction. By using it in our pipeline, we ensure that the catalogs produced by our code are compatible with JYPE's catalogs. The methodology applied when using this method is described in detail in Appendix \ref{sec:methods_gcfinder_appendix}.

By adopting adequate values of BACK$\_$SIZE, BACKFITLER$\_$SIZE, and PHOT\_AUTOPARAMS, we realize that Source Extractor does not consider the extended light profile of the galactic halo in its detection, making it possible to recover the GC candidates in this region. This methodology works because the background subtraction of \texttt{Source Extractor} first estimates the mode using the median of each cell of size BACK$\_$SIZE and then computes the median from those modes in larger cells of size BACK$\_$FILTERSIZE $\cdot$ BACK$\_$FITLERSIZE. This final combination of mode and median filtered image is then subtracted from the original one before the detection procedure. Therefore, the \texttt{Source Extractor} method is, in a sense, a combination of mode and median smoothing procedure.

\subsection{Comparison among methods}\label{comparison_of_methods_1}

As mentioned before in this article, NGC\,1023 has a companion that appears overlapped to it in the image, which makes this galaxy challenging to model.  We consider GALFITM the software more suitable for our proposes since it is more automatic than ISOFIT and ELLIPSE and because it builds the profile of the galaxies in all bands together. On the other hand, GALFITM needs considerable time to run. Due to the speed and ease of the median smoothing technique and the Source Extractor method, we decide to investigate them based on their efficiency in detecting GCs.

\begin{figure*}
\centering
\includegraphics[width=7.5cm]{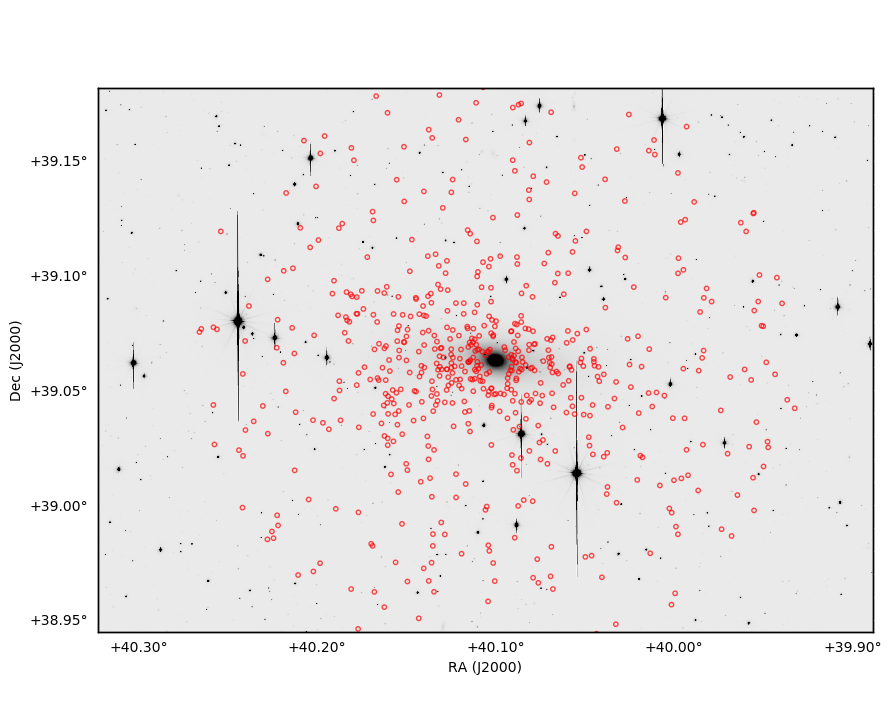}
\includegraphics[width=7.5cm]{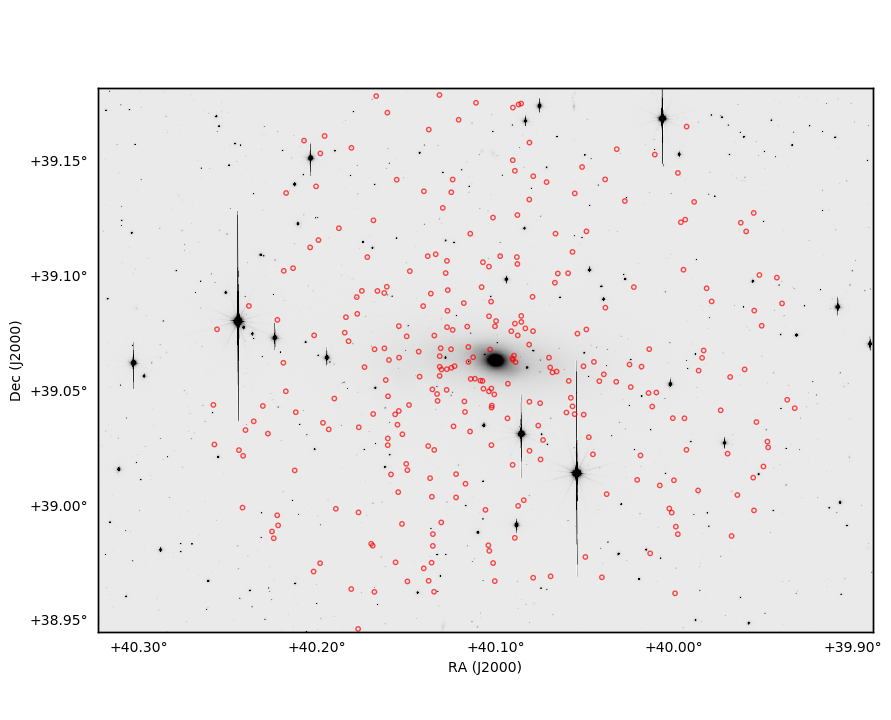}
\includegraphics[width=7.5cm]{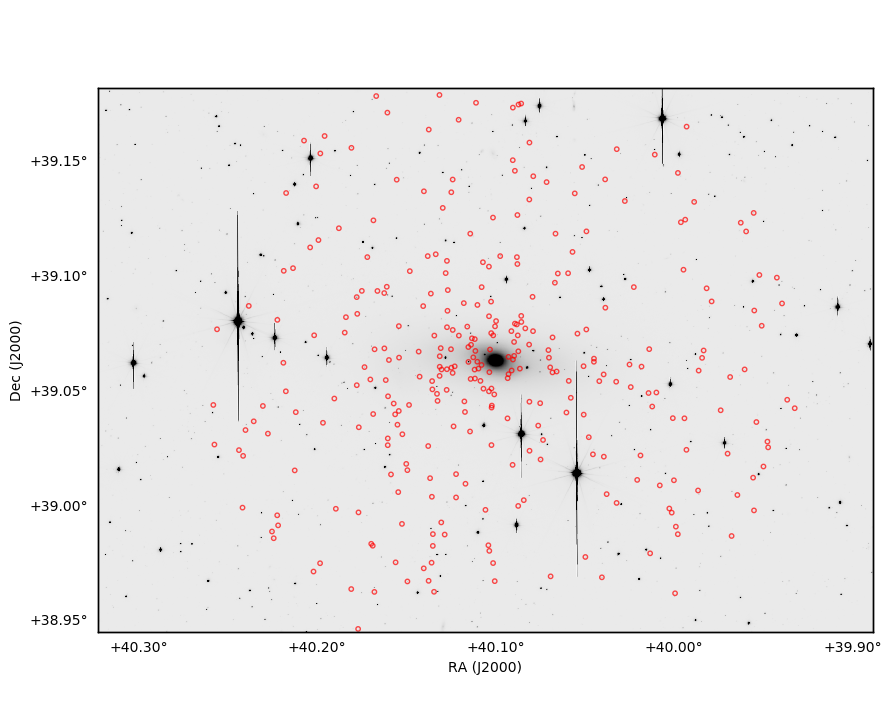}
\includegraphics[width=7.5cm]{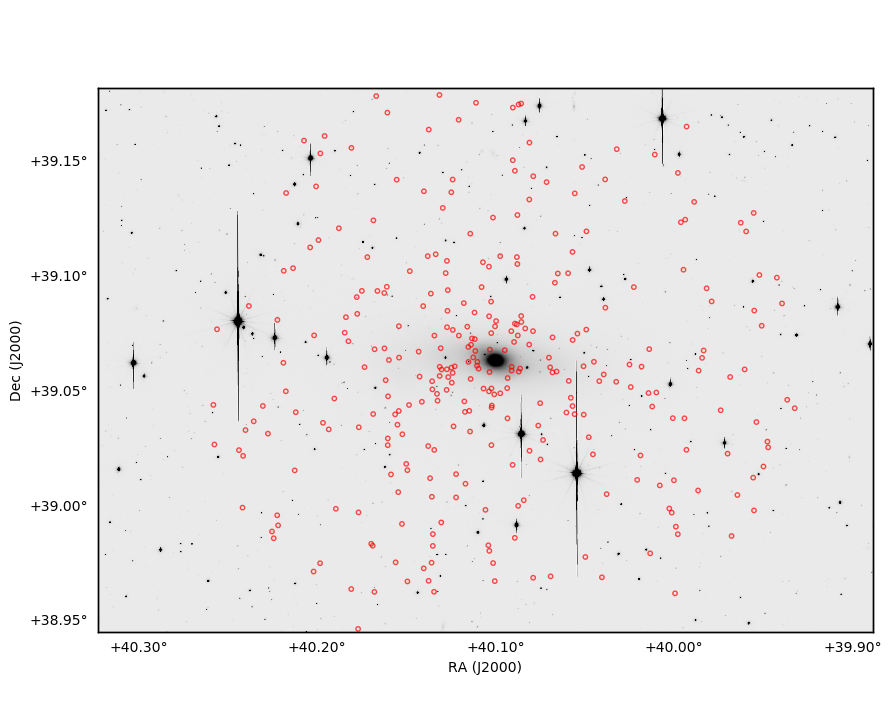}
\caption{Top left panel: GCs present in the reference catalog from \cite{Kartha2014}. Top right panel: GCs detected using the Source Extractor method that are also presented in \cite{Kartha2014}. Bottom left panel: GCs detected using the median smoothing method that are also found in \cite{Kartha2014}. Bottom right panel: Objects detected using ISOFIT and CMODEL method \citep{Ciambur2015} that are also reported in \cite{Kartha2014}. In the background we show the white image of  NGC\,1023 used for the detection of sources. The FoV is $\approx$ 0.1 $\deg^2$.}
\label{fig:detection_methods_example}
\end{figure*}

The detection efficiency consists of how many GCs the method can recover in the extended light region of the host galaxy's halo. To perform this comparison, we use the catalog of GC candidates by \cite{Kartha2014}, with 627 objects. We use the results obtained with ISOFIT and CMODEL \citep{Ciambur2015} to represent the traditional methods of modeling and removing the light profile of the galaxy, as the residual images obtained with this method had higher quality. We observe that when removing the light profile from NGC\,1023 using ISOFIT and CMODEL we can detect 314 GCs from the reference catalog, when we do median smoothing we find 317 objects, and finally when we use the Source Extractor method we detect 297 objects in the white image. These numbers correspond to detections made in the white image, without doing any selection using \texttt{GCFinder}. 
We observe that the difference in GC candidates detected in the different methods is less than 7\%. The undetected GC candidates are found near the center of the galaxy in the image, a region where, modeling or not the light profile of the galaxy, we are not able to detect many GCs, as presented in Figure \ref{fig:detection_methods_example}. Therefore, we conclude that the \texttt{Source Extractor} method is more adequate for our objectives of developing a semiautomatic pipeline that can be used to study big amounts of data since it is the simplest method and the loss of GC candidates is small.

We note that masking the objects in the image would allow us to improve significantly the models built using the tested packages (therefore we would obtain better residual images for detection of objects), but we choose to not perform this step in our analysis to simplify the steps done and also because it would be challenging to implement this step in a semiautomatic pipeline. For more details about the effect of using masks, please see \cite{varela2009wings}.

\section{Color bimodality analysis: BIC and ICL values}\label{appendix:bicicl}

Table \ref{tab:bicaic_info} show the BIC and ICL values for GMM with one and two components, following the results presented in Section \ref{sec:photometry_methodology}.

\begin{table*}
\caption{BIC and ICL values for GMM with one and two components for each color.} 
\label{tab:bicaic_info}
\centering
\begin{tabular}{l c c c c }
\hline\hline
  \multicolumn{1}{c}{Color} &
  \multicolumn{1}{c}{BIC - 1 component} &
  \multicolumn{1}{c}{BIC - 2 components} &
  \multicolumn{1}{c}{ICL - 1 component} &
  \multicolumn{1}{c}{ICL - 2 components} \\ \hline
  u-g & 1268 & 1252 & 1268 & 1277\\
  u-r & 1282 & 1262 & 1282 & 1286\\
  u-i & 1312 & 1304 & 1312 & 1331\\
  u-z & 1313 & 1311 & 1313 & 1341\\
  g-r & -56 & -92 & -56 & -90\\
  g-i & 613 & 617 & 613 & 627\\
  g-z & 946 & 968 & 946 & 981\\
  r-i & 343 & 341 & 343 & 350\\
  r-z & 792 & 750 & 792 & 759\\
  i-z & 793 & 738 & 793 & 747\\
  J0378-J0410 & 944 & 963 & 944 & 998\\
  J0378-J0430 & 1077 & 1062 & 1077 & 1088\\
  J0378-J0515 & 1211 & 1174 & 1211 & 1196\\
  J0378-J0660 & 1332 & 1300 & 1332 & 1327\\
  J0378-J0861 & 1268 & 1226 & 1268 & 1250\\
  J0410-J0430 & 993 & 990 & 993 & 1018\\
  J0410-J0515 & 1108 & 1092 & 1108 & 1115\\
  J0410-J0660 & 1186 & 1206 & 1186 & 1241\\
  J0410-J0861 & 1165 & 1162 & 1165 & 1185\\
  J0430-J0515 & 1278 & 1262 & 1278 & 1289\\
  J0430-J0660 & 1323 & 1279 & 1323 & 1301\\
  J0430-J0861 & 1271 & 1239 & 1271 & 1261\\
  J0515-J0660 & 1375 & 1290 & 1375 & 1307\\
  J0515-J0861 & 1309 & 1205 & 1309 & 1222\\
  J0660-J0861 & 1232 & 1061 & 1232 & 1068\\
\hline
\end{tabular}
\end{table*}

\end{appendix}
\end{document}